\documentclass[english]{iopart}

\usepackage[dvips]{graphicx}
\usepackage{iopams}
\usepackage{babel}
\usepackage{units}
\usepackage{amsbsy}
\usepackage{amssymb}
\usepackage{esint}
\usepackage{hyperref}

\renewcommand{\vec}[1]{\mathbf{#1}}

\makeatother

\begin{document}

\title[Flat surface bands in 3DTI in a ferromagnetic exchange field]{Appearance of flat
  surface bands in three-dimensional topological insulators in a ferromagnetic exchange field}

\author{Tomi Paananen, Henning Gerber, Matthias G\"otte, and Thomas Dahm}

\address{Universit\"{a}t Bielefeld, Fakult\"{a}t f\"{u}r Physik, 
Postfach 100131, D-33501 Bielefeld, Germany}

\date{\today}

\ead{paananen@physik.uni-bielefeld.de}

\ead{thomas.dahm@uni-bielefeld.de}

\begin{abstract}
We study the properties of the surface states in three-dimensional topological
insulators in the presence of a ferromagnetic exchange field. 
We demonstrate that for layered
materials like Bi$_2$Se$_3$ the surface states on the top surface behave
qualitatively different than the surface states at the side surfaces.
We show that the group velocity of the surface states can be tuned by
the direction and strength of the exchange field. If the exchange field
becomes larger than the bulk gap of the material, a phase transition
into a topologically nontrivial semimetallic state occurs. In particular,
the material becomes a Weyl semimetal, if the exchange field possesses
a non-zero component perpendicular to the layers. Associated with
the Weyl semimetallic state we show that Fermi arcs appear at the
surface. Under certain circumstances either one-dimensional or
even two-dimensional surface flat bands can appear. We show that the
appearence of these flat bands is related to chiral symmetries
of the system and can be understood in terms
of topological winding numbers.
In contrast to previous systems that have been suggested to
possess surface flat bands, the present system has a much
larger energy scale, allowing the observation of surface
flat bands at room temperature. The flat bands are tunable in the
sense that they can be turned on or off by rotation of the
ferromagnetic exchange field.
Our findings are supported by both
numerical results on a finite system as well as approximate analytical
results.
\end{abstract}



\maketitle

\section{Introduction}

A topological insulator is a material with an insulating
energy gap in its bulk, but possesses conducting surface states
due to significant spin-orbit coupling.
The existence of these surface states is guaranteed by a
topological invariant making them particularly robust
against time-reversal invariant perturbations or disorder. 
Due to spin-orbit coupling the surface states are 
spin-momentum coupled allowing for interesting potential
spintronics applications. The dispersion of the surface
states forms a Dirac cone, i.e. the conduction electrons
at the surface are effectively massless. This peculiar state of
matter has first been suggested theoretically \cite{Bernevig,Fu} and 
afterwards confirmed experimentally
\cite{Koenig,Hsieh1,Chen,Xia,Hsieh2,Kuroda}.
The number of materials identified as three dimensional topological insulators
(3DTI) is steadily increasing
\cite{Chen,Xia,Hsieh2,Kuroda,Chadov,Lin,Hasan,Ando}.

In the present work we study 3DTI in the presence of a ferromagnetic exchange 
field. Experimentally, it has been demonstrated that such fields can be 
introduced into topological insulators either by doping with ferromagnetic 
dopants~\cite{Yu:Science10,Chen2,Hor} or by proximity to ferromagnetic 
materials \cite{Moodera}.
As the ferromagnetic exchange field breaks time-reversal symmetry, it 
allows for
controlled modification or removal of the surface states and
could lead to interesting effects or devices
\cite{Garate,Yokoyama,Yu:Science10,BlackSchaffer}.
In previous work it has been pointed out that depending on the relative
orientation of the exchange field with respect to the surface of
the topological insulator, the surface states may open an
energy gap or remain intact \cite{RLChu,Honolka,Scholz,Lunde}. 
However, only small exchange splittings have been considered. 
In the present work we are considering
exchange splittings up to the order of the gap of the topological
insulator, which is up to 0.3~eV for present materials.
Exchange splittings of the order of 1~eV can be reached with ferromagnetic 
materials.

In a recent work we have studied a two dimensional thin strip of a 
particle-hole symmetric topological insulator and found that under 
exchange fields of such strength
an edge state flat band can appear \cite{PD}. Such flat bands are of
particular interest, because the group velocity vanishes 
allowing highly localized wave packets. Flat bands have been found
previously in other condensed matter systems like graphene,
superfluid $^3$He, or unconventional superconductors 
\cite{Nakada,Machida,Silaev,SchnyderTimmPRL,BrydonNJP,PALee,Tewari,Lau,Hu,Tanaka,Kashiwaya,Golovik,Volovik}.
In particular, the appearence of flat bands in $d$-wave superconductors
as surface Andreev bound states has been studied well in the past both
theoretically and experimentally
\cite{Hu,Tanaka,Kashiwaya,RyuHatsugai,Sato,Fogelstroem,Walter,Aprili,Krupke,Chesca1,Chesca2}.
Such surface flat bands have been shown to lead to an enhanced barrier
for vortex entry \cite{Iniotakis,Graser} or increased nonlinear
electromagnetic reponse \cite{Barash,Zare,Zhuvarel}.
However, it remained an open question whether flat bands
may also appear in three dimensional topological insulators
under sufficiently strong ferromagnetic exchange fields.

In the present work, we will present a systematic investigation of the possible
surface states of 3DTI and their behavior under a ferromagnetic exchange field.
We will show under which circumstances surface flat bands appear.
In particular, we identify a case in which a two dimensional flat
band can be generated. We demonstrate that the appearence of our flat bands 
can be understood in terms of a classification recently
proposed by Matsuura et al \cite{Matsuura} using a topological
invariant in the presence of a chiral symmetry.
We will also show that the exchange field can produce highly anisotropic
Dirac cones, i.e. that the group velocity is different in different directions.
In this case the velocity can be tuned by rotation of the exchange field,
i.e. rotation of the remanent magnetization of the ferromagnetism.

Recently the possibility of realizing a Weyl semimetallic state
in pyrochlore iridates has raised a lot of interest \cite{Wan}.
This state is a generalization of the two-dimensional 
Dirac electrons in graphene to a three-dimensional bulk system.
In a Weyl semimetal conduction band and valence band touch
each other only at a finite number of points. These so-called
Weyl nodes are exceptionally stable for topological reasons.
Of particular interest are the surface states of a Weyl semimetal
which may form open Fermi ``arcs'' \cite{Wan,Balents,Burkov}. 
In the present work we show that a 3DTI with a sufficiently strong ferromagnetic exchange field
becomes a Weyl semimetal in most cases. The surface flat bands are
directly related to the appearence of surface Fermi arcs in this system.

In contrast to previous proposals for surface flat bands in other systems
like graphene, superfluid $^3$He, or unconventional superconductors
the present system has the advantage that the relevant energy
scale is much larger ($\sim$ 0.3~eV). This allows observation
of the flat bands at room temperature, while for all other
previous proposals cryogenic temperatures are necessary.
In addition, the surface states
can be tuned by rotation of the exchange field. For example, the
flat band can be turned on and off by a rotation of the remanent
magnetization of a ferromagnet. We demonstrate that such behavior
can be achieved for realistic material parameters leading to
new possible spintronic devices.
 
\section{Models}

As a starting point we consider the generic effective two-orbital
Hamiltonian for a three dimensional topological insulator suggested already
in several previous works \cite{Zhang:NPhys09,Liu:PRB10,Li:NPhys10,Shan}. To facilitate
numerical calculations we choose the lattice regularized version that has
been suggested by Li et al \cite{Li:NPhys10}:
\begin{equation} \label{eq:hamiltonian}
	H(\mathbf k)=\epsilon_0(\mathbf k) \mathbb{I}_{4 \times 4} +
\sum_{i=0}^3 m_i(\mathbf k) \Gamma^i + \sum_{\alpha \in \{x,y,z \}} V_{\alpha}\Gamma_{\alpha}
\end{equation}
Here, $\epsilon_0(\mathbf k) = C + 2 D_2 ( 1- \cos k_x )+ 2 D_2 ( 1- \cos k_y )+ 2 D_1 ( 1- \cos k_z )$,
$m_0(\mathbf k) = M - 2 B_2 ( 1- \cos k_x ) - 2 B_2 ( 1- \cos k_y )- 2 B_1 ( 1- \cos k_z )$,
$m_1(\mathbf k) = 2 A_2 \sin k_x$, $m_2(\mathbf k) = 2 A_2 \sin k_y$, and $m_3(\mathbf k) = 2 A_1 \sin k_z$.
The Dirac $\Gamma$ matrices are represented by
$\Gamma^{0,1,2}=( \mathbb{I}_{2 \times 2}\otimes \tau_x , \sigma_x \otimes \tau_z, 
\sigma_y \otimes \tau_z)$ in the spin-orbit basis. 
Here, the Pauli matrices in orbital space
are denoted by $\tau_i$ and the ones in spin space by $\sigma_i$.
As has been pointed out by Hao and Lee~\cite{Hao:PRB11}, there exist
the following two different choices for $\Gamma^3$:
$\Gamma^3_I=\mathbb{I}_{2 \times 2}\otimes \tau_y$ and $\Gamma^3_{II}=\sigma_z \otimes \tau_z$.
These correspond to two different types of spin-orbit coupling in $z$-direction.
We follow the convention of Hao and Lee and denote these two choices as 
model I and model II, respectively. 
Note that if $k_z=0$ there is no difference between the two models. 
Model I is isotropic within the $xy$-plane, but the coupling in $z$-direction
is different. Thus one has qualitatively different behavior of surface states 
at a $z$-boundary than at an $x$- or $y$-boundary. For model II the spin-orbit
coupling is isotropic and it is sufficient to consider surface states
at one selected boundary, because the qualitative behavior is the same 
in all three spatial directions. It has been discussed in
Ref.~\cite{Liu:PRB10} that in the absence of an exchange field model I and 
model II are related by a unitary transformation and a 90 degree rotation
within the $xy$-plane. However, this is not true anymore in the
presence of an exchange field, because the spin operators are mapped to
pseudospin operators under the unitary transformation as has been
pointed out in Ref.~\cite{Brouwer}. Thus, for the purpose of the
present work the two models become different in the presence of an
exchange field. Model I is appropriate for Bi$_2$Se$_3$ and its
relatives.

The components of the ferromagnetic exchange field in $x$, $y$, and 
$z$-direction are denoted by $V_{x,y,z}$, respectively, and are modeled
by Zeeman terms in the Hamiltonian Eq.~(\ref{eq:hamiltonian}). 
The matrices for the exchange field components are given 
by $\Gamma_{\alpha}=\sigma_{\alpha} \otimes  \mathbb{I}_{2 \times 2}$.
 The parameters $A_1$, $A_2$, $B_1$, $B_2$, $C$,
$D_1$, $D_2$, and $M$ have been derived from bandstructure calculations
for the Bi$_2$Se$_3$ family of materials in
Refs.~\cite{Zhang:NPhys09,Liu:PRB10}.
In our numerical calculations we will consider the case $A_i \ge M >0$ and
$B_i \ge M$ as is relevant for these materials.

\section{Symmetry considerations}
\label{Secsymmetries}

Let us first discuss certain symmetries of the Hamiltonian
(\ref{eq:hamiltonian}) that will be of particular importance in this work.
To begin with we focus on the particle-hole symmetric case $C=D_1=D_2=0$,
in which the effects become particularly clear and the topological invariant
proposed by Matsuura et al \cite{Matsuura} can be used. In Section
\ref{Secbrokenph} we will discuss the modifications that appear when
particle-hole symmetry is slightly broken, as is the case in the
Bi$_2$Se$_3$ family of materials.

For $C=D_1=D_2=0$ the four bulk bands for model I can be found by
analytical diagonalization of the $4 \times 4$ matrices and are given by
\begin{equation} \label{eq:model1bulk}
	E_i^I(\mathbf k)=\pm \sqrt{m^2+V^2 \pm
        2 \sqrt{\left( m_0^2+m_3^2 \right) V^2
        + \left( m_1 V_x + m_2 V_y \right)^2 }}
\end{equation}
while for model II we have
\begin{equation} \label{eq:model2bulk}
	E_i^{II}(\mathbf k)=\pm \sqrt{m^2+V^2 \pm
        2 \sqrt{m_0^2 V^2
          + \left( m_1 V_x + m_2 V_y + m_3 V_z \right)^2 }}
\end{equation}
where $V^2=V_x^2+V_y^2+V_z^2$ and $m^2=m_0^2+m_1^2+m_2^2+m_3^2$.
As is clear from these expressions, the bulk spectrum is fully symmetric around
energy $E=0$ for both models.

The $\Gamma$ matrices
introduced in the previous section respect the following
commutation and anti-commutation relations
\begin{eqnarray}
& \{\Gamma^i,\Gamma^j\}=2\delta_{ij},\\
& [\Gamma_x,\Gamma^0]=[\Gamma_x,\Gamma_I^3]=\{\Gamma_x,\Gamma_{II}^3\}=[\Gamma_x,\Gamma^1]=\{\Gamma_x,\Gamma^2\}=0,\\
& [\Gamma_y,\Gamma^0]=[\Gamma_y,\Gamma_I^3]=\{\Gamma_y,\Gamma_{II}^3\}=[\Gamma_y,\Gamma^2]=\{\Gamma_y,\Gamma^1\}=0,\\
& [\Gamma_z,\Gamma^0]=[\Gamma_z,\Gamma_I^3]=[\Gamma_z,\Gamma_{II}^3]=\{\Gamma_z,\Gamma^1\}=\{\Gamma_z,\Gamma^2\}=0,
\label{eq:commu_anti_commu_mod_I}
\end{eqnarray}
where $\delta_{ij}$ is Kronecker's delta symbol.

In the absence of an exchange field time reversal symmetry 
is respected. However, application of an exchange field in any
direction breaks time reversal symmetry. According to the
classification by Schnyder et al \cite{Schnyderpt,Ryu} a system
with broken time reversal symmetry may still be topologically
nontrivial, if a chiral symmetry is present. Let $\Theta$ be
a chiral symmetry operator, which by definition anticommutes
with the Hamiltonian, i.e.
\begin{equation} \label{eq:anticomm1}
	\left\{ H(\mathbf k), \Theta \right\}=0
\end{equation}
If such a symmetry exists, the system falls into the AIII
chiral symmetry class \cite{Schnyderpt,Ryu,Altland}.

Let us consider the symmetry operator $\Theta_1=\sigma_x \otimes \tau_z$,
which happens to be identical to the operator $\Gamma^1$.
$\Theta_1$ anticommutes with $\Gamma^0$, $\Gamma^2$, $\Gamma_I^3$,
$\Gamma_{II}^3$, $\Gamma_y$, and $\Gamma_z$, but commutes with
$\Gamma^1$ and $\Gamma_x$. Thus, for $k_x=0$ and
an exchange field within the $y$-$z$-plane Hamiltonian
(\ref{eq:hamiltonian}) with $C=D_1=D_2=0$
possesses the chiral symmetry $\Theta_1$.

Similarly, we can consider the symmetry operator 
$\Theta_2=\sigma_y \otimes \tau_z=\Gamma^2$.
$\Theta_2$ anticommutes with $\Gamma^0$, $\Gamma^1$, $\Gamma_I^3$,
$\Gamma_{II}^3$, $\Gamma_x$, and $\Gamma_z$, but commutes with
$\Gamma^2$ and $\Gamma_y$. Thus, for $k_y=0$ and
an exchange field within the $x$-$z$-plane Hamiltonian
(\ref{eq:hamiltonian}) with $C=D_1=D_2=0$
possesses the chiral symmetry $\Theta_2$.

Alternatively, we may also consider the symmetry operator 
$\Theta_3=\sigma_z \otimes \tau_z$,
which happens to be identical to the operator $\Gamma_{II}^3$.
$\Theta_3$ anticommutes with $\Gamma^0$, $\Gamma^1$, $\Gamma^2$, $\Gamma_I^3$,
$\Gamma_x$, and $\Gamma_y$, but commutes with
$\Gamma_{II}^3$ and $\Gamma_z$. Thus, for model I with exchange field
within the $x$-$y$-plane Hamiltonian
(\ref{eq:hamiltonian}) with $C=D_1=D_2=0$
possesses the chiral symmetry $\Theta_3$.
For model II this symmetry is respected for $k_z=0$.

From these considerations we see that Hamiltonian
(\ref{eq:hamiltonian}) under certain circumstances falls into the
AIII chiral symmetry class and we have identified three important
symmetries.

\section{Nonequivalent surface boundaries}

Our aim is to calculate the energy dispersion of the surface states
for both models, for all possible nonequivalent surfaces perpendicular to
$x$-, $y$-, and $z$-directions, and for the corresponding directions of the exchange
field. In this way we will determine all possible types of
surface states that can appear in a 3DTI in a ferromagnetic exchange field.

For all cases we will present numerical calculations based
on an exact diagonalization of Hamiltonian (\ref{eq:hamiltonian}) 
on a finite size lattice of dimension $500 \times 500 \times 200$.
Periodical boundary conditions are employed parallel to the surface
with 500 $k$-modes in both directions, while open boundary conditions are used
perpendicular to the surface on 200 real space points.
Our numerical results are compared with approximate analytical results
for a continuous half space using a small $k$ expansion of
Hamiltonian (\ref{eq:hamiltonian}) near the $\Gamma$ point $\vec{k}=0$.
For the appearance of the flat bands we will check our results using the 
topological winding number proposed by Matsuura et al \cite{Matsuura}.

In total we find that we need to consider seven nonequivalent cases:
for model II the spin-orbit coupling in $z$-direction is of the
same type as in $x$- and $y$-direction. For that reason it is sufficient
to study a single boundary direction, which we choose to be a $y$-boundary,
i.e. a boundary with $y=$~const. As regards the direction of the
exchange field we have to distinguish two nonequivalent cases here: parallel and
perpendicular to the surface, i.e. $V_x \neq 0$ and $V_y \neq 0$.
In contrast, for model I we have five nonequivalent cases. For model I
the spin-orbit coupling in $z$-direction is of a different
type than in $x$- and $y$-direction, but the in-plane coupling is still
isotropic.
Therefore we need to distinguish a $z$-boundary and a $y$-boundary.
For the $z$-boundary there are again two nonequivalent directions
for the exchange field: $V_x \neq 0$ and $V_z \neq 0$. For the
$y$-boundary, however, all field directions are nonequivalent
and we have three cases here.

We will see that among these seven cases there are three in which
flat surface bands appear. One dimensional flat bands are
found for model II with $y$-boundary and exchange field in $x$-direction
as well as for model I with $y$-boundary and exchange field in $z$-direction.
A two dimensional flat band is found for model I with $y$-boundary and 
exchange field in $x$-direction.

\section{Model I}

In this section we discuss the five nonequivalent cases for the
particle-hole symmetric model I. We start with the more
interesting case of a $y$-boundary.

\subsection{Boundary perpendicular to the $y$-direction with finite $V_y$}
\label{sec:mIyVy}

In this case with $V_x=V_z=0$ the bulk energy bands Eq.~(\ref{eq:model1bulk})
simplify to the following expression:
\begin{equation} \label{eq:model1bulka}
	E_i^I(\mathbf k)=\pm \sqrt{m_1^2+\left( V_y \pm
        \sqrt{m_0^2+m_2^2 + m_3^2} \right)^2 }
\end{equation}
In the absence of an exchange field this band structure usually possesses a gap,
because $m_0$, $m_1$, $m_2$, and $m_3$ do not become zero simultaneously.
Thus, the system is insulating. However, 
the gap closes when $V_y$ reaches a critical value $V_{cr}$, which is
derived in appendix I and is of the order of $M$.
The Fermi surface at zero energy is then defined by the
two equations $m_1=0$ and $V_y^2= m_0^2+m_2^2 + m_3^2$.
These two equations define a line in three dimensional 
$(k_x,k_y,k_z)$ space. Therefore the Fermi surface is one-dimensional
and the system has entered a semimetallic state. If one looks at
the $\Gamma$ point $k_x=k_y=k_z=0$, where $m_1=m_2=m_3=0$ and
$m_0=M$ it is clear that the semimetallic
state is entered at $V_y=M$ or closely below. Thus, the parameter
$M>0$ sets the scale for the exchange field, at least for the
range of the parameters $A_i$ and $B_i$ considered here.
In appendix I we derive the ranges of the exchange field
under which the system becomes semimetallic.

To find approximate analytical solutions for the surface states
we expand Hamiltonian
(\ref{eq:hamiltonian}) up to second order in $k_y$. 
If we assume a boundary in $y$-direction the momentum $k_y$ has
to be replaced by the momentum operator $-i \partial_y$.
To find the surface states we search for nontrivial solutions 
of the Schr\"odinger equation that vanish both at $y=0$ and for
$y \rightarrow \infty$. In this case the Hamiltonian can be written as
\begin{eqnarray}
H(\bf k)&=H_0(\mathbf k)+H'(\mathbf k),
\label{eq:Ham_B_y_V_y_I}
\end{eqnarray}
where
\begin{eqnarray}
\label{eq:H_0_H'_B_y_V_y}
H_0(\mathbf k)&=(\tilde m_0(\mathbf k)+B_2\partial_y^2)\Gamma^0-i2A_2\partial_y\Gamma^2,\\
H'(\mathbf k)&=m_1(\mathbf k)\Gamma^1+m_3(\mathbf k)\Gamma_I^3+V_y\Gamma_y.
\label{eq:H_0_H'_B_y_V_y_2}
\end{eqnarray}
Here, $\tilde m_0(\mathbf k)=M - 2 B_2 ( 1- \cos k_x ) - 2 B_1 ( 1- \cos k_z
)$ and $\mathbf k=(k_x,k_z)$. $\Gamma_I^3$ anticommutes with $H_0$.
$\Gamma_y$ commutes with both $H_0$ and $\Gamma_I^3$, and
$\Gamma^1$ anticommmutes with $H_0$, $\Gamma_I^3$, and $\Gamma_y$. 
In this case the eigenstates of $H({\bf k})$ are linear combinations of (up to four) 
eigenstates of $H_0({\bf k})$ (see appendix II for a more detailed explanation). 
Surface states of $H$ are superpositions of surface states of $H_0$. 
Following the general procedure from appendix II we will first determine 
the surface states of $H_0$ and then deduce the ones of $H$ from them.

For a system of finite width in $y$-direction,
the energy of the surface states of $H_0$ behaves like $e^{-L}$ as a 
function of the system size $L$ in the $y$-direction.
Let us consider a half infinite system with a single boundary at $y=0$. 
In this case the surface states of $H_0$ have zero energy.
To find these zero energy eigenstates we can exploit that $H_0$ commutes with
the operator $\Theta_4=\sigma_x \otimes \tau_x$. Then there exist
simultaneous eigenstates of $H_0$ and $\Theta_4$. The eigenstates of
$\Theta_4$ are $(1,0,0,1)^T$ and $(0,1,1,0)^T$ with eigenvalue $+1$
and $(1,0,0,-1)^T$ and $(0,1,-1,0)^T$ with eigenvalue $-1$.
We try the following two ans\"atze:
\begin{eqnarray}
\label{eq:sur_s_B_y_V_y}
\psi_{1,\mathbf k}(y)=C(1,0,0,1)^Tf_{\mathbf k}(y),\\
\psi_{2,\mathbf k}(y)=C(0,1,-1,0)^Tf_{\mathbf k}(y),
\label{eq:sur_s_B_y_V_y_2}
\end{eqnarray}
where $C$ is a normalization constant and $f_{\mathbf k}(y)$ is solution of the equation
\begin{equation}
\label{eq:solu_f_B_y_V_y}
[\tilde m_0(\mathbf k)+B_2\partial_y^2+2A_2\partial_y]f_{\mathbf k}(y)=0,
\end{equation}
The other two eigenstates of $\Theta_4$ lead to exponentially increasing
functions
with $y$ and thus cannot fulfil the boundary condition for $y \rightarrow \infty$.
Solving the differential equation (\ref{eq:solu_f_B_y_V_y}) we
find that $f_{\mathbf k}(y)$ is given by
\begin{equation}
\label{eq:solu_f_B_y_V_y_2}
f_{\mathbf k}(y)=e^{-\frac{A_2}{B_2}y}\sinh\left(\sqrt{\frac{A_2^2}{B_2^2}-\frac{\tilde m_0(\mathbf k)}{B_2}} \;
y\right).
\end{equation}
This solution can only fulfil the boundary condition for $y \rightarrow
\infty$, if $\tilde m_0(\mathbf k) > 0$. For those $\mathbf k$ values
where this condition is not fulfilled anymore, a surface state
does not exist.

Having determined the surface states of $H_0$ we can now infer the ones
of $H$ by noting that $H'$ just couples the two solutions
Eq.~(\ref{eq:sur_s_B_y_V_y}) and (\ref{eq:sur_s_B_y_V_y_2}) as
\begin{eqnarray}
\label{eq:sur_s_B_y_V_y_3}
\Psi_{\mathbf k}(y)=(u_{\mathbf k},v_{\mathbf k},-v_{\mathbf k},u_{\mathbf k})^Tf_{\mathbf k}(y).
\end{eqnarray}
Here, $u$ and $v$ are the components of the solution spinor 
$\xi_{\mathbf k}=(u_{\mathbf k},v_{\mathbf k})^T$ of the following eigenequation
\begin{eqnarray}
\label{eq:sur_s_spin_equ_B_y_V_y_2}
\left[(V_y-m_3(\mathbf k))\sigma_y+m_1(\mathbf k)\sigma_x\right]
\xi_{\mathbf k}=E\xi_{\mathbf k}.
\end{eqnarray}
The full surface state solutions are then given by
\begin{eqnarray}
\label{eq:sur_s_B_y_V_y_4}
\Psi_{\pm,\mathbf k}(y)=\frac{C}{\sqrt{2}}(1,\pm e^{i\theta_{\mathbf k}},\mp e^{i\theta_{\mathbf k}} ,1)^Tf_{\mathbf k}(y),
\end{eqnarray} 
where
\[
\sin \theta_{\mathbf k}=\frac{V_y-2A_1\sin k_z}{\sqrt{(V_y-2A_1\sin k_z
    )^2+4A_2^2\sin^2 k_x}},
\]
with the corresponding eigenenergies
\begin{eqnarray}
\label{eq:eigen_energy_B_y_V_y_surf}
E_{\pm}(\mathbf k)&=\pm\sqrt{(V_y-2A_1\sin k_z )^2+4A_2^2\sin^2 k_x }.
\end{eqnarray}
These solutions show that we have surface states as long as $\tilde m_0(\mathbf k)>0$.
For small momenta the dispersion Eq.~(\ref{eq:eigen_energy_B_y_V_y_surf}) shows that the presence of
the exchange field in $y$-direction shifts the surface Dirac cone
in $k_z$-direction. The Dirac cone remains ungapped and its velocity is unchanged.

\begin{figure}
\begin{minipage}{0.495\textwidth}
\centering
\includegraphics[width=\textwidth]{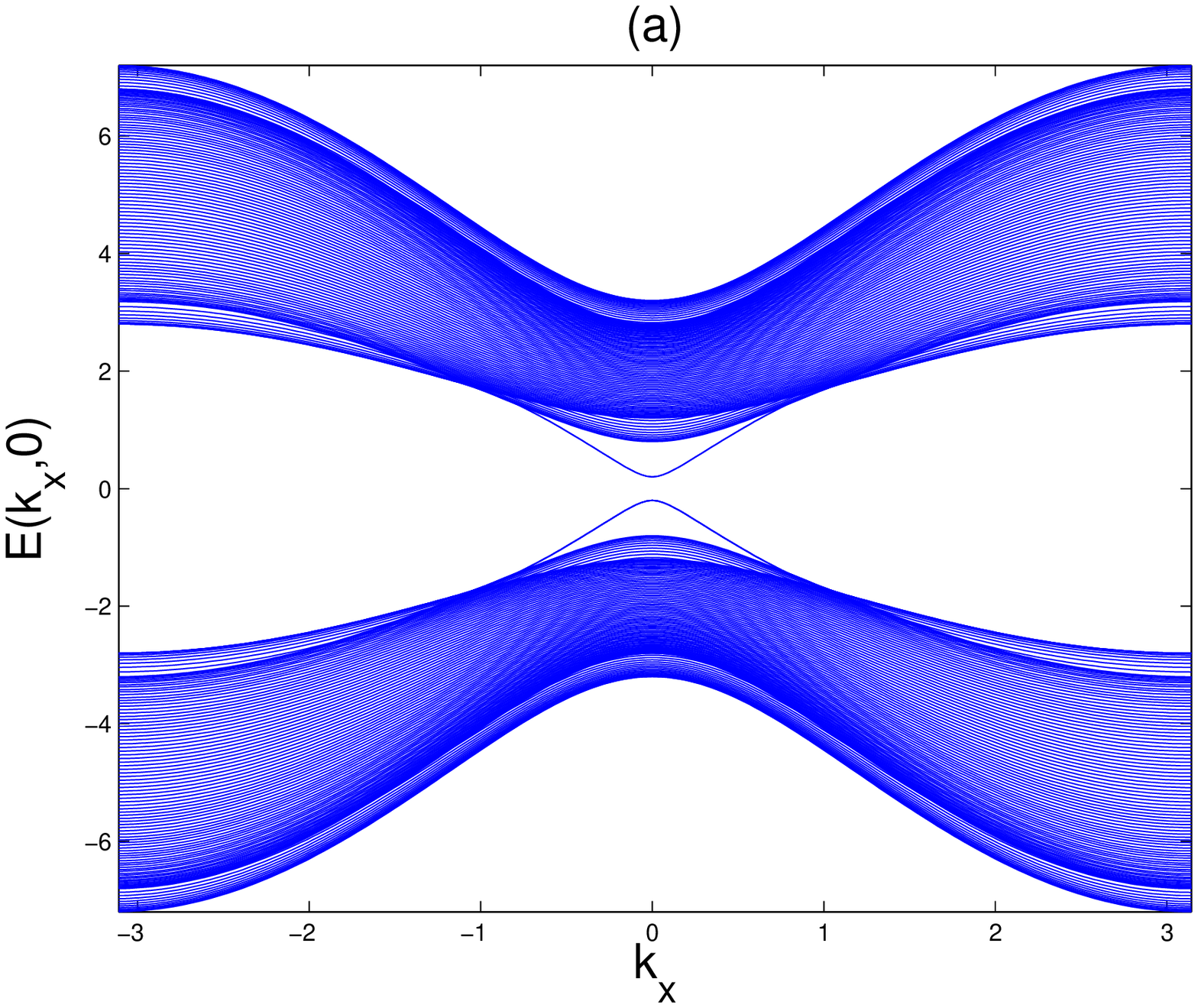}
\end{minipage}
\begin{minipage}{0.495\textwidth}
\centering
\includegraphics[width=\textwidth]{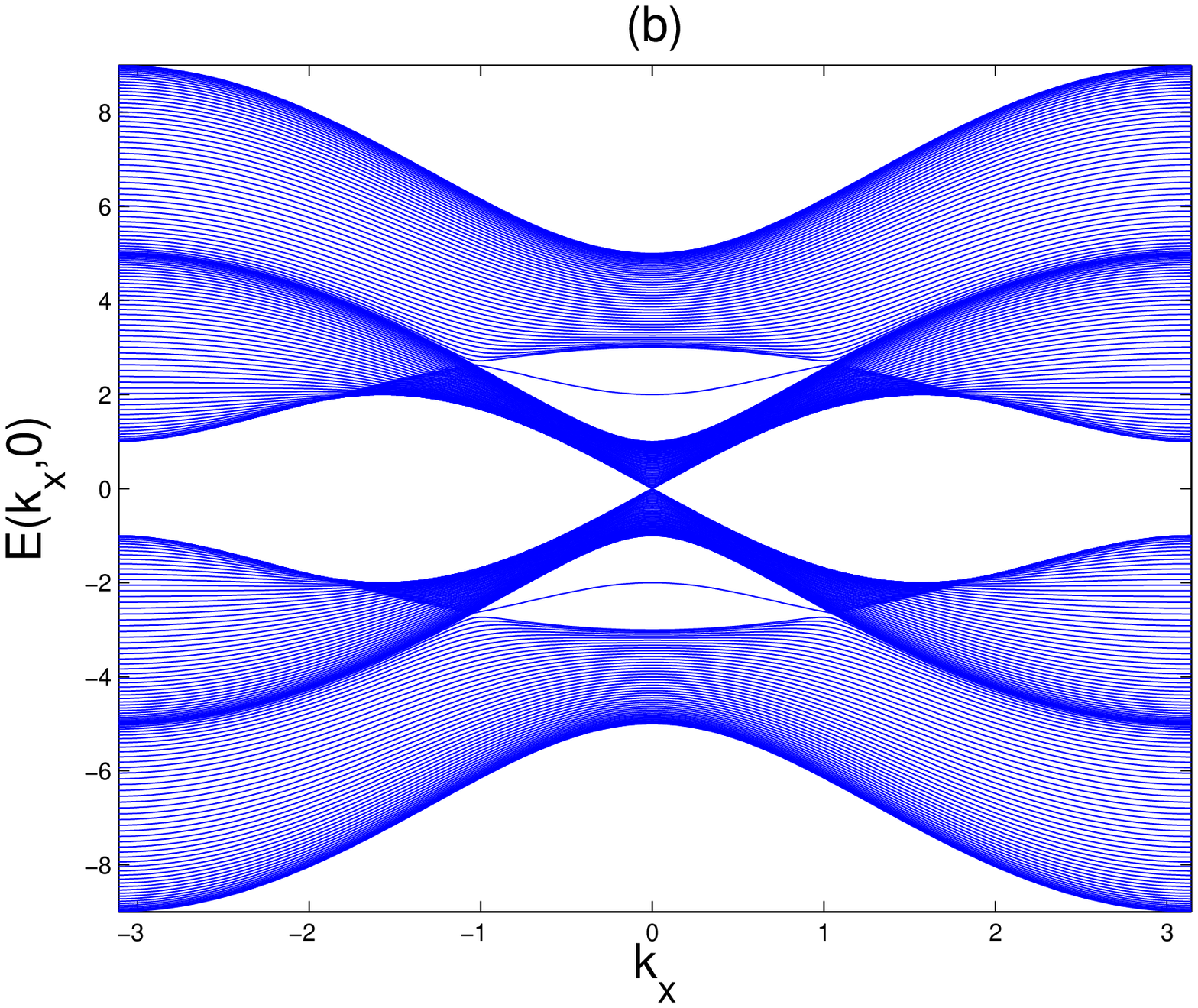}
\end{minipage}
\begin{minipage}{0.495\textwidth}
\centering
\includegraphics[width=\textwidth]{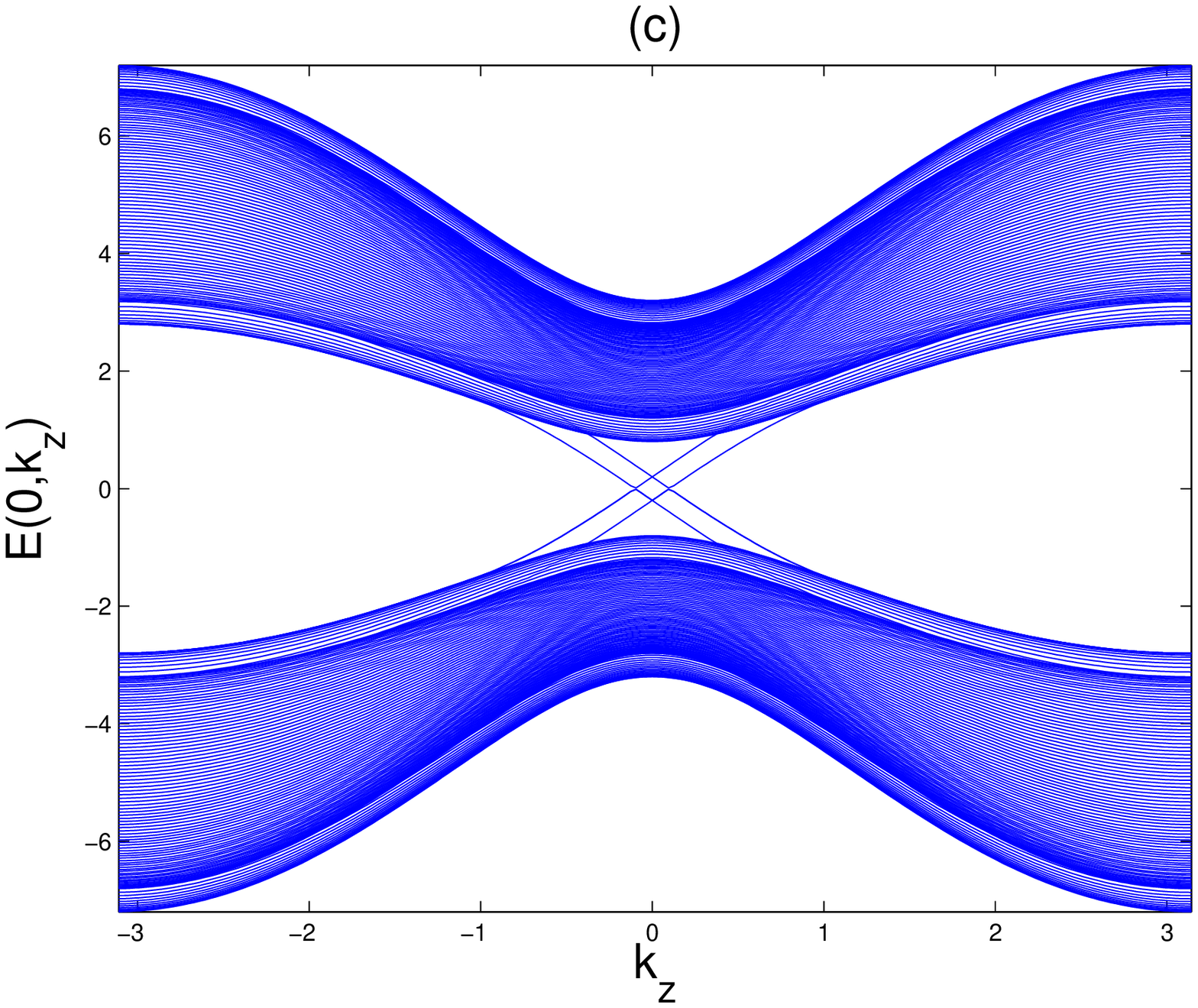}
\end{minipage}
\begin{minipage}{0.495\textwidth}
\centering
\includegraphics[width=\textwidth]{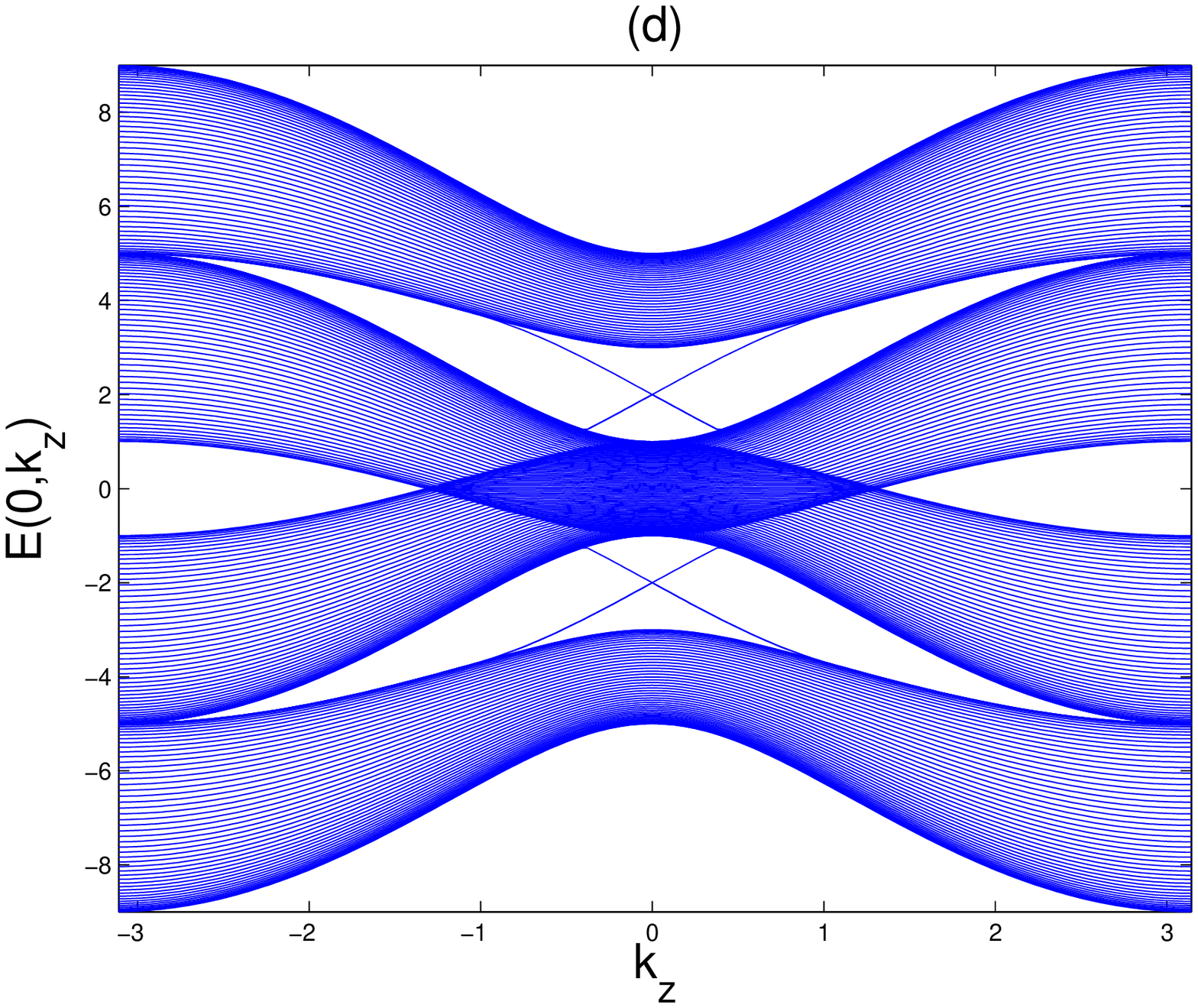}
\end{minipage}
\caption{\label{Fig1}
Numerical dispersions of bulk and surface states for model I with
$V_y/M=0.2$ for (a) and (c) and $2.0$ for (b) and (d).
In (a) and (b) $k_z=0$, and in (c) and (d) $k_x=0$.
We have used $B_1=B_2=A_1=A_2=M=1$, and $C=D=0$.
}
\end{figure}

In Fig.~\ref{Fig1} we show results obtained from numerical calculation of
the eigenenergies of a finite slab on a finite size lattice for the present 
case showing nice agreement with the analytical result. Here and in the
following we are showing results for the parameter choice
$A_1=A_2=B_1=B_2=M=1$ and $C=D_1=D_2=0$. Results for parameters appropriate
for Bi$_2$Se$_3$ are discussed in section~\ref{Secbrokenph}. 
Fig.~\ref{Fig1}(a) (for $k_z=0$) and
(c) (for $k_x=0$) show the dispersion in the insulating state for a small exchange field of $V_y/M=0.2$.
In Fig.~\ref{Fig1}(b) and (d) the semimetallic state with $V_y/M=2.0$ is shown.
In the present case no surface flat band occurs. Note, that in 
Fig.~\ref{Fig1}(c) four surface state dispersions are seen. Only two
of them correspond to Eq.~(\ref{eq:eigen_energy_B_y_V_y_surf}). The other
two dispersions are localized on the opposite surface, which is
present in the numerical calculation. These states are related by parity
to the ones found analytically above. Their dispersion is thus obtained
from Eq.~(\ref{eq:eigen_energy_B_y_V_y_surf}) by changing
$(k_x,k_z) \rightarrow (-k_x,-k_z)$, i.e.
\begin{eqnarray}
\label{eq:eigen_energy_B_y_V_y_surf_opp}
E_{\pm,2}(\mathbf k)&=\pm\sqrt{(V_y+2A_1\sin k_z )^2+4A_2^2\sin^2 k_x }.
\end{eqnarray}

We note that the surface states Eq.~(\ref{eq:sur_s_B_y_V_y_4})
possess an interesting nontrivial spin texture.
To see this we evaluate the expectation value of the
spin components in the two orbitals. For orbital 1 the
spin matrices can be written in the form
\begin{equation}
 \hat{s}_{1,i} = \sigma_i \otimes \frac{1}{2} 
\left( \mathbb{I}_{2 \times 2} + \tau_z \right)
\end{equation}
where $\sigma_i$ for $i \in \{ x,y,z \}$ are the spin Pauli matrices.
For orbital 2 we have analogously
\begin{equation}
 \hat{s}_{2,i} = \sigma_i \otimes \frac{1}{2} 
\left( \mathbb{I}_{2 \times 2} - \tau_z \right)
\end{equation}
Using Eq.~(\ref{eq:sur_s_B_y_V_y_4}) we find the following
expectation values in orbital 1:
\begin{eqnarray}
\nonumber
\left\langle \Psi_{\pm,\mathbf k} \left| \hat{s}_{1,x} \right|
\Psi_{\pm,\mathbf k} \right\rangle &= \pm \frac{1}{2} \cos \theta_{\mathbf k} \\
\nonumber
\left\langle \Psi_{\pm,\mathbf k} \left| \hat{s}_{1,y} \right|
\Psi_{\pm,\mathbf k} \right\rangle &= \pm \frac{1}{2} \sin \theta_{\mathbf k} \\
\nonumber
\left\langle \Psi_{\pm,\mathbf k} \left| \hat{s}_{1,z} \right|
\Psi_{\pm,\mathbf k} \right\rangle &= 0
\end{eqnarray}
and in orbital 2:
\begin{eqnarray}
\nonumber
\left\langle \Psi_{\pm,\mathbf k} \left| \hat{s}_{2,x} \right|
\Psi_{\pm,\mathbf k} \right\rangle &= \mp \frac{1}{2} \cos \theta_{\mathbf k} \\
\nonumber
\left\langle \Psi_{\pm,\mathbf k} \left| \hat{s}_{2,y} \right|
\Psi_{\pm,\mathbf k} \right\rangle &= \pm \frac{1}{2} \sin \theta_{\mathbf k} \\
\nonumber
\left\langle \Psi_{\pm,\mathbf k} \left| \hat{s}_{2,z} \right|
\Psi_{\pm,\mathbf k} \right\rangle &= 0
\end{eqnarray}
From these expressions we see that the spin rotates within the $x$-$y$
plane. The spin direction of the two surface states is always
opposite. The spin-$x$-component is opposite in the two orbitals,
while the spin-$y$-component is the same in the two orbitals. Therefore, the total
spin points in $y$-direction, perpendicular to the surface:
\begin{eqnarray}
\nonumber
\left\langle \Psi_{\pm,\mathbf k} \left| \Gamma_x \right|
\Psi_{\pm,\mathbf k} \right\rangle &= 0 \\
\nonumber
\left\langle \Psi_{\pm,\mathbf k} \left| \Gamma_y \right|
\Psi_{\pm,\mathbf k} \right\rangle &= \pm \sin \theta_{\mathbf k} \\
\nonumber
\left\langle \Psi_{\pm,\mathbf k} \left| \Gamma_z \right|
\Psi_{\pm,\mathbf k} \right\rangle &= 0
\end{eqnarray}
Note, that while the total spin is perpendicular to the
surface momentum, the partial spins in the two orbitals are not.

\subsection{Boundary perpendicular to the $y$-direction with finite $V_x$}

\label{subsec1vxy}
Let us consider next the case that both $V_x$ and $V_y$ are nonzero, but
still $V_z=0$. It is useful to go over to polar coordinates in this
case and write $V_x=V_0 \cos \vartheta$ and $V_y=V_0 \sin \vartheta$.
The bulk energy bands Eq.~(\ref{eq:model1bulk})
can then be brought into the following form:
\begin{eqnarray} 
\nonumber
	E_i^I(\mathbf k)&=\pm \Big\{ \left(m_1 \sin \vartheta - m_2 \cos
          \vartheta \right)^2+\\ \label{eq:model1bulkb}
&+ \left( V_0 \pm \sqrt{m_0^2 + m_3^2 + \left(m_1 \cos \vartheta + m_2 \sin
          \vartheta \right)^2} \right)^2 \Big\}^{1/2}
\end{eqnarray}
Compared with Eq.~(\ref{eq:model1bulka}) this corresponds to a rotation
of the exchange field within the $x$-$y$-plane. Again, the system is
insulating in the absence of an exchange field and the gap closes,
when $V_0$ reaches the critical value $V_{cr}$. In the semimetallic
state the Fermi surface is defined by the two equations
$m_1 \sin \vartheta - m_2 \cos \vartheta = 0$ and
$V_0^2 = m_0^2 + m_3^2 + \left(m_1 \cos \vartheta + m_2 \sin
          \vartheta \right)^2 $.
Therefore, the Fermi surface is still one-dimensional.

Determination of the surface states becomes more difficult now, because 
$\Gamma_x$ neither commutes nor anticommutes with $H_0$ in 
Eq.~(\ref{eq:H_0_H'_B_y_V_y}). 
As a result, it affects the spatial part of the surface states.
We can, however, determine the surface states of the Hamiltonian 
\begin{equation}
\label{eq:H'_0_V_x}
H'_0(\mathbf k)=H_0(\mathbf k)+V_x\Gamma_x \, .
\end{equation}
$H'_0(\mathbf k)$ still commutes with $\Theta_4=\sigma_x \otimes \tau_x$
and we can thus look for zero energy states of $H'_0(\mathbf k)$ 
using the same ansatz Eq.~(\ref{eq:sur_s_B_y_V_y}) and (\ref{eq:sur_s_B_y_V_y_2}) as before.
The surface state solutions of $H'_0$ (with a boundary at $y=0$ and 
a half infinite system as before) are then found to be
\begin{eqnarray}
\label{eq:surf_solu_B_y_V_x_k_0}
\psi_{1,\mathbf k}(y)&=C(1,0,0,1)^Te^{-\frac{A_2}{B_2}
  y}\sinh\left(\sqrt{\frac{A_2^2}{B_2^2}-\frac{\tilde m_0(\mathbf k)+V_x}{B_2}} \; y\right),\\
\psi_{2,\mathbf k}(y)&=C'(0,1,-1,0)^Te^{-\frac{A_2}{B_2}
  y}\sinh\left(\sqrt{\frac{A_2^2}{B_2^2}-\frac{\tilde m_0(\mathbf k)-V_x}{B_2}} \; y\right),
\label{eq:surf_solu_B_y_V_x_k_0_2}
\end{eqnarray}
where $C$ and $C'$ are normalization constants. It is clear that state 
$1$ exists only if $\tilde m_0(\mathbf k)+ V_x>0$ and
state $2$ exists only if $\tilde m_0(\mathbf k)- V_x>0$. 

The matrices $\Gamma^1$, $\Gamma_I^3$, or $\Gamma_y$ neither commute nor 
anticommute with $H'_0$. Thus they affect the spatial part of the 
surface states as well as the spin part.
In this case we cannot separate the Hamiltonian into parts. However, 
if $|\mathbf k|$ and $V_y$ are small we can treat $H'$
Eq.~(\ref{eq:H_0_H'_B_y_V_y_2}) as a perturbation using degenerate
perturbation theory.
We assume that the perturbation only couples the two surface states $1$ and $2$ 
to each other and neglect coupling to bulk states.
This is a reasonable assumption, as the surface states are well localized and
in most cases well separated in energy from the bulk states. In this case 
the overlap between the bulk states and the surface states is very small. 
Also, due to the symmetric energy spectrum around $E=0$ for each bulk state
with energy $E$ there exists another one with energy $-E$. Thus, their
contributions tend to cancel each other in perturbation theory.

\begin{figure}[t]
\centering
\includegraphics[width=0.7\textwidth]{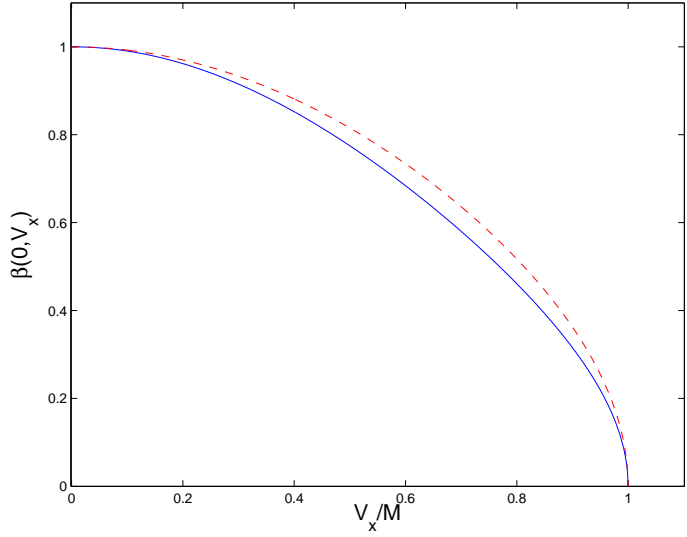}
\caption{\label{Fig2}
The parameter $\beta$ Eq.~(\ref{eq:bertu_etu_k_B_y_V_y}) as a function of
$V_x/M$ for two sets of parameters:
dashed red line for $A_1=A_2=B_1=B_2=M=1$, solid blue line for the 
Bi$_2$Se$_3$ set of parameters in section~\ref{Secbrokenph}.
}
\end{figure}

\begin{figure}[t]
\begin{minipage}{0.495\textwidth}
\centering
\includegraphics[width=\textwidth]{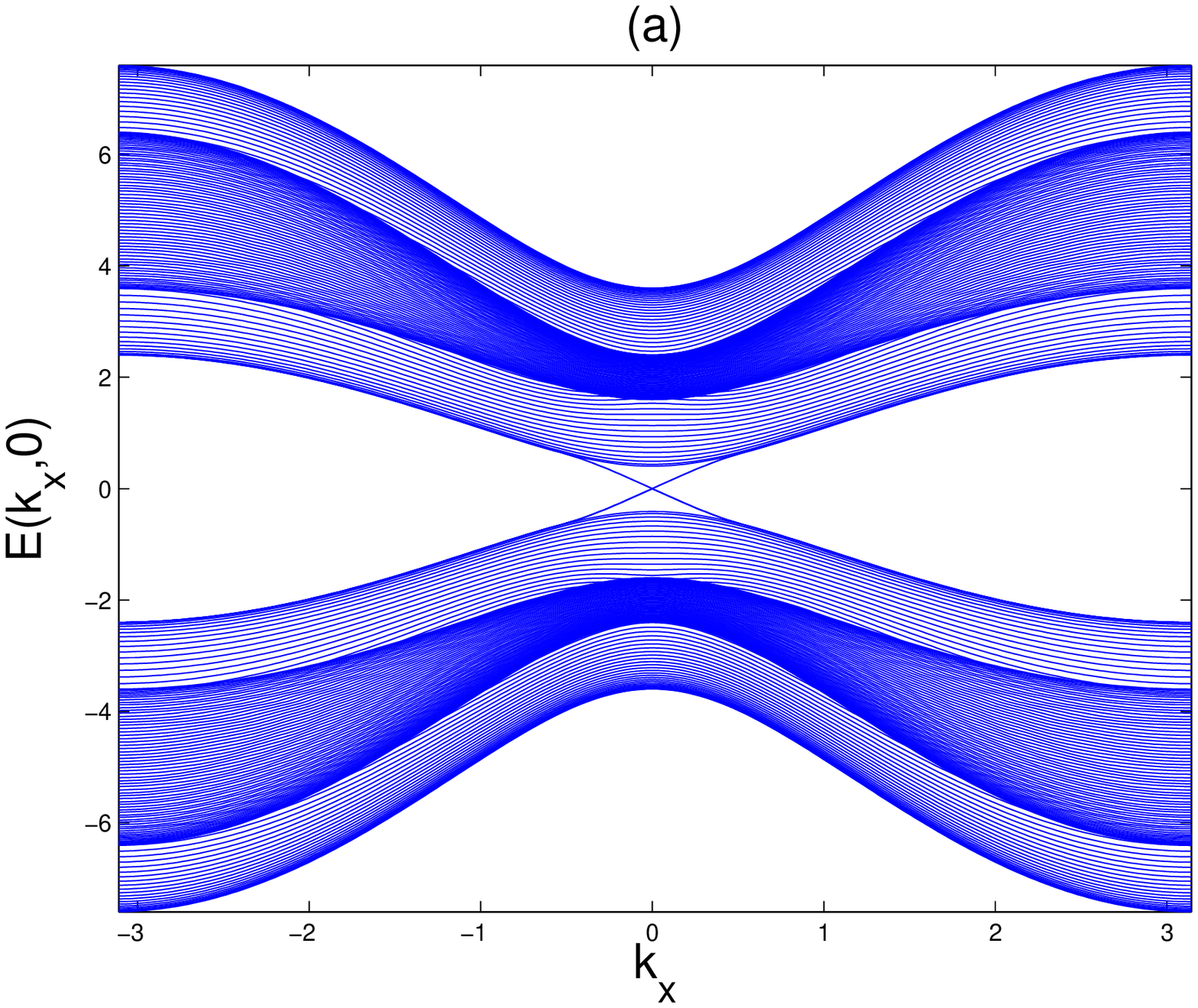}
\end{minipage}
\begin{minipage}{0.495\textwidth}
\centering
\includegraphics[width=\textwidth]{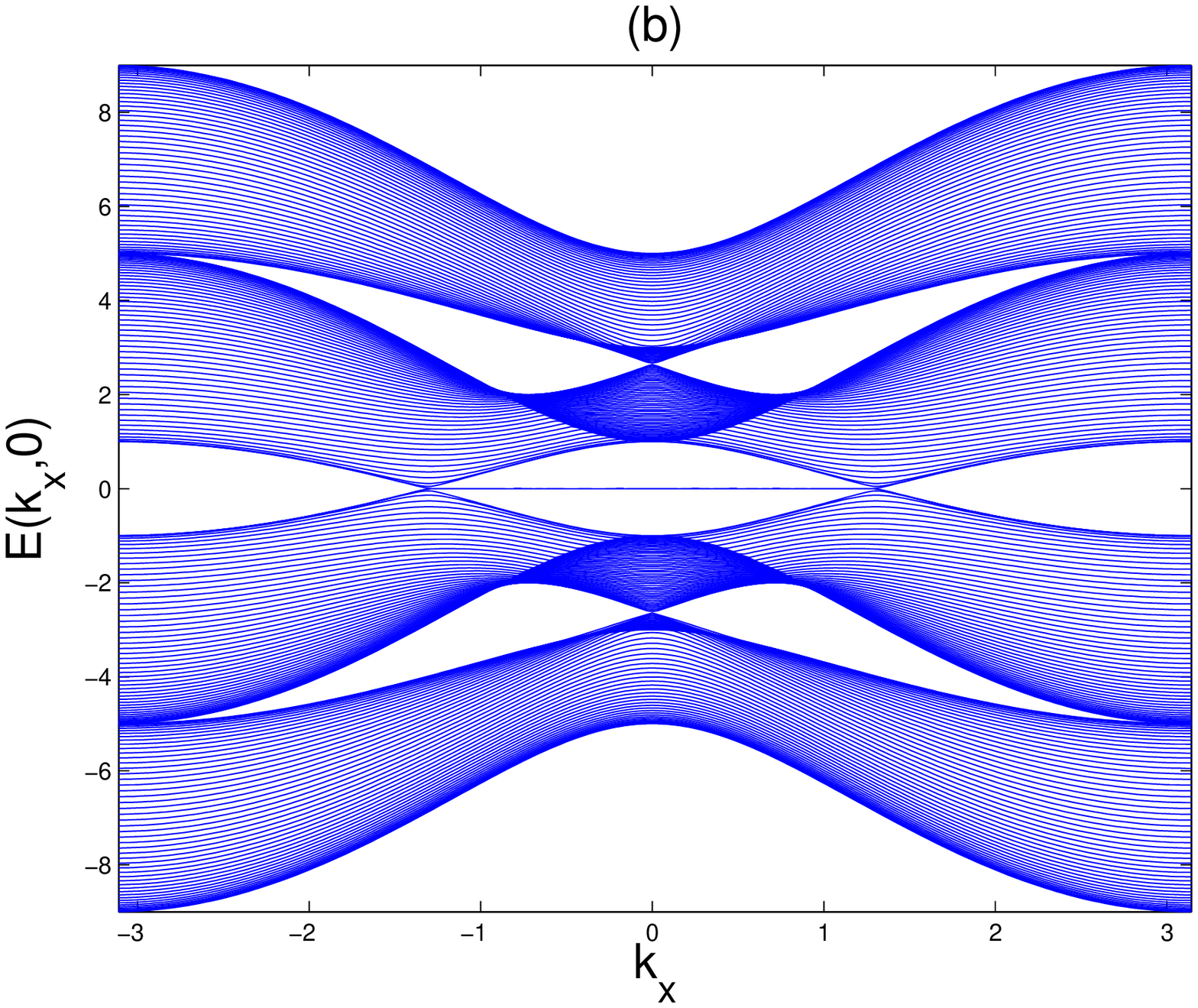}
\end{minipage}
\begin{minipage}{0.495\textwidth}
\centering
\includegraphics[width=\textwidth]{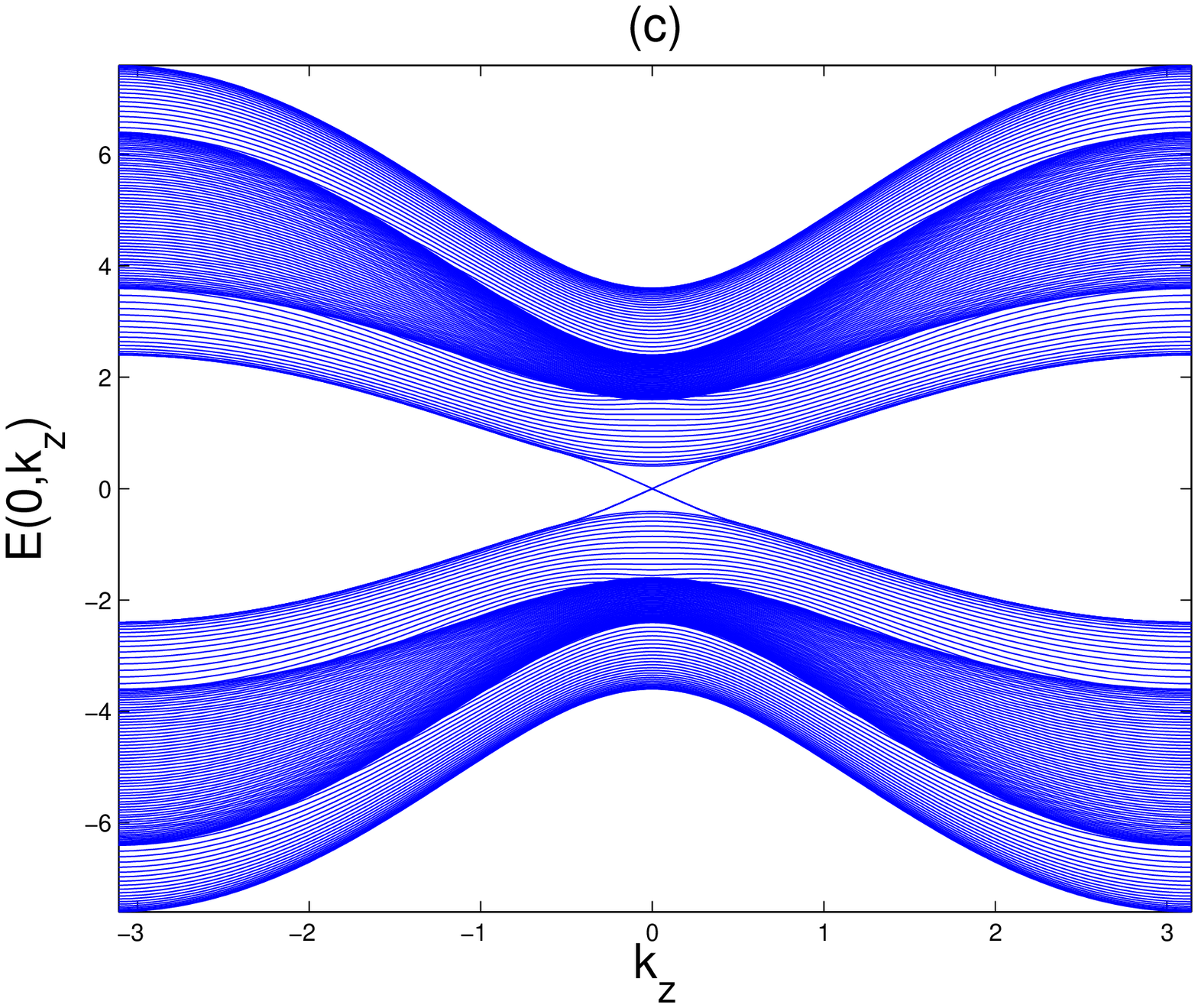}
\end{minipage}
\begin{minipage}{0.495\textwidth}
\centering
\includegraphics[width=\textwidth]{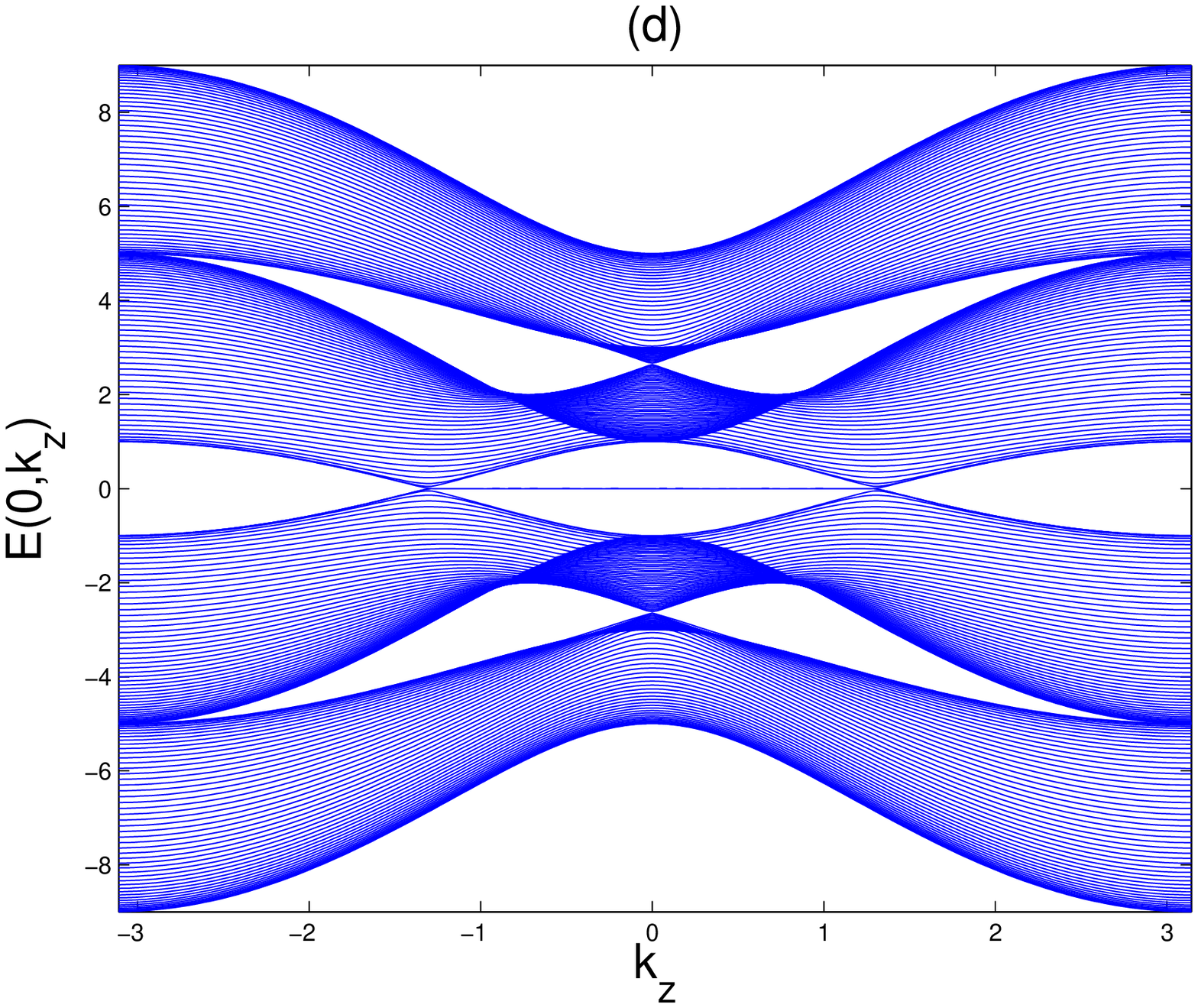}
\end{minipage}
\caption{\label{Fig3}
Numerical dispersions of bulk and surface states for model I with
$V_x/M=0.6$ for (a) and (c) and $2.0$ for (b) and (d).
In (a) and (b) $k_z=0$, and in (c) and (d) $k_x=0$.
The other parameters are same as in figure~\ref{Fig1}.
}
\end{figure}

If both states $1$ and $2$ exist, the surface states for the full 
Hamiltonian $H$ are then approximately given by
\begin{eqnarray}
\label{eq:solu_bertu_B_y_V_x}
\Psi_{\pm,\mathbf k}(y)&=\frac{1}{\sqrt{2}}(\psi_{1,\mathbf k}(y)\pm e^{i\theta_{\mathbf k}}\psi_{2,\mathbf k}(y)), 
\end{eqnarray}
where 
\[
\sin \theta_{\mathbf k} =\frac{2A_1\sin k_z -V_y}{\sqrt{(2A_1\sin k_z
    -V_y)^2+4A_2^2\sin^2 k_x }}.
\]
The surface state eigenenergies in this case are found to be
\begin{eqnarray}
\label{eq:ener_bertu_B_y_V_x}
E_{\pm}(\mathbf k)&=\pm\beta(V_x,\mathbf k)\sqrt{(2A_1\sin k_z
  -V_y)^2+4A_2^2\sin^2 k_x },
\end{eqnarray}
where
\begin{eqnarray}
\label{eq:bertu_etu_k_B_y_V_y}
\nonumber \beta(V_x,\mathbf k)=&CC'\int_0^{\infty}\,dy\,\bigg[ e^{-2\frac{A_2}{B_2} y}\sinh\left(\sqrt{\frac{A_2^2}{B_2^2}-\frac{\tilde m_0(\mathbf k)}{B_2}+\frac{V_x}{B_2}} \; y\right)\\
\nonumber &\times\sinh\left(\sqrt{\frac{A_2^2}{B_2^2}-\frac{\tilde m_0(\mathbf k)}{B_2}+\frac{V_x}{B_2}} \; y\right) \bigg]\\
& =\frac{4 A_2^2\sqrt{{\tilde m}^2_0(\mathbf k)-V_x^2}}{4 
  A_2^2 \tilde m_0(\mathbf k)+B_2 V_x^2} 
\end{eqnarray}
is the spatial overlap of the two states Eq.~(\ref{eq:surf_solu_B_y_V_x_k_0})
and (\ref{eq:surf_solu_B_y_V_x_k_0_2}).
The dependence of $\beta$ as a function of $V_x$ at $k=0$ is
shown in Fig.~\ref{Fig2}.

For small momenta the dispersion Eq.~(\ref{eq:ener_bertu_B_y_V_x}) shows that 
the presence of
an exchange field within the $xy$-plane does not affect the presence of
the surface Dirac cone. 
The Dirac cone is just shifted in $k_z$-direction and remains ungapped.
However, the velocity of the Dirac cone is isotropically suppressed
by the factor $\beta(V_x,\mathbf k)$. Thus, the presence of an
exchange field component in $x$-direction allows tuning of the
group velocity of the surface states, as shown in Fig.~\ref{Fig2}. 

The spin texture of the surface states Eq.~(\ref{eq:solu_bertu_B_y_V_x})
turns out to be the same as in the previous section, except for the
fact that all spin components are suppressed by the factor
$\beta(V_x,\mathbf k)$. Thus, the presence of the $x$-component
of the exchange field $V_x$ leads to a suppression of the spin
polarization of the surface states.

Figure~\ref{Fig3} shows dispersions with exchange field $(V_x,0,0)$. One can see 
from the figure that in this case $k_x$- and $k_z$-direction are equivalent.
This is due to the fact that $\Gamma_x$ commutes with both $\Gamma^1$ and
$\Gamma^3$. 
In Fig.~\ref{Fig1}~(b) and (d) we see that a flat band appears, if $V_x>M$. 
This flat band is apparently two dimensional, as it stays flat in both
$k_x$ and $k_z$ direction. It still exists if we add a finite
$V_y$. 
The existence of this flat band goes beyond the perturbative
treatment above.
In the next subsection we discuss the existence of this flat band
in terms of a topological invariant recently proposed by
Matsuura et al.\cite{Matsuura}

\subsection{Existence of a two-dimensional flat band}
\label{subsecflat1}
Matsuura et al.\cite{Matsuura} presented a general classification
of the gapless topological phases like in semimetals or 
nodal superconductors. They showed that a generalized
bulk-boundary correspondence exists that relates the
topological properties of the Fermi surface to
the presence of protected flat bands at the
surface of the system. In particular, it was found
that the dimension of the surface flat band is
always given by the dimension of the Fermi surface
plus 1, if it exists (see Table V in Ref.\cite{Matsuura}).

In the present case the system is topologically nontrivial
and belongs to class AIII as discussed in section~\ref{Secsymmetries}.
The Fermi surface is one-dimensional and we may thus expect the
appearance of a two-dimensional flat band.

The presence or absence of the flat band can be classified by
a topological winding number. To construct this winding
number, we use
the chiral symmetry $\Theta_3=\sigma_z \otimes \tau_z$ that
was discussed in section~\ref{Secsymmetries} and is valid
in the present case. Whenever a chiral symmetry is
present, the Hamiltonian anticommutes with the
symmetry operator. In this case the bulk Hamiltonian
can be brought into off-diagonal block form
by transforming to the eigenbasis of $\Theta_3$:
\begin{eqnarray}
\label{eq:Schnyder_form_1}
\nonumber H(\mathbf k)
&=\left(\begin{array}{cc}
0 & D^{\dagger}(\mathbf k)\\
D(\mathbf k) & 0
\end{array}\right),
\end{eqnarray}
where the block $D(\mathbf k)$ is found to be
\begin{equation}
\label{eq:Schnyder_form_2}
\fl D(\mathbf k)=\left(\begin{array}{cc}
m_0(\mathbf k)+im_3(\mathbf k) & -m_1(\mathbf k)+im_2(\mathbf k)+V_x-iV_y \\
m_1(\mathbf k)+im_2(\mathbf k)+V_x+iV_y & m_0(\mathbf k)-im_3(\mathbf k)
\end{array}\right).
\end{equation}
From the block $D(\mathbf k)$ we can define a winding number \cite{Schnyder,Matsuura}
\begin{eqnarray}
\label{eq:WN_schnyder}
w &=\frac{1}{2\pi} \textrm{Im}\int \, dk_\perp\,
\textrm{Tr}\left(D^{-1}(\mathbf k) \partial_{k_\perp}
D(\mathbf k) \right) \nonumber \\
&=\frac{1}{2\pi} \textrm{Im}\int \, dk_\perp\,
\textrm{Tr}\left(\partial_{k_\perp}
\ln D(\mathbf k) \right) \nonumber \\
&=\frac{1}{2\pi} \textrm{Im}\int \, dk_\perp\,
\partial_{k_\perp} \ln \textrm{det} D(\mathbf k) \\
&=\frac{1}{2\pi} \textrm{Im}\int \, dk_\perp\,
\left( \textrm{det} D(\mathbf k) \right)^{-1} \partial_{k_\perp} \textrm{det} D(\mathbf k).
\nonumber 
\end{eqnarray}
Here, $k_\perp$ is the momentum component perpendicular to the surface. The
winding number $w$ is always an integer and depends on the momentum
components parallel to the surface. It measures the phase change of the
complex number $\textrm{det} \, D(\mathbf k)$, when $k_\perp$ runs through 
the Brillouin zone. If $w(k_{||})=0$, there exists no zero energy
state for the momentum $k_{||}$ at the surface. If $w(k_{||})$ is nonzero,
a zero energy surface state exists. In the present case, where we
consider a boundary in $y$-direction we have
\begin{equation}
\label{eq:WN_schnyder_2}
w(k_x,k_z)=\frac{1}{2\pi}\textrm{Im}\int_{-\pi}^{\pi}\, dk_y\, \partial_{k_y} \ln \textrm{det} D(\mathbf k).
\end{equation}

\begin{figure}
\centering
\includegraphics[width=0.7\textwidth,angle=270]{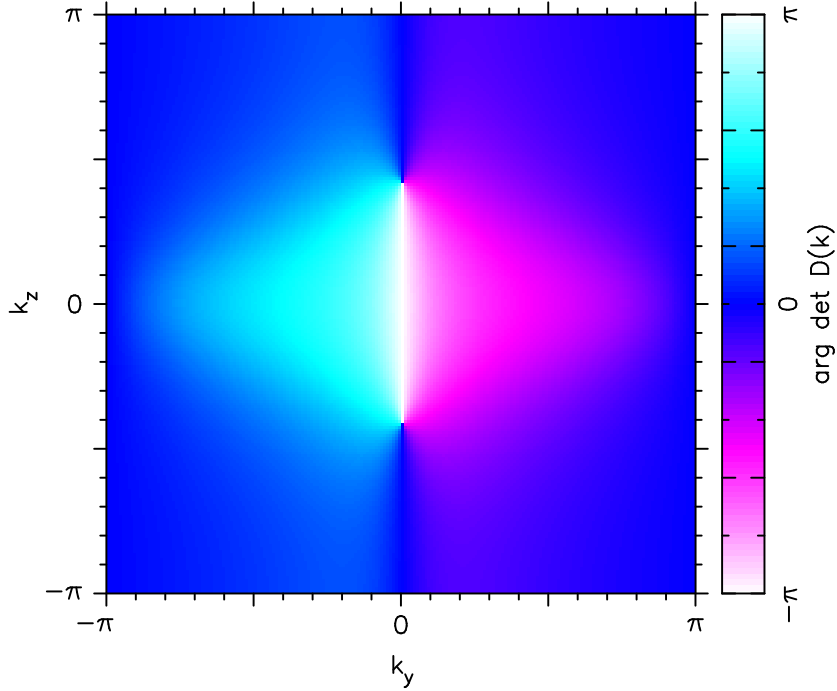}
\caption{\label{Fig4}
The phase of $\textrm{det} \, D(\mathbf k)$
for $k_x=0$ as a function of $k_y$ and $k_z$ in
a color coded scale. The parameters are the same
as in Fig.~\ref{Fig3}~(b) and (d).
}
\end{figure}
Fig.~\ref{Fig4} shows the phase of $\textrm{det} \, D(\mathbf k)$
for $k_x=0$ as a function of $k_y$ and $k_z$ in
a color coded scale for the same set of parameters as in
Fig.~\ref{Fig3}~(b) and (d).
Blue color corresponds to phase 0 and white to phase $\pm \pi$.
From the figure one notices that there exist a momentum space
vortex and an anti-vortex at the positions $(k_y,k_z)=(0,\pm 1.318)$,
around which the phase winds by $2\pi$. These positions are
points on the bulk Fermi surface. Note that $\textrm{det} \, D(\mathbf k)=0$
on the Fermi surface and the phase becomes singular there.
From Fig.~\ref{Fig4} it becomes clear that the winding number
(\ref{eq:WN_schnyder_2}) becomes 1, when $k_z \in [-1.318,1.318]$
and 0 outside. This is just the momentum range of the flat band
seen in Fig.~\ref{Fig3}~(b). This example demonstrates that the winding
number (\ref{eq:WN_schnyder_2}) correctly predicts the presence
of the flat band.

\begin{figure}[t]
\begin{minipage}{0.495\textwidth}
\centering
\includegraphics[width=\textwidth]{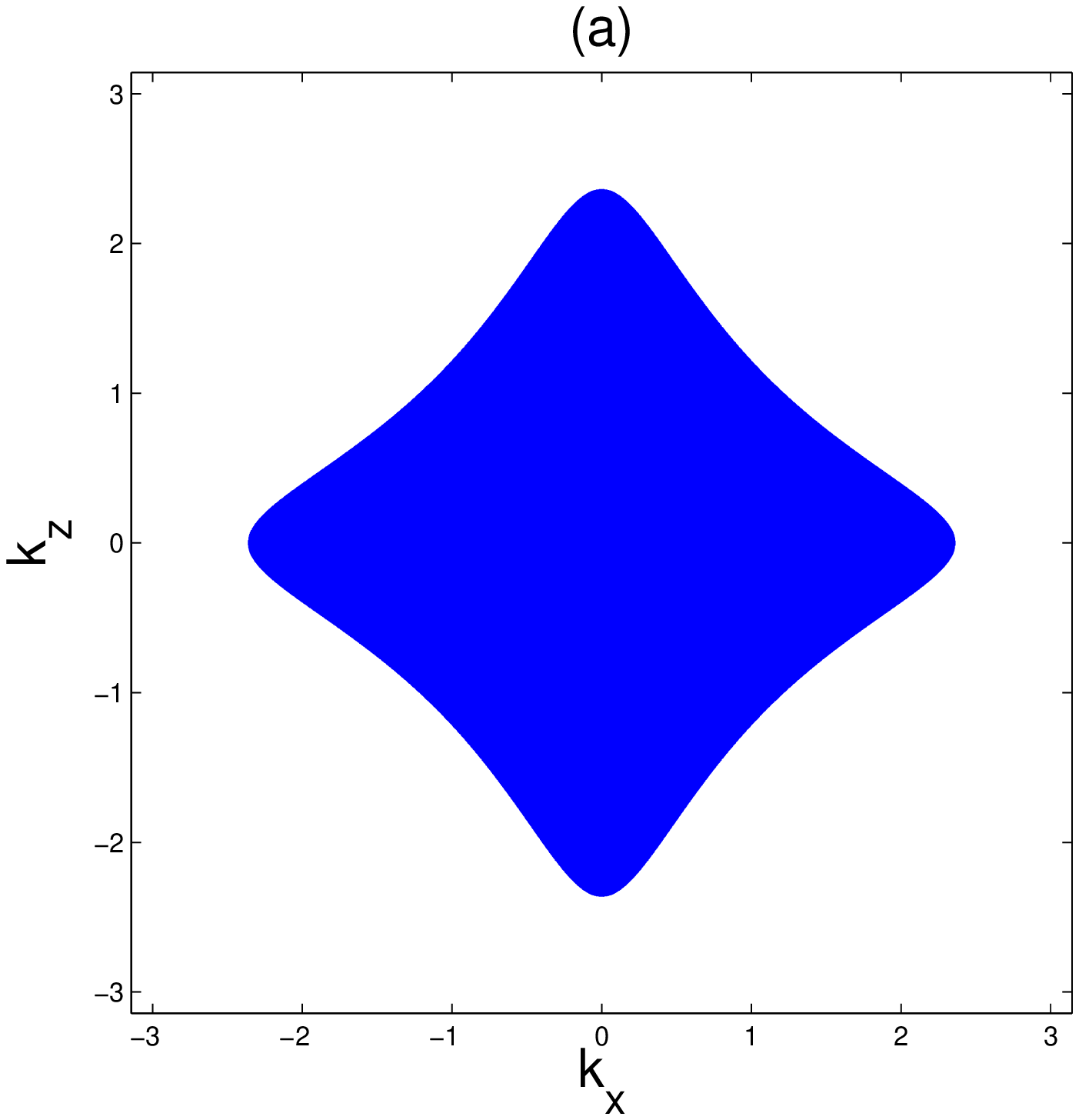}
\end{minipage}
\begin{minipage}{0.495\textwidth}
\centering
\includegraphics[width=\textwidth]{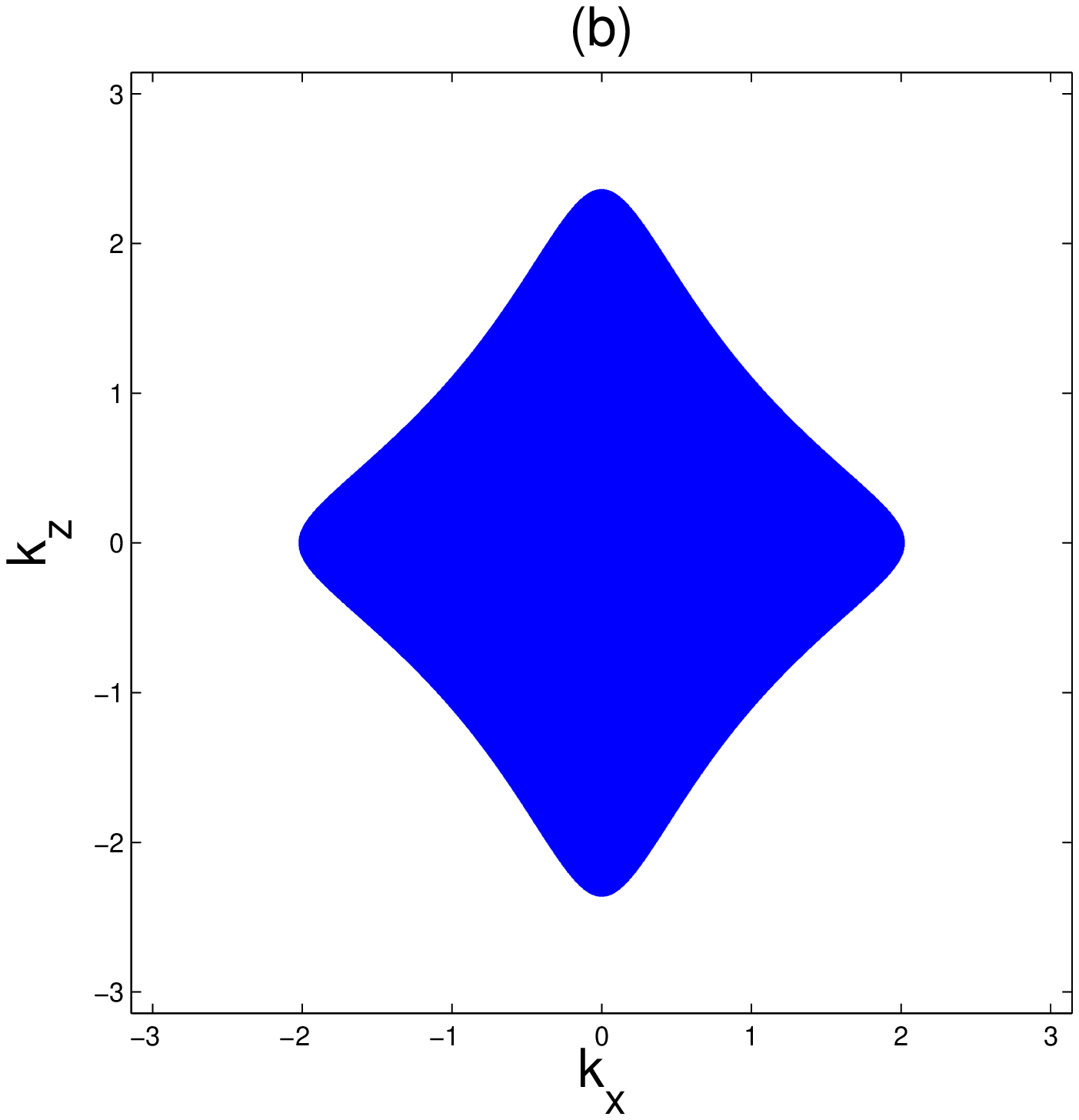}
\end{minipage}
\begin{minipage}{0.495\textwidth}
\centering
\includegraphics[width=\textwidth]{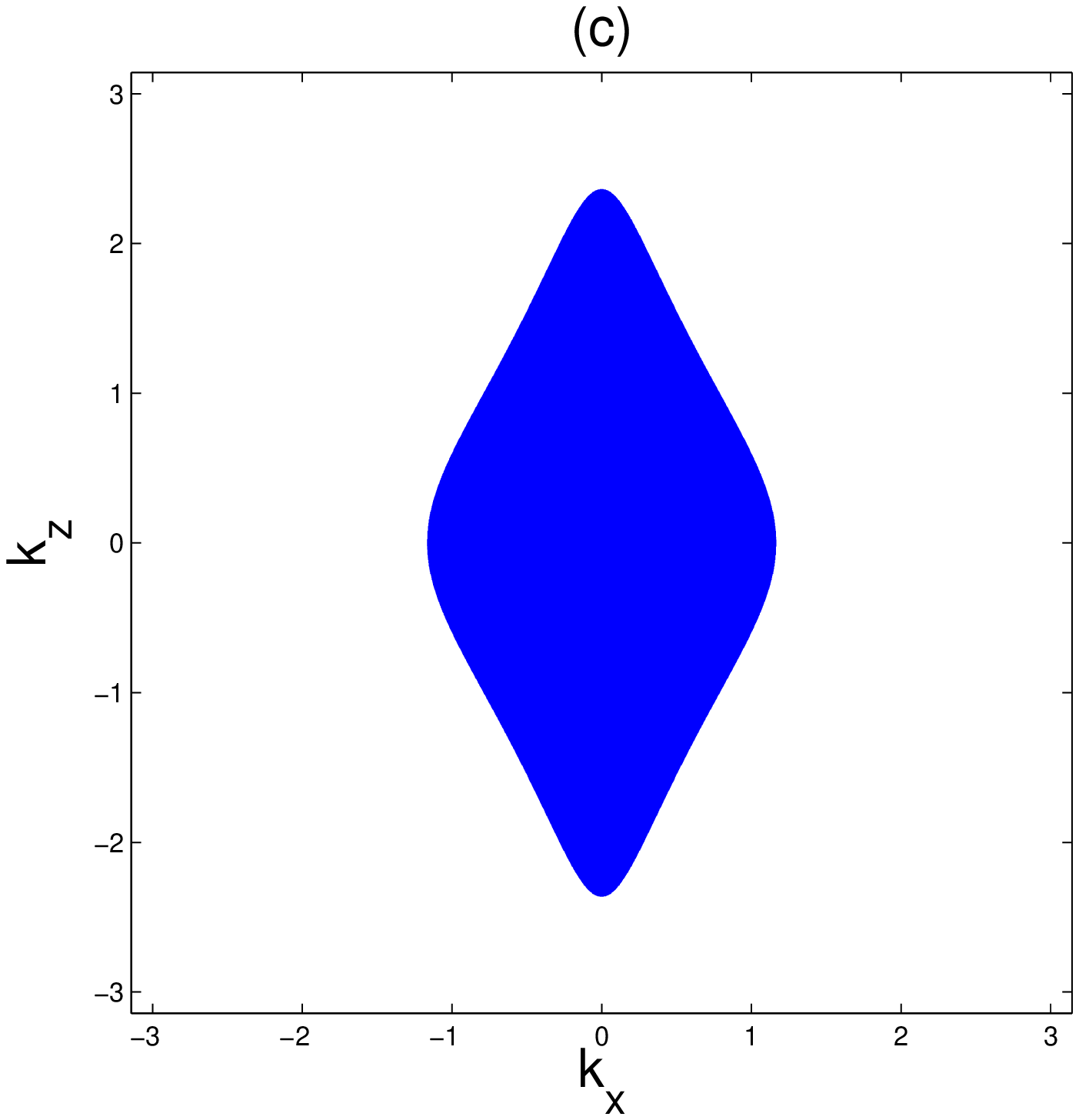}
\end{minipage}
\begin{minipage}{0.495\textwidth}
\centering
\includegraphics[width=\textwidth]{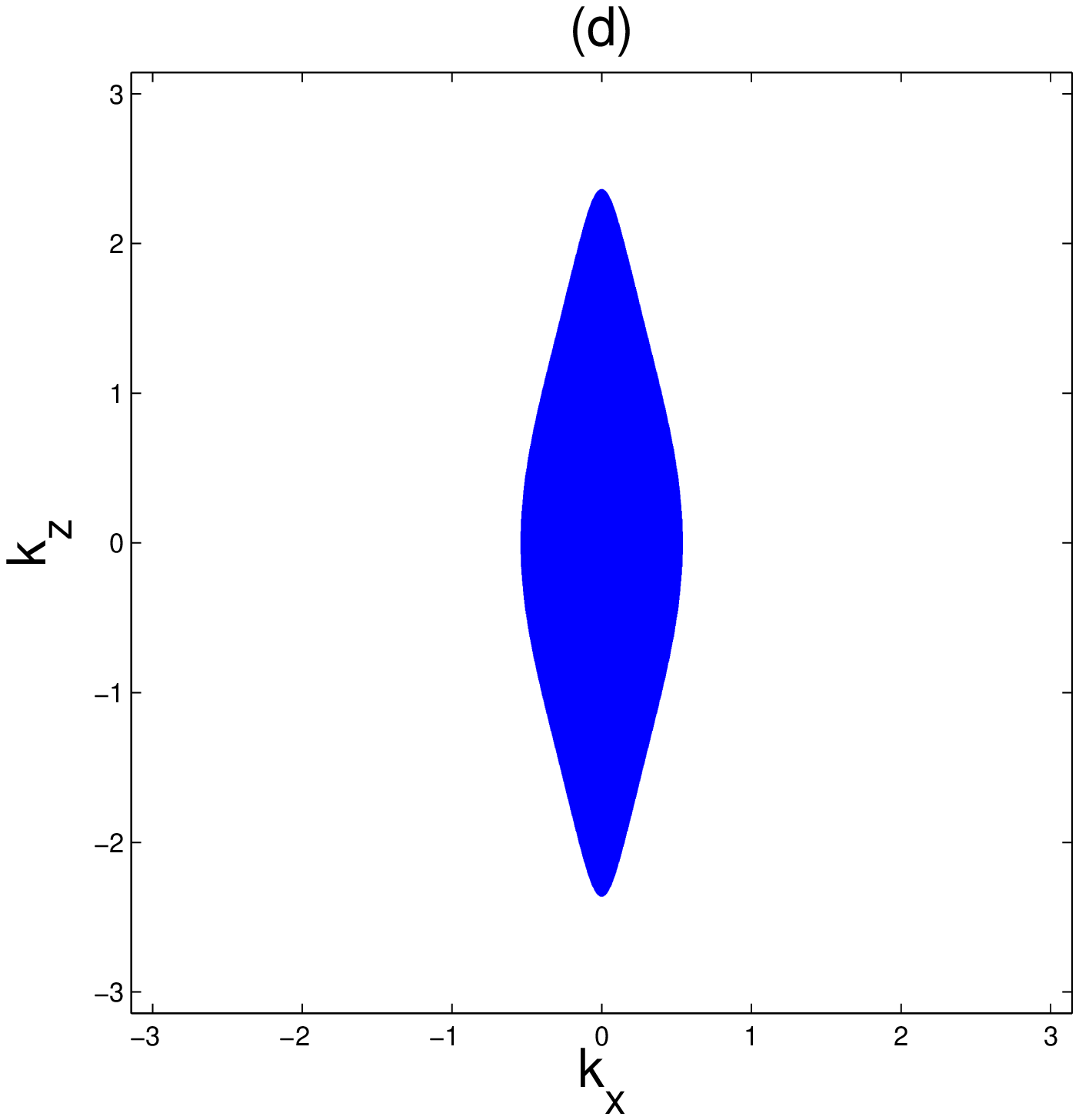}
\end{minipage}
\caption{\label{Fig5}
The flat band area within the $(k_x,k_z)$-plane for different directions of
the exchange field.
The field $(V_x,V_y)$ is given by $(V_0,0.0)$ (a),
$V_0(\cos \pi/8,\sin \pi/8 )$ (b), 
$V_0/\sqrt{2}(1,1)$ (c), and $V_0(\sin \pi/8 ,\cos \pi/8 )$ (d), respectively.
Here, $V_0/M=2.8$, $B_i=A_i=M=1$, and $C=D=0$. The size of the flat band
area is reduced when the exchange field is rotated towards the $y$-direction.
}
\end{figure}

Using Eq.~(\ref{eq:WN_schnyder_2}) we can derive an analytical
condition for the existence of the flat band, which is given
in appendix III.
In Figures~\ref{Fig5} and \ref{Fig6} we show the regions in $(k_x,k_z)$-space
in which the two-dimensional flat band on the $y=0$ surface appears for
different sets of parameters. These areas were calculated using the
analytical condition from appendix III. Note, that the boundaries
of these areas are just the projections of the one-dimensional Fermi surface
onto the $(k_x,k_z)$-plane, as shown in appendix III.

Figure~\ref{Fig5} illustrates the effect of a rotation of the exchange field
within the $xy$-plane on the flat band area. The flat band area is
shown for exchange fields $V_x=V_0 \cos \vartheta$ and $V_y=V_0 \sin \vartheta$
with $V_0=2.8 M$ and four angles of rotation $\vartheta$.
When the direction of the field is rotated from the $x$-direction into 
$y$-direction the size of the flat band area is reduced in $x$-direction 
but remains unchanged in $z$-direction. When the exchange field points
in $y$-direction, the flat band finally disappears.

\begin{figure}
\begin{minipage}{0.495\textwidth}
\centering
\includegraphics[width=\textwidth]{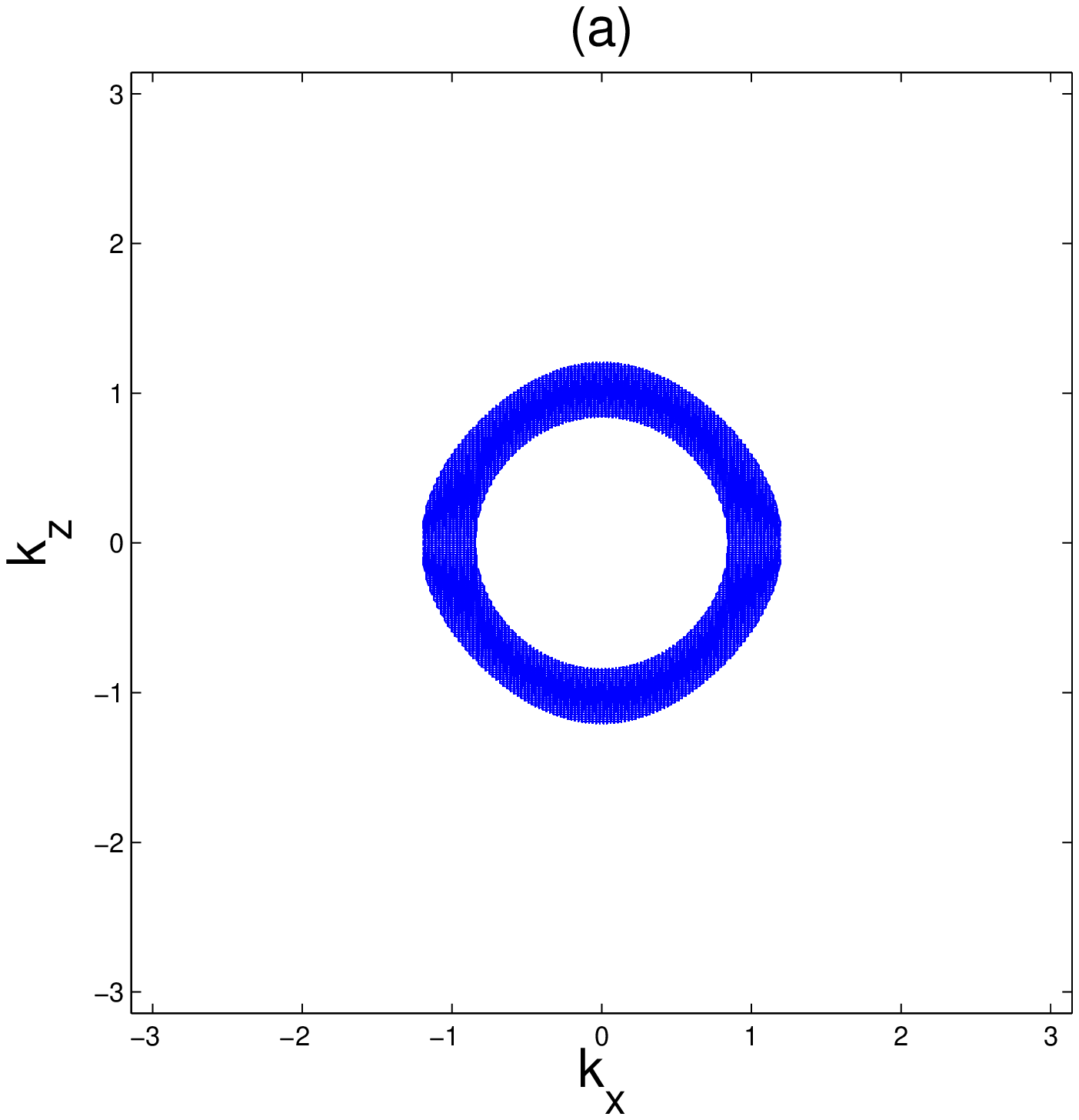}
\end{minipage}
\begin{minipage}{0.495\textwidth}
\centering
\includegraphics[width=\textwidth]{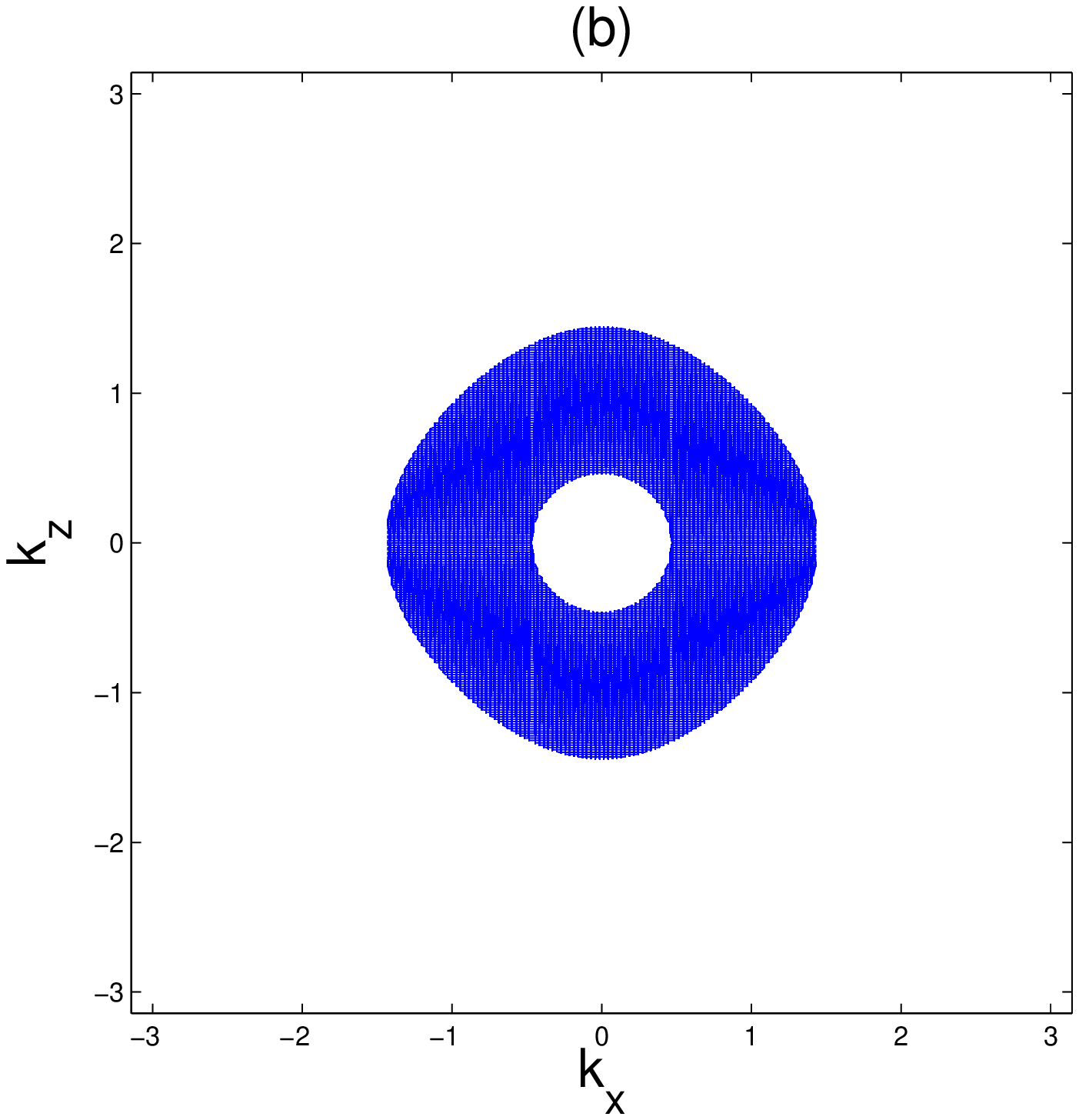}
\end{minipage}
\begin{minipage}{0.495\textwidth}
\centering
\includegraphics[width=\textwidth]{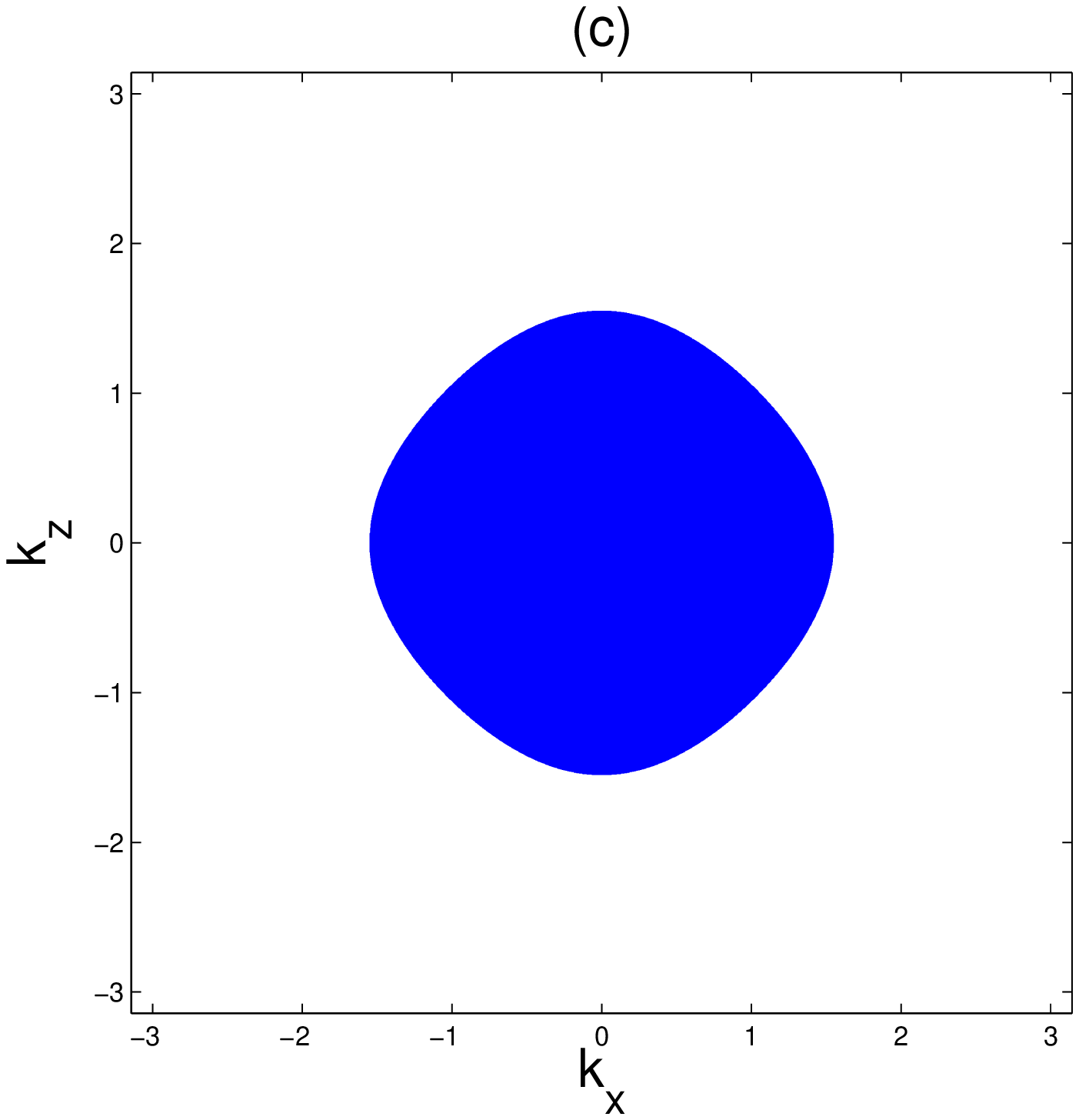}
\end{minipage}
\begin{minipage}{0.495\textwidth}
\centering
\includegraphics[width=\textwidth]{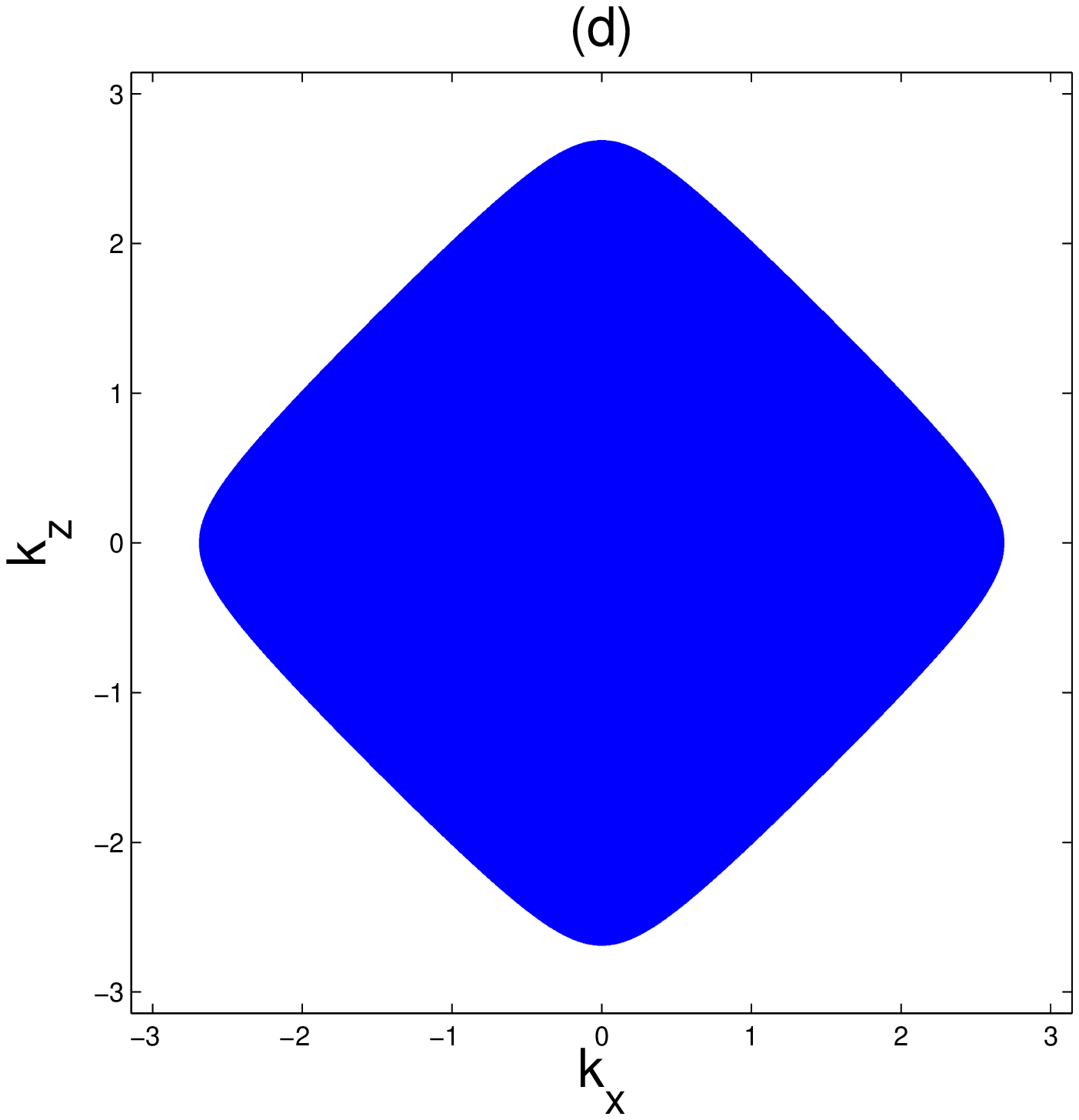}
\end{minipage}
\caption{\label{Fig6}
The flat band area within the $(k_x,k_z)$-plane for different values of the exchange
field in $x$-direction. $V_x/M=0.4$ (a) , $V_x/M=0.8$ (b), $V_x/M=1.0$ (c),
and $V_x/M=2.8$ (d).
Here, we have used the parameters $V_y=V_z=0$, $B_i=M=1$, and $A_i=0.15$.
}
\end{figure}

In appendix I we show that for
$M<2A_1^2/B_1$ the minimal strength of the exchange field to create
a semimetallic state and thus a two-dimensional surface flat band for
the present case is given by $V_{cr}=M$.
However, if $M>2A_1^2/B_1$ (implicitly assuming $B\geq A$ and $B\geq M$), the minimal 
exchange field strength is given by the more complicated expression
\begin{equation}
\label{eq:min_V_flat_band}
V_{cr}=\frac{A_1\sqrt{4B_1M-4A_1^2-M^2}}{\sqrt{B_1^2-A_1^2}}.
\end{equation}
which is smaller than $M$. In this case
we can have the situation that the flat band area is not simply
connected anymore as shown in figures~\ref{Fig6}~(a) and (b).
This case appears, when Eq.~(\ref{eq:c12model1}) from appendix I
possesses two solutions instead of just one.

\subsection{Boundary perpendicular to the $y$-direction with finite $V_z$}
\label{subsec1vz}

Next, we consider the case that $V_z$ is nonzero and $V_x=0$.
The component $V_y$ will be treated perturbatively like in
section~\ref{subsec1vxy}.
The bulk energy bands Eq.~(\ref{eq:model1bulk}) in this case simplify to
\begin{equation} \label{eq:model1bulkc}
	E_i^I(\mathbf k)=\pm \sqrt{m_1^2 + m_2^2 +\left[ V_z \pm
        \sqrt{m_0^2 + m_3^2} \right]^2 }
\end{equation}
Again, the system is
insulating in the absence of an exchange field and the gap closes,
when $V_z$ reaches the critical value $V_{cr}$. In the semimetallic
state the Fermi surface is defined now by three instead of two equations, 
$m_1 = m_2 = 0$ and $V_z^2 = m_0^2 + m_3^2$.
For this reason, the Fermi surface becomes zero-dimensional,
i.e. there are point nodes.

In this case the chiral symmetry $\Theta_3$ does not hold anymore, because
$\Theta_3$ commutes with $\Gamma_z$. However, as discussed in
section~\ref{Secsymmetries} for the special case $k_x=0$ the system
possesses the chiral symmetry $\Theta_1$. For that reason we may
expect a one-dimensional flat band with $k_x=0$ in this case.

To determine the surface states in second order in $k_y$ one first notices that
$\Gamma_z$ neither commutes nor anticommutes with $H_0$. Therefore, it affects the 
spatial part of the surface states. However, similarly as in section~\ref{subsec1vxy},
we can determine the zero energy surface states of the Hamiltonian
\begin{equation} \label{eq:model1hpp}
H''_0(\mathbf k)=H_0(\mathbf k)+V_z\Gamma_z = (\tilde m_0(\mathbf
k)+B_2\partial_y^2)\Gamma^0-i2A_2\partial_y\Gamma^2 + V_z\Gamma_z \; . 
\end{equation}
This Hamiltonian commutes with the symmetry operator $\Theta_5=\sigma_z \otimes \tau_x$
and it anticommutes with $\Theta_1$. As both $\Theta_1$ and $\Theta_5$ commute
with each other, it is useful to look for surface state solutions among the
common eigenstates of $\Theta_1$ and $\Theta_5$.

These eigenstates are
$(-1,1,1,1)^T$, $(1,-1,1,1)^T$, $(1,1,-1,1)^T$, and $(1,1,1,-1)^T$.
We thus try the following two ans\"atze:
\begin{eqnarray}
\label{eq:sur_s_B_y_V_z}
\psi_{1,\mathbf k}(y)=(1,-1,1,1)^Tf_{\mathbf k}(y),\\
\psi_{2,\mathbf k}(y)=(1,1,-1,1)^Tf_{\mathbf k}(y),
\label{eq:sur_s_B_y_V_z_2}
\end{eqnarray}
(the other two eigenstates lead to exponentially increasing functions again).
We find that $f_{\mathbf k}(y)$ is solution of the equations
\begin{equation}
\label{eq:solu_f_B_y_V_z}
[\tilde m_0(\mathbf k)+B_2\partial_y^2+2A_2\partial_y \pm V_z]f_{\mathbf k}(y)=0,
\end{equation}
where the plus sign holds for $\psi_{1,\mathbf k}$ and the minus sign
for $\psi_{2,\mathbf k}$.

Solving the differential equation (\ref{eq:solu_f_B_y_V_z}) we
find the solutions
\begin{eqnarray}
\label{eq:surf_solu_B_y_V_z_k_0}
 \psi_{1,\mathbf k}(y)&=C(1,-1,1,1)^Te^{-\frac{A_2}{B_2}
   y}\sinh\left(\sqrt{\frac{A_2^2}{B_2^2}-\frac{\tilde m_0(\mathbf k)+V_z}{B_2}} \; y\right), \\
 \psi_{2,\mathbf k}(y)&=C'(1,1,-1,1)^Te^{-\frac{A_2}{B_2}
   y}\sinh\left(\sqrt{\frac{A_2^2}{B_2^2}-\frac{\tilde m_0(\mathbf k)-V_z}{B_2}} \; y\right). 
\label{eq:surf_solu_B_y_V_z_k_0_2}
\end{eqnarray}
It is clear that state 
$1$ exists only if $\tilde m_0(\mathbf k)+ V_z>0$ and
state $2$ exists only if $\tilde m_0(\mathbf k)- V_z>0$. 

Having determined the zero energy surface states of $H''_0$ we can now try to
obtain the ones of $H=H''_0 + m_1 \Gamma^1 + m_3 \Gamma_I^3 + V_y \Gamma_y$ from them.
First, one notices that $\Gamma^1=\Theta_1$ anticommutes with $H''_0$
and the states (\ref{eq:surf_solu_B_y_V_z_k_0}) and
(\ref{eq:surf_solu_B_y_V_z_k_0_2}) are already eigenstates of $\Gamma^1$.
Thus, these eigenstates are also eigenstates of $H''_0 + m_1 \Gamma^1$
with energies $\mp 2 A_2 \sin k_x$. The operators $\Gamma_I^3$
and $\Gamma_y$ neither commute nor anticommute with $H''_0 + m_1 \Gamma^1$.
However, if $k_z$ and $V_y$ are small we can treat the terms $m_3 \Gamma_I^3+ V_y \Gamma_y$
perturbatively, again. For the same reasons as in section~\ref{subsec1vxy},
we may assume that the perturbation only couples the two surface
states 1 and 2 to each other. The surface states for the
full Hamiltonian are then found to be of the form 
\begin{eqnarray}
\label{eq:solu_bertu_B_y_V_z}
\Psi_{+,\mathbf k}(y)&=i \sin \frac{\theta_{\mathbf k}}{2} \psi_{1,\mathbf
  k}(y) + \cos \frac{\theta_{\mathbf k}}{2} \psi_{2,\mathbf k}(y) \\
\Psi_{-,\mathbf k}(y)&=i \cos \frac{\theta_{\mathbf k}}{2} \psi_{1,\mathbf
  k}(y) - \sin \frac{\theta_{\mathbf k}}{2} \psi_{2,\mathbf k}(y) 
\end{eqnarray}
where 
\[
\sin \theta_{\mathbf k} =\frac{2A_1\sin k_z -V_y}{\sqrt{(2A_1\sin k_z
    -V_y)^2+4A_2^2\sin^2 k_x }}.
\]
The energies are given by
\begin{eqnarray}
\label{eq:ener_bertu_B_y_V_z}
E_{\pm}(\mathbf k)&=\pm\sqrt{\beta(V_z,\mathbf k)^2 \left( 2A_1\sin
  k_z-V_y \right)^2+ 4A_2^2 \sin^2k_x }.
\end{eqnarray}
Here, the spatial overlap $\beta(V_z,\mathbf k)$ has the same functional form as 
in Eq.~(\ref{eq:bertu_etu_k_B_y_V_y}). For small momenta this 
dispersion shows that the $y$-component of the exchange field shifts
the surface Dirac cone in $k_z$-direction.
The Dirac cone remains ungapped by the exchange field.
The velocity of the Dirac cone is suppressed by the
$z$-component of the exchange field only in
$k_z$-direction, but not in $k_x$-direction. Thus,
in the present case the exchange field can tune the
group velocity of the surface electrons in an anisotropical way.

To determine the spin texture of the surface states
we evaluate the spin expectation values in the two
orbitals. For orbital 1 we find
\begin{eqnarray}
\nonumber
\left\langle \Psi_{\pm,\mathbf k} \left| \hat{s}_{1,x} \right|
\Psi_{\pm,\mathbf k} \right\rangle &= \pm \frac{1}{2} \cos \theta_{\mathbf k} \\
\nonumber
\left\langle \Psi_{\pm,\mathbf k} \left| \hat{s}_{1,y} \right|
\Psi_{\pm,\mathbf k} \right\rangle &= \mp \frac{\beta(V_z,\mathbf k)}{2} \sin \theta_{\mathbf k} \\
\nonumber
\left\langle \Psi_{\pm,\mathbf k} \left| \hat{s}_{1,z} \right|
\Psi_{\pm,\mathbf k} \right\rangle &= 0
\end{eqnarray}
and in orbital 2:
\begin{eqnarray}
\nonumber
\left\langle \Psi_{\pm,\mathbf k} \left| \hat{s}_{2,x} \right|
\Psi_{\pm,\mathbf k} \right\rangle &= \mp \frac{1}{2} \cos \theta_{\mathbf k} \\
\nonumber
\left\langle \Psi_{\pm,\mathbf k} \left| \hat{s}_{2,y} \right|
\Psi_{\pm,\mathbf k} \right\rangle &= \mp \frac{\beta(V_z,\mathbf k)}{2} \sin \theta_{\mathbf k} \\
\nonumber
\left\langle \Psi_{\pm,\mathbf k} \left| \hat{s}_{2,z} \right|
\Psi_{\pm,\mathbf k} \right\rangle &= 0
\end{eqnarray}
As in the previous cases the spin rotates within the $x$-$y$
plane. The spin direction of the two surface states is always
opposite. The spin-$x$-component is opposite in the two orbitals,
while the spin-$y$-component is identical. With increasing
$V_z$ the spin-$y$-component is suppressed by the
$\beta(V_z,\mathbf k)$ factor. The total
spin points in $y$-direction again, perpendicular to the surface:
\begin{eqnarray}
\nonumber
\left\langle \Psi_{\pm,\mathbf k} \left| \Gamma_x \right|
\Psi_{\pm,\mathbf k} \right\rangle &= 0 \\
\nonumber
\left\langle \Psi_{\pm,\mathbf k} \left| \Gamma_y \right|
\Psi_{\pm,\mathbf k} \right\rangle &= \mp \beta(V_z,\mathbf k) \sin \theta_{\mathbf k} \\
\nonumber
\left\langle \Psi_{\pm,\mathbf k} \left| \Gamma_z \right|
\Psi_{\pm,\mathbf k} \right\rangle &= 0
\end{eqnarray}

\begin{figure}[t]
\begin{minipage}{0.495\textwidth}
\centering
\includegraphics[width=\textwidth]{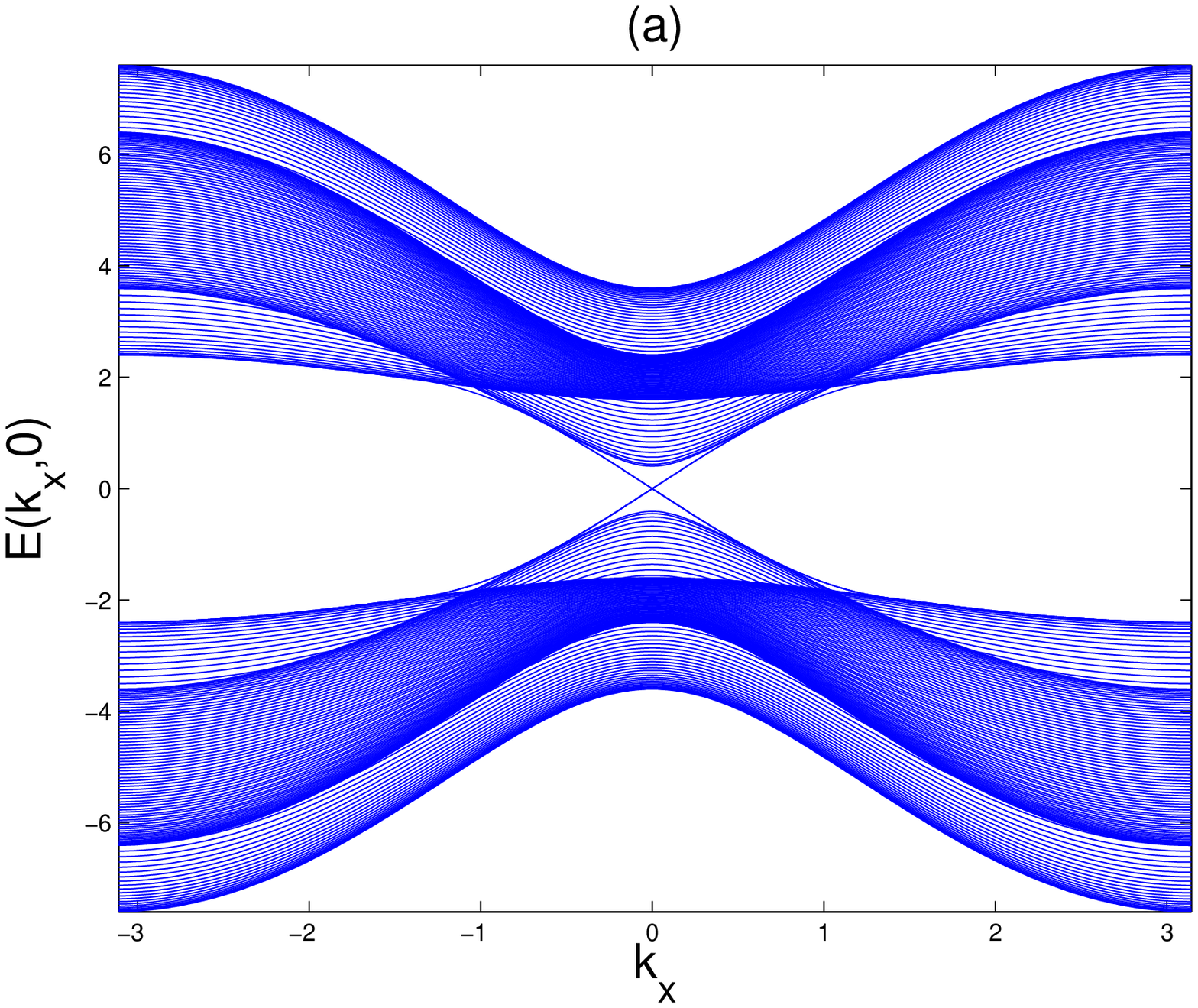}
\end{minipage}
\begin{minipage}{0.495\textwidth}
\centering
\includegraphics[width=\textwidth]{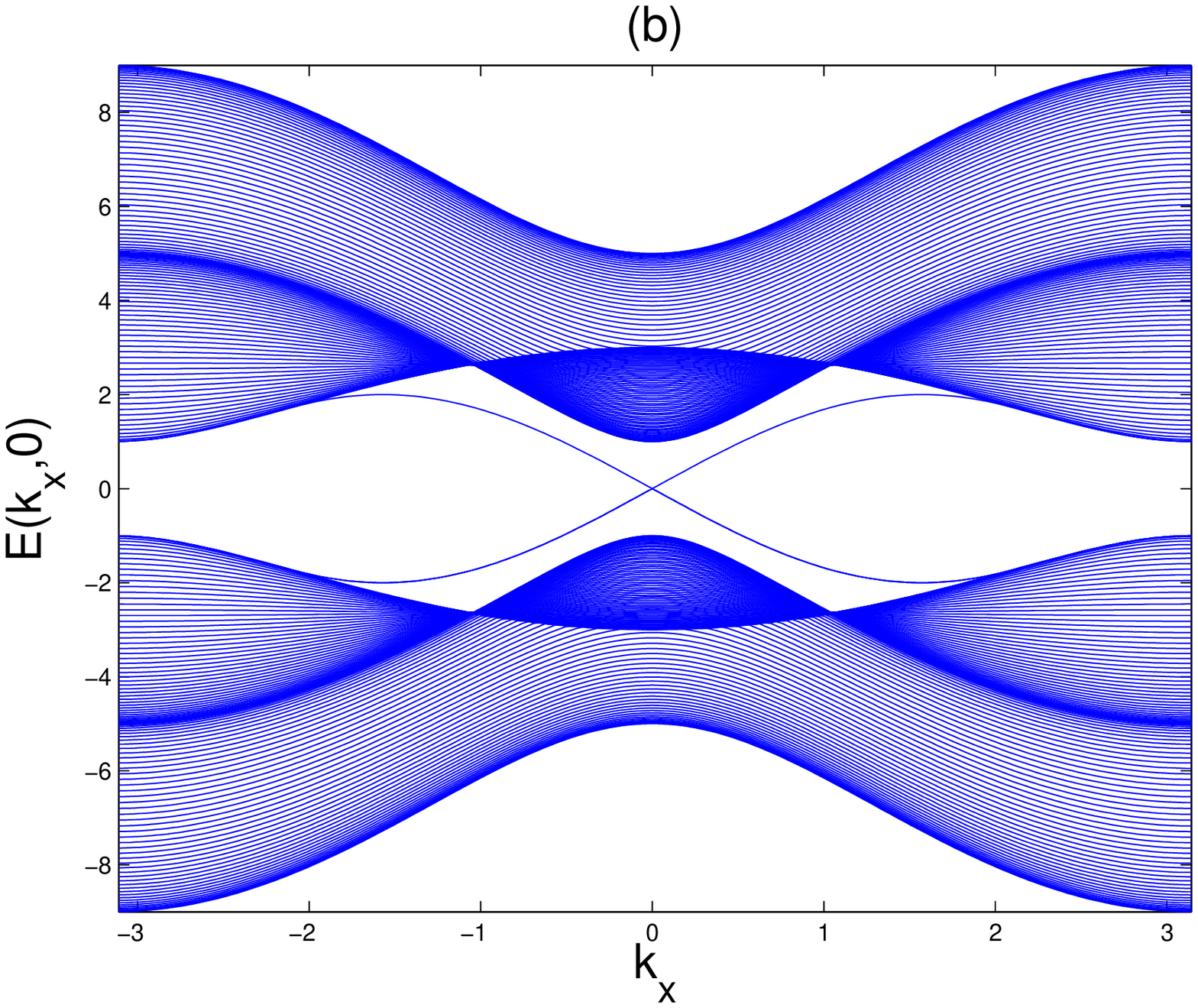}
\end{minipage}
\begin{minipage}{0.495\textwidth}
\centering
\includegraphics[width=\textwidth]{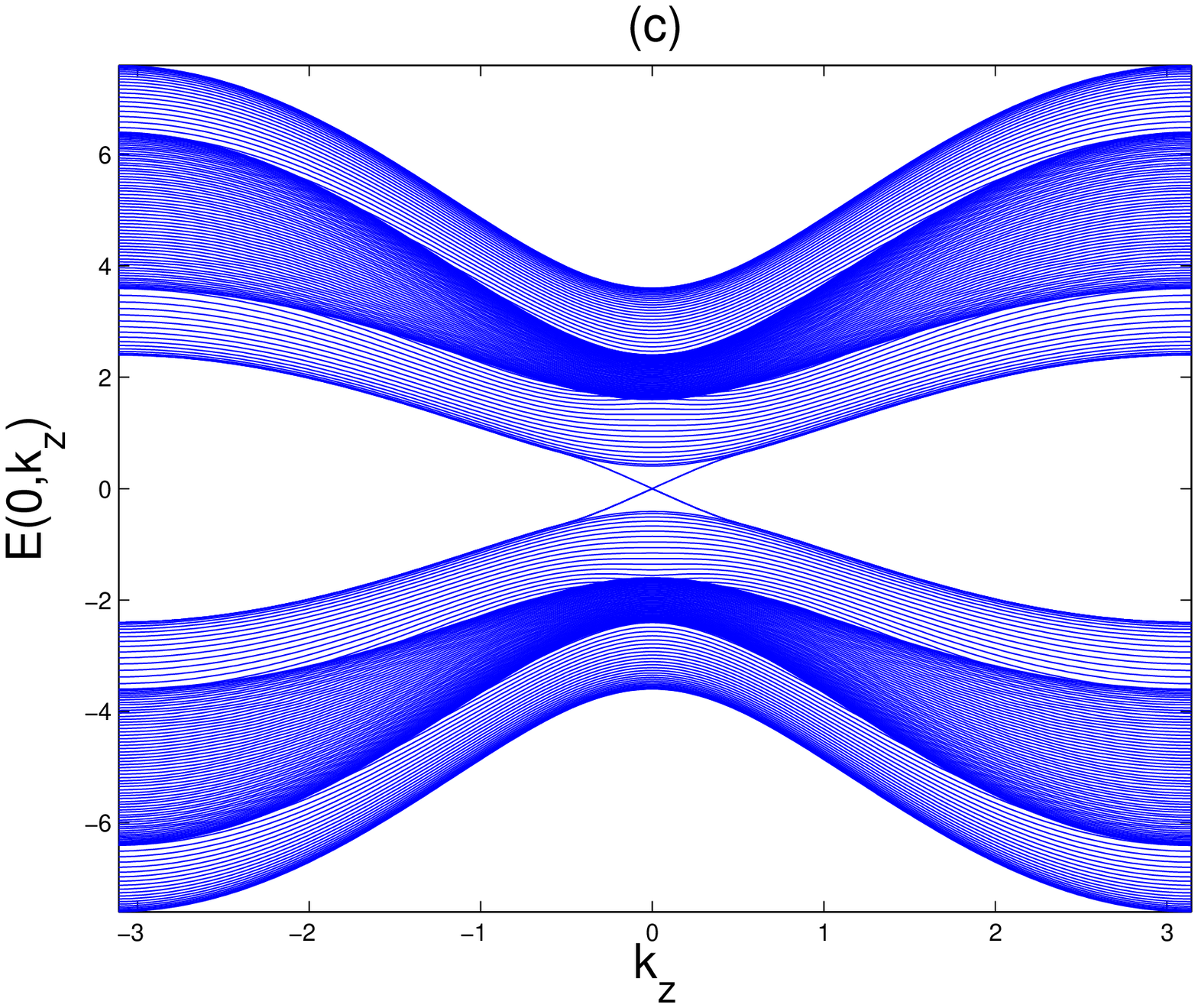}
\end{minipage}
\begin{minipage}{0.495\textwidth}
\centering
\includegraphics[width=\textwidth]{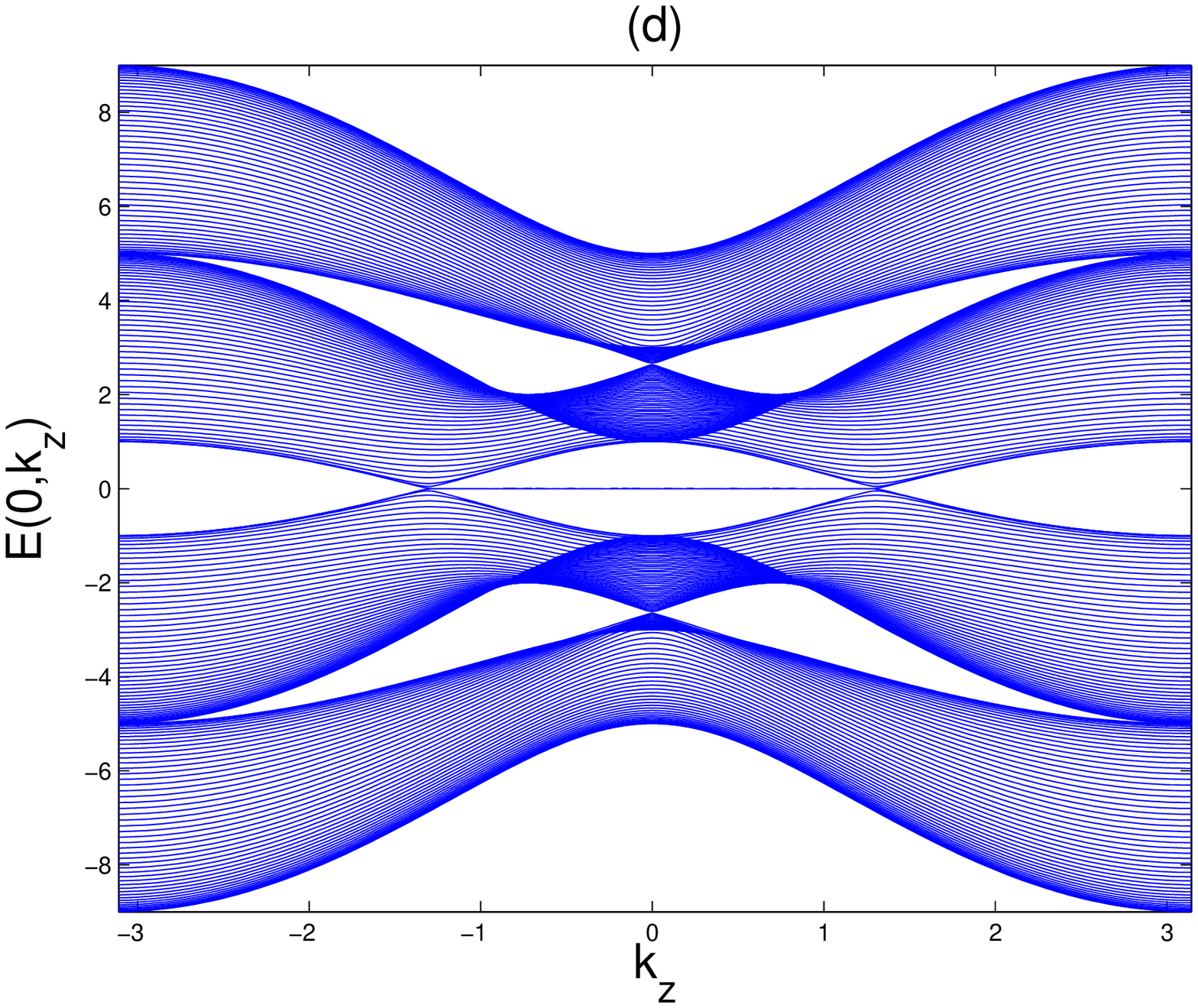}
\end{minipage}
\caption{\label{Fig7}
Numerical dispersions of bulk and surface states for model I with
$V_z/M=0.6$ for (a) and (c) and $2.0$ for (b) and (d).
In (a) and (b) $k_z=0$, and in (c) and (d) $k_x=0$.
The other parameters are same as in figure~\ref{Fig1}}
\end{figure}

Figure~\ref{Fig7} shows results for the energy dispersions with finite $V_z$
obtained from numerical calculations on a finite size system. 
One can see from the figure that the directions $k_x$ and $k_z$ differ. This is again due to the fact
that $\Gamma_z$ commutes with $\Gamma_I^3$ but anticommutes with
$\Gamma^1$. We see from the figure that 
a one dimensional flat band appears in $k_z$-direction if $V_z$ exeeds $M$. 

\subsection{Existence of a one-dimensional flat band}
\label{subsecflat2}

Similarly as in the case with nonzero $V_x$ the appearance of this
one-dimensional flat band can be understood using the classification
of Matsuura et al.\cite{Matsuura} For that purpose we consider
the Hamiltonian without the $\Gamma^1$ term:
\begin{equation}
\label{eq:H_flat_band_cond_3_z}
	H_1(\mathbf k)=m_0(\mathbf k) \Gamma^0 + m_2(\mathbf k) \Gamma^2+
m_3(\mathbf k) \Gamma_I^3+ V_z \Gamma_z
\end{equation}
If one considers this Hamiltonian for $k_x=0$ as a function of the
two coordinates $k_y$ and $k_z$, its Fermi surface will be a point
node in the two-dimensional Brillouin zone. However, one can also 
consider this Hamiltonian as a function of
the three coordinates $k_x$, $k_y$, and $k_z$, keeping the
$k_x$ dependence in $m_0(\mathbf k)$. Then, its Fermi surface
will be a line node in the three-dimensional Brillouin zone.
In any case, $H_1$ belongs to class AIII due to the
chiral symmetry $\Theta_1$.
Using the chiral symmetry $\Theta_1$ we can bring $H_1$ into
off-diagonal block form, similarly as in section~\ref{subsecflat1}:
\begin{eqnarray}
\label{eq:Schnyder_form_1b}
\nonumber H_1(\mathbf k)
&=\left(\begin{array}{cc}
0 & D_1^{\dagger}(\mathbf k)\\
D_1(\mathbf k) & 0
\end{array}\right),
\end{eqnarray}
where the block $D_1(\mathbf k)$ is found to be
\begin{equation}
\label{eq:Schnyder_form_2b}
D_1(\mathbf k)=\left(\begin{array}{cc}
m_0(\mathbf k)+im_3(\mathbf k) & -im_2(\mathbf k)-V_z \\
-im_2(\mathbf k)-V_z & m_0(\mathbf k)-im_3(\mathbf k)
\end{array}\right).
\end{equation}
Using this block, we can again define the winding number
\begin{equation}
\label{eq:WN_schnyder_3}
w(k_x,k_z)=\frac{1}{2\pi}\textrm{Im}\int_{-\pi}^{\pi}\, dk_y\, \partial_{k_y} \ln \textrm{det} D_1(\mathbf k).
\end{equation}
Following the method outlined in appendix~III we can
derive an analytical condition for the existence of
a flat band of $H_1$ which reads
\begin{equation}
\label{eq:flat_band_cond_3_z}
|V_z|>\sqrt{\tilde m_0(\mathbf k)^2+4A_1^2\sin^2(k_z)}.
\end{equation}
For $k_x=0$ this yields a range of $k_z$ values
for which a one-dimensional flat band exists, consistent with the
numerical result in Fig.~\ref{Fig7}.
If we consider $H_1$ as a function of three coordinates
$k_x$, $k_y$, and $k_z$, this condition tells us that
$H_1$ actually possesses a two-dimensional surface flat band
within a certain area in $(k_x,k_z)$-space.
Now, $H=H_1 + m_1\Gamma^1$ and $\Gamma^1$ anticommutes
with $H_1$. This means that the zero energy states
of $H_1$ are eigenstates of $\Gamma^1=\Theta_1$, too.
As a result, the dispersion of the surface flat band
of the full Hamiltonian $H$ is given by
\begin{equation}
\label{eq:flat_band_disp_z}
E_{\pm}(k_x,k_z)=\pm 2A_2\sin k_x.
\end{equation}
Thus, the surface state dispersions become highly anisotropic, being flat in
$k_z$-direction, but dispersive in $k_x$-direction.

\begin{figure}
\begin{minipage}{0.495\textwidth}
\centering
\includegraphics[width=\textwidth]{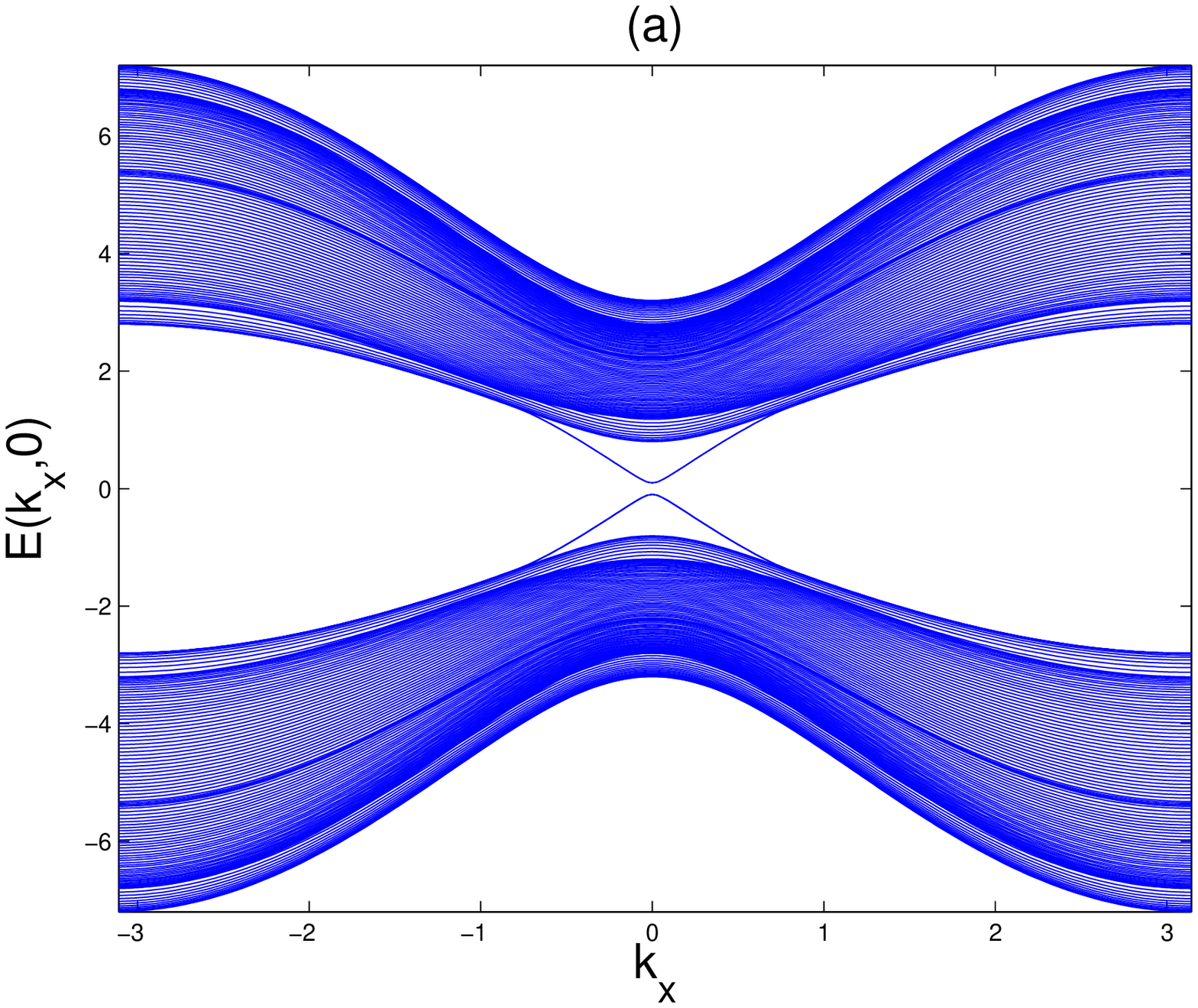}
\end{minipage}
\begin{minipage}{0.495\textwidth}
\centering
\includegraphics[width=\textwidth]{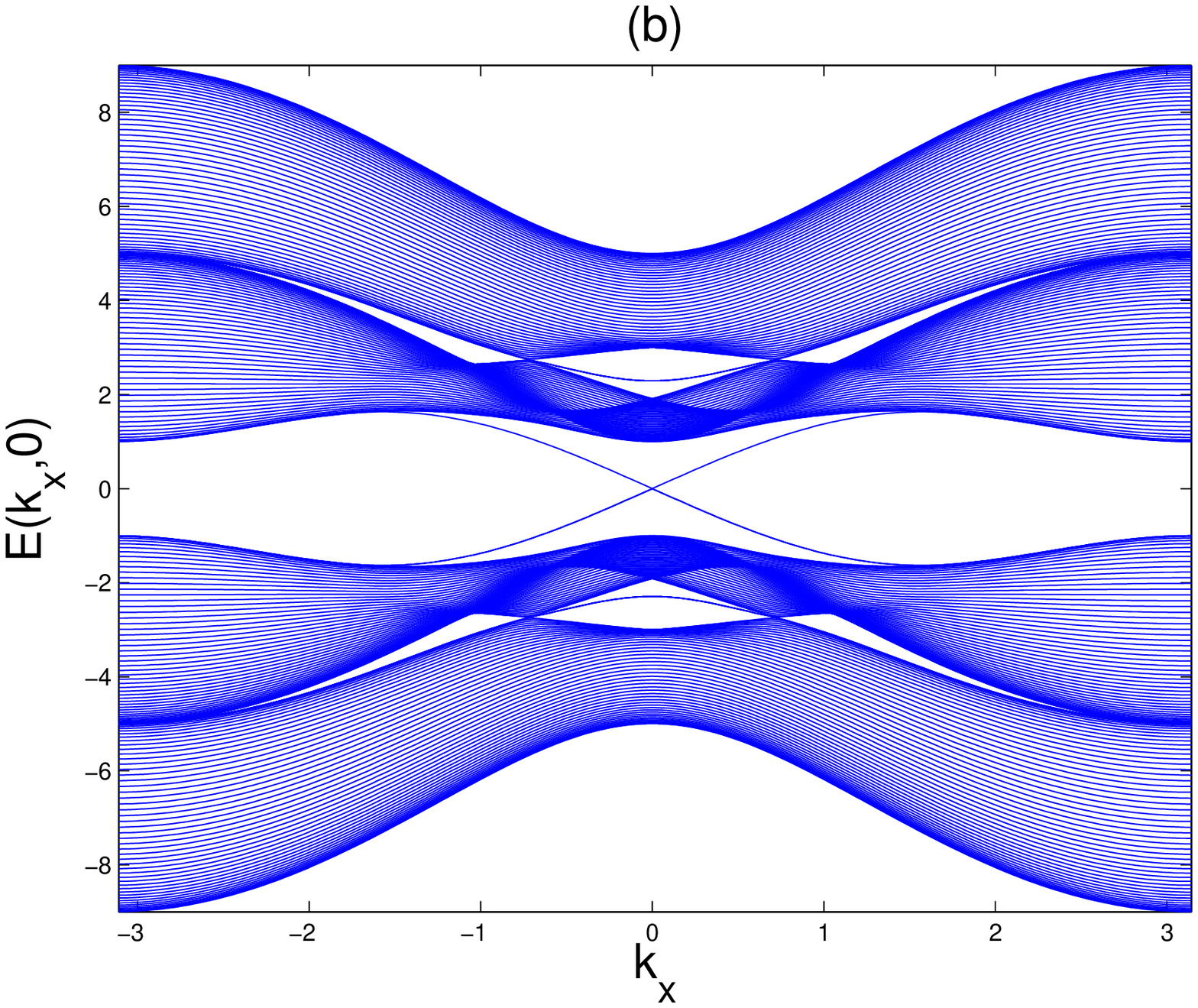}
\end{minipage}
\begin{minipage}{0.495\textwidth}
\centering
\includegraphics[width=\textwidth]{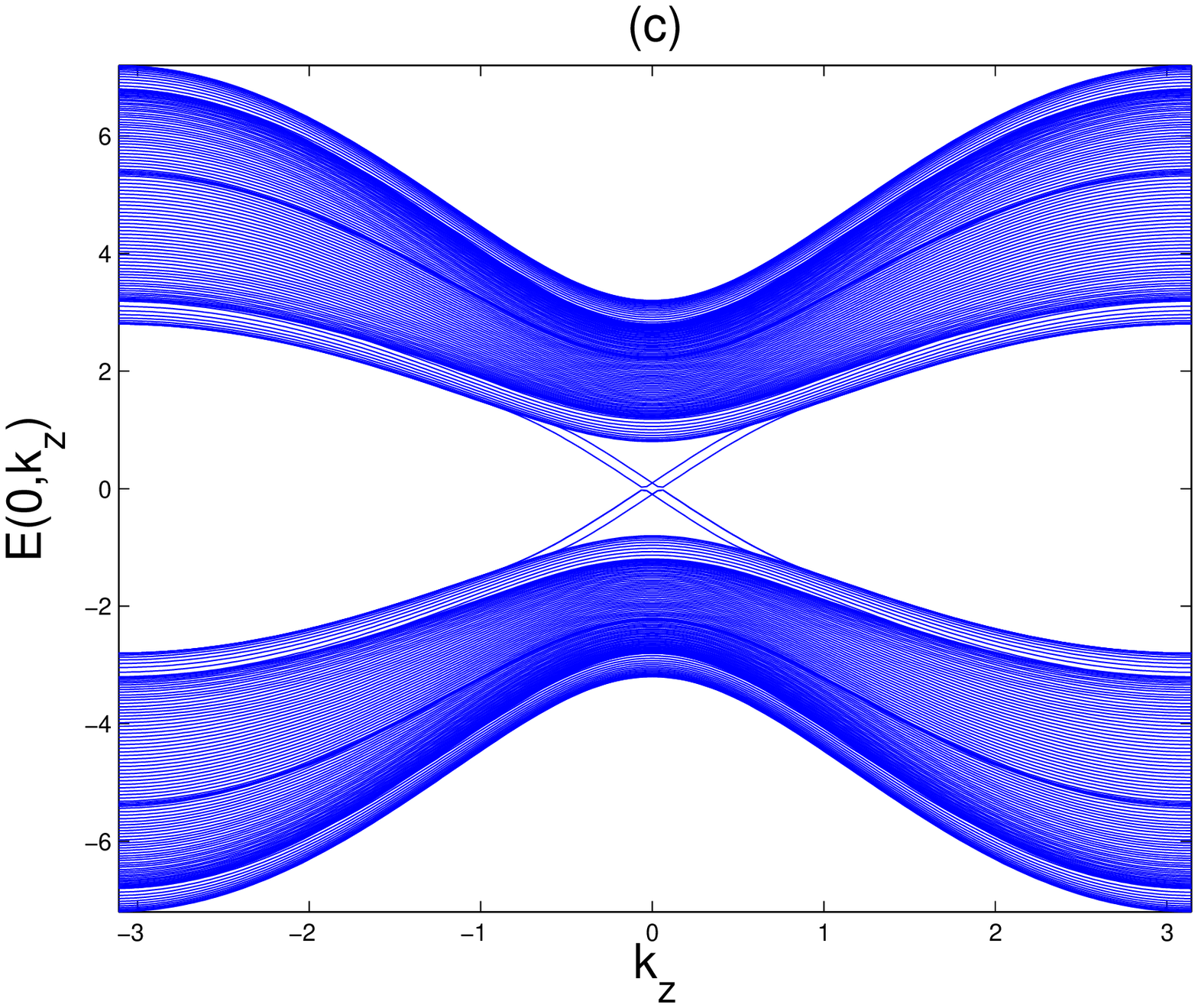}
\end{minipage}
\begin{minipage}{0.495\textwidth}
\centering
\includegraphics[width=\textwidth]{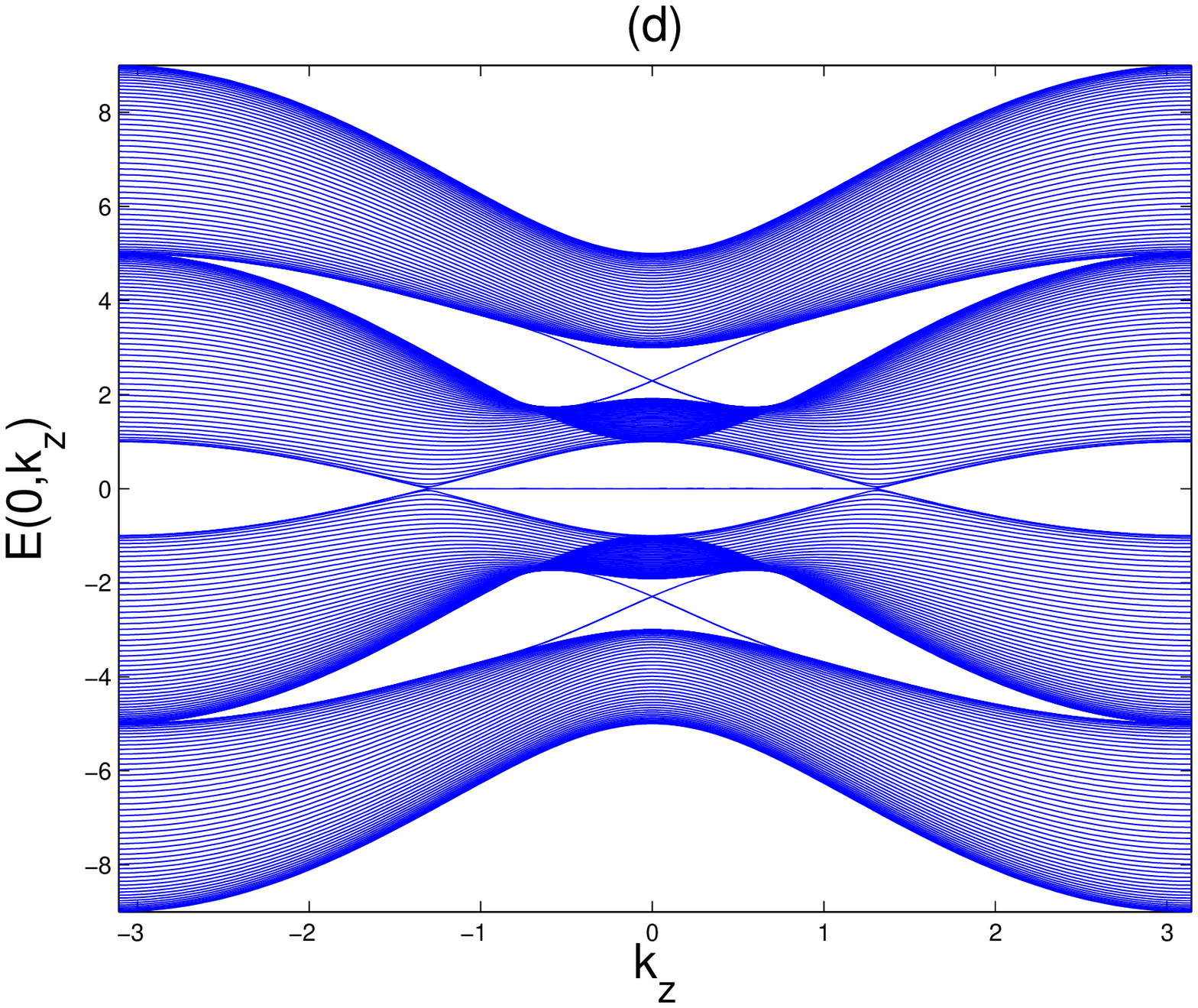}
\end{minipage}
\caption{\label{Fig8}
Numerical dispersions of bulk and surface states for model I with exchange fields 
$(V_x,V_y,V_z)=0.2M(1/2,1/2,1/\sqrt{2})$ for (a) and (c) and $2.0M(1/2,1/2,1/\sqrt{2})$ for (b) and (d).
In (a) and (b) $k_z=0$, and in (c) and (d) $k_x=0$.
The other parameters are same as in figure~\ref{Fig1}
}
\end{figure}

In Figure~\ref{Fig8} we illustrate an example of the general case, where all
three components of the exchange field are nonzero.
One sees from the figure that at small fields the surface states are splitted,
similarly as in Figure~\ref{Fig1}~(a) and (c). However, when the magnitude of the exchange
field exceeds $M$, a one-dimensional flat band appears in $k_z$ direction,
similarly as in Figure~\ref{Fig7}~(b) and (d).

In order to understand the appearance of a flat band in
this general case let us consider the
following partial bulk Hamiltonian:
\begin{equation}
\label{eq:H_I_0_general}
	H_0(\mathbf k)=m_0({\mathbf k}) \Gamma^0 + m_2({\mathbf k}) \Gamma^2+
V_x \Gamma_x + V_y \Gamma_y + V_z \Gamma_z
\end{equation}
We first construct a symmetry operator $\Theta_{13}$ that anticommutes
with this $H_0$. We first note that the
operator $\Theta_1$ anticommutes with $\Gamma^0$, $\Gamma^2$, $\Gamma_y$, and
$\Gamma_z$, but commutes with $\Gamma_x$, while the 
operator $\Theta_3$ anticommutes with $\Gamma^0$, $\Gamma^2$, $\Gamma_x$, and
$\Gamma_y$, but commutes with $\Gamma_z$. If we choose, however, the
following linear superposition of $\Theta_1$ and $\Theta_3$
\begin{equation}
\label{eq:theta_13}
 \Theta_{13} = \frac{V_z}{\sqrt{V_x^2 + V_z^2}} \Theta_1 -
\frac{V_x}{\sqrt{V_x^2 + V_z^2}} \Theta_3
= \sin \chi \, \Theta_1 - \cos \chi \, \Theta_3
\end{equation}
it is easy to see that $\Theta_{13}$ anticommutes with the four
operators $\Gamma^0$, $\Gamma^2$, $\Gamma_y$, and $V_x\Gamma_x + V_z
\Gamma_z$.
Here, we defined the angle $\chi$ via $V_x=V_0 \cos \chi$
and $V_z=V_0 \sin \chi$ with $V_0=\sqrt{V_x^2+V_z^2}$.
Thus, the partial Hamiltonian $H_0$ possesses a chiral symmetry $\Theta_{13}$,
which depends on the direction of the exchange field.
For $V_z=0$ the symmetry operator reduces to $\Theta_3$, corresponding to
the case in section~\ref{subsecflat1} and for $V_x=0$ the
symmetry reduces to the case discussed in section~\ref{subsecflat2}.

Using the chiral symmetry $\Theta_{13}$ we can bring $H_0$ into
off-diagonal block form by transforming to the eigenbasis of
$\Theta_{13}$.
There are two eigenvectors of $\Theta_{13}$ with
eigenvalue -1, which are
$\eta_1=(0,0,\sin \frac{\chi}{2}, \cos \frac{\chi}{2})^T$ and $\eta_2=(
\cos \frac{\chi}{2}, -\sin \frac{\chi}{2},0,0)^T$
and two eigenvectors with eigenvalue +1, which are
$\eta_3=(\sin \frac{\chi}{2}, \cos \frac{\chi}{2},0,0)^T$ and 
$\eta_4=(0,0,\cos \frac{\chi}{2}, -\sin \frac{\chi}{2})^T$.
In this basis $H_0$ becomes block off-diagonal
\begin{eqnarray}
\label{eq:Block_form_general_1}
\nonumber H_0(\mathbf k)
&=\left(\begin{array}{cc}
0 & D_2^{\dagger}(\mathbf k)\\
D_2(\mathbf k) & 0
\end{array}\right),
\end{eqnarray}
where the block $D_2(\mathbf k)$ is
\begin{equation}
\label{eq:Block_form_general_2}
D_2(\mathbf k)=\left(\begin{array}{cc}
m_0 & V_0 + i (m_2 + V_y)  \\
V_0 + i (m_2 - V_y ) & m_0
\end{array}\right)
\end{equation}
and its determinant
\begin{equation}
\label{eq:Block_form_general_det}
\mathrm{det} \; D_2(\mathbf k)=
m_0^2 +m_2^2 - V_0^2 - V_y^2 - 2 i m_2 V_0
\end{equation}
The Hamiltonian $H_0$ possesses a two-dimensional zero energy surface flat
band, if the corresponding winding number becomes nonzero. Following 
appendix III we find the criterion that
\begin{equation}
\label{eq:criterion_general_det}
\left( V^2 - \tilde m_0^2 \right) \left( V^2 - \left( \tilde m_0 - 4 B_2
\right)^2 \right) < 0
\end{equation}
where $\tilde m_0(\mathbf k)=M - 2 B_2 ( 1- \cos k_x ) - 2 B_1 ( 1- \cos k_z
)$. In the vicinity of $k_x=k_z=0$ this criterion can usually be fulfilled
for $V>M$, if $V$ does not become too large.

Having seen that the partial Hamiltonian $H_0$ possesses a two-dimensional 
zero energy surface flat band we can apply
\begin{equation}
\label{eq:H_I_prime_general}
	H'(\mathbf k)=m_1({\mathbf k}) \Gamma^1 + m_3({\mathbf k}) \Gamma_I^3
\end{equation}
as a perturbation to find the approximate surface state dispersion
for the full Hamiltonian. For that purpose we need the surface state
wave function $\psi_s$. As the zero energy state is a common eigenstate
of $H_0$ and $\Theta_{13}$, $\psi_s$ will be a superposition of
two eigenvectors of $\Theta_{13}$ with the same eigenvalue, like
\begin{equation}
\label{eq:psis_general}
\psi_s(y) = f_1(y) \eta_1 + f_2(y) \eta_2
\end{equation}
The two functions $f_1(y)$ and $f_2(y)$ have to be determined from the
differential equation $H_0(k_y \rightarrow - i \partial_y) \psi_s = 0$ 
and are found to be superpositions of
three exponentially decaying functions. An analogous superposition
of the +1 eigenstates of $\Theta_{13}$ leads to exponentially
increasing functions, which cannot fulfil the boundary conditions.
They correspond to solutions localized at the opposite boundary. 
Both functions $f_1(y)$ and $f_2(y)$ fulfil
$f_1(y=0)=f_2(y=0)=0$ and are normalized such that
\begin{equation}
\label{eq:psis_norm}
\int_0^\infty dy \left| f_1(y) \right|^2 +
\int_0^\infty dy \left| f_2(y) \right|^2 = 1 \; .
\end{equation}

The energy of the perturbed system is given by
\begin{equation}
\label{eq:E_perturbed}
E = \left\langle \psi_s \left| m_1 \Gamma^1 + m_3 \Gamma_I^3 \right| \psi_s \right\rangle
\end{equation}
By direct calculation we find that
\begin{equation}
\label{eq:Gamma3I_matrix_elements}
\left\langle \eta_i \left| \Gamma_I^3 \right| \eta_j \right\rangle = 0
\quad \mbox{for} \quad i,j \in \left\{ 1,2 \right\}
\end{equation}
and
\begin{equation}
\label{eq:Gamma1_matrix_elements}
\left\langle \eta_i \left| \Gamma^1 \right| \eta_j \right\rangle = - \sin \chi
\delta_{ij} = - \frac{V_z}{\sqrt{V_x^2+V_z^2}} \delta_{ij}
\quad \mbox{for} \quad i,j \in \left\{ 1,2 \right\} \; .
\end{equation}
As a result we find for the energy of the surface states
of the full Hamiltonian:
\begin{equation}
\label{eq:flat_band_disp_A}
E(k_x,k_z)=- m_1 \frac{V_z}{\sqrt{V_x^2+V_z^2}} =
- \frac{V_z}{\sqrt{V_x^2+V_z^2}} 2A_2 \sin k_x
\end{equation}
From this expression we see that for $V_z \neq 0$ we always have a
one-dimensional flat band in $k_z$-direction.
The group velocity in $x$-direction can be tuned by rotating 
the exchange field within the $xz$-plane and vanishes when
$V_z$ becomes zero. This expression shows how the
one-dimensional flat band develops into
the two-dimensional flat band for exchange field within the $xy$-plane.

\subsection{Weyl semimetal}
\label{Weyl}

In this section we show that the semimetallic state of model I for
an exchange field $V>V_{cr}$ and $V_z \neq 0$ is actually a realization
of a Weyl semimetal. The Weyl semimetallic phase can be viewed as
a three-dimensional generalization of the two-dimensional Dirac
electrons in graphene, as has been pointed out recently 
\cite{Wan,Balents,Burkov}. In contrast to graphene, just two
linearly dispersing adjacent bands touch at a finite number of points
in the three-dimensional Brillouin zone. In the vicinity of these
Weyl nodes, which have also been termed ``diabolic'' points \cite{Murakami}, 
the effective two-band Hamiltonian can be written in the form
\begin{eqnarray}
H(\bf k)&=\hbar v_F \left( k_x \sigma_x + k_y \sigma_y +
k_z \sigma_z \right)
\label{eq:Ham_weyl}
\end{eqnarray}
where the Pauli matrices $\sigma_i$ do not need to refer to the
spin degree of freedom. Such a diabolic point is exceptionally stable
due to topology, as arbitrary perturbations cannot remove it unless
an annihilation with another diabolic point occurs. A recent theoretical 
work proposed the appearence of a Weyl semimetallic phase in 
pyrochlore iridates \cite{Wan}. It was shown that the surface states
of the Weyl semimetal may form open Fermi ``arcs'', i.e. Fermi lines
which terminate at the projection of the diabolic points onto
the surface Brillouin zone.

In appendix I we showed that for $V>V_{cr}$ model I enters a semimetallic
phase. If $V_z \neq 0$ the bulk spectrum indeed possesses either two
or four Fermi points on the $k_z$-axis, where two bands touch each other.
We still need to show that around these points the Hamiltonian
can be written in a form like Eq.~(\ref{eq:Ham_weyl}). In order to
do so we use a similar technique as we have used for determination
of the dispersion of the surface states. Let ${\mathbf k}_0 =
(0,0,k_{z,0})$ be the position of a Fermi point. We first consider
the partial bulk Hamiltonian at ${\mathbf k}_0$
\begin{equation}
\label{eq:H_weyl_0}
	H_0(\mathbf k)=m_0({\mathbf k}_0) \Gamma^0 + m_3({\mathbf k}_0) \Gamma_I^3+
V_x \Gamma_x + V_y \Gamma_y + V_z \Gamma_z
\end{equation}
and construct a symmetry operator $\Theta_{12}$ that anticommutes with it.
The zero energy eigenstates of $H_0$ are then simultaneous eigenstates
of $\Theta_{12}$. From the eigenstates of $\Theta_{12}$ we
determine the two zero energy eigenstates of $H_0$ at the Fermi point.
We then expand the full Hamiltonian to lowest order in $k_x$, $k_y$, and $k_z-k_{z,0}$
around the Fermi points. This leads to a perturbation of the form
\begin{eqnarray}
	H'(\mathbf k)&=- 2 B_1 \left( k_z-k_{z,0}\right) \sin k_{z,0} \Gamma^0
+ 2 A_2 k_x \Gamma^1 + 2 A_2 k_y \Gamma^2 + \nonumber \\
& + 2 A_1 \left( k_z-k_{z,0}\right) \cos k_{z,0} \Gamma_I^3 
\label{eq:H_weyl_1}
\end{eqnarray}
The effective low energy $2 \times 2$ Hamiltonian near the Fermi points
is obtained by degenerate perturbation theory within the subspace
of the two said eigenstates.

To construct the symmetry operator $\Theta_{12}$ we first note that the
operator $\Theta_1$ anticommutes with $\Gamma^0$, $\Gamma_I^3$, $\Gamma_y$, and
$\Gamma_z$, but commutes with $\Gamma_x$, while the 
operator $\Theta_2$ anticommutes with $\Gamma^0$, $\Gamma_I^3$, $\Gamma_x$, and
$\Gamma_z$, but commutes with $\Gamma_y$. If we choose the
following linear superposition of $\Theta_1$ and $\Theta_2$
\begin{equation}
\label{eq:theta_12}
 \Theta_{12} = \frac{V_y}{\sqrt{V_x^2 + V_y^2}} \Theta_1 -
\frac{V_x}{\sqrt{V_x^2 + V_y^2}} \Theta_2
= \sin \varphi \, \Theta_1 - \cos \varphi \, \Theta_2
\end{equation}
it is easy to see that $\Theta_{12}$ anticommutes with the four
operators $\Gamma^0$, $\Gamma_I^3$, $\Gamma_z$, and $V_x\Gamma_x + V_y \Gamma_y$.
There are two eigenvectors of $\Theta_{12}$ with
eigenvalue -1, which are
$\eta_1=(0,0,i e^{-i \varphi},1)^T$ and $\eta_2=(-i e^{-i \varphi},1,0,0)^T$
and two eigenvectors with eigenvalue +1, which are
$\eta_3=(0,0,-i e^{-i \varphi},1)^T$ and $\eta_4=(i e^{-i \varphi},1,0,0)^T$.
One of the two zero energy eigenstates of $H_0$ is a linear combination
of $\eta_1$ and $\eta_2$, while the other one is a linear combination
of $\eta_3$ and $\eta_4$. After a staightforward calculation which
exploits the fact that at the Fermi points we have
$m_0^2+m_3^2 = V_x^2 + V_y^2 + V_z^2$ (see appendix I), we find the
following eigenstates of $H_0$:
\begin{eqnarray}
\psi_1 &= \frac{1}{2} \left( -i e^{-i \varphi},1,-e^{i(\chi-\varphi-\vartheta)}, i
  e^{i(\chi-\vartheta)}\right)^T \\
\psi_2 &= \frac{1}{2}  \left( - e^{-i (\varphi+\vartheta)},ie^{-i\vartheta},-i e^{i(\chi-\varphi)}, 
  e^{i\chi}\right)^T 
\label{eq:psi_12_weyl}
\end{eqnarray}
where
\begin{equation}
e^{i\chi} = \frac{m_0 + i m_3}{\sqrt{m_0^2+m_3^2}}
\quad \mbox{and} \quad
e^{i\vartheta} = \frac{\sqrt{V_x^2 + V_y^2} + i V_z}{\sqrt{V_x^2 + V_y^2 + V_z^2}}
\end{equation}
To find the effective low energy $2 \times 2$ Hamiltonian near the Fermi
points for convenience we use the two basis states
\begin{equation}
\psi_1' = \frac{1}{\sqrt{2}} \left( \psi_1 + \psi_2 \right)
\quad \mbox{and} \quad
\psi_2' = \frac{1}{\sqrt{2}} \left( \psi_1 - \psi_2 \right)
\end{equation}
In this basis the Hamiltonian $H'$ becomes
\begin{eqnarray}
	H'&= \left[ - 2 B_1 \sin k_{z,0} \cos \chi
+ 2 A_1 \cos k_{z,0} \sin \chi \right] \left( k_z-k_{z,0}\right) \sigma_z
+ \nonumber \\
&  + 2 A_2 k_x \left[ - \sin \varphi \, \sigma_x + \cos \varphi \sin \vartheta \,
  \sigma_y \right] + \nonumber \\
& 2 A_2 k_y \left[ \cos \varphi \, \sigma_x + \sin \varphi \sin \vartheta \,
  \sigma_y \right]
\label{eq:H_weyl_2}
\end{eqnarray}
Apparently, this is an anisotropic Weyl-type Hamiltonian. For example
for $V_x=0$ this expression simplifies to
\begin{eqnarray}
	H'&= \left[ - 2 B_1 \sin k_{z,0} \cos \chi
+ 2 A_1 \cos k_{z,0} \sin \chi \right] \left( k_z-k_{z,0}\right) \sigma_z
+ \nonumber \\
&  - 2 A_2 k_x \, \sigma_x  +  2 A_2 k_y \sin \vartheta \, \sigma_y 
\label{eq:H_weyl_3}
\end{eqnarray}

Having seen that model I for $V>V_{cr}$ and $V_z \neq 0$ is 
a Weyl semimetal we can see now that the one-dimensional
surface flat band from the previous section 
Eq.~(\ref{eq:flat_band_disp_z}) is just
a Fermi ``arc'' in the sense of Ref.~\cite{Wan}.
It exists only in a limited range of $k_z$ values
given by Eq.~(\ref{eq:flat_band_cond_3_z}).
The end points of the Fermi arc are just
the projections of the Weyl nodes onto the
surface Brillouin zone $(k_x,k_z)$, as one sees
by setting $k_x=0$ in Eq.~(\ref{eq:flat_band_cond_3_z})
and comparison with Eqs.~(\ref{eq:fc}) and (\ref{eq:fcv0})
in appendix I.

\subsection{Boundary perpendicular to the $z$-direction}

The case with a $z$-boundary is much easier to treat
than the case with a $y$-boundary.
In this case we can decompose the Hamiltonian
in the following way:
\begin{eqnarray}
\nonumber H(\bf k)&=H_0(\mathbf k)+H'(\mathbf k),
\label{eq:Ham_B_z_I}
\end{eqnarray}
where
\begin{eqnarray}
\label{eq:H_0_H'_B_z}
H_0(\mathbf k)&=(\tilde m_0(\mathbf k)+B_1\partial_z^2)\Gamma^0+2A_1\partial_z\Gamma_I^3,\\
H'(\mathbf k)&=m_1(\mathbf k)\Gamma^1+m_2(\mathbf k)\Gamma^2+V_x\Gamma_x
+V_y\Gamma_y+V_z\Gamma_z.
\label{eq:H_0_H'_B_z_2}
\end{eqnarray}
Here, $\tilde m_0(\mathbf k)=M - 2 B_2 ( 1- \cos k_x ) - 2 B_2 ( 1- \cos k_y
)$ and $\mathbf k=(k_x,k_y)$ now. 

We are now in the lucky situation that
$\Gamma_x$, $\Gamma_y$ and $\Gamma_z$ all
commute with $H_0$, and both $\Gamma^1$ and $\Gamma^2$ anticommute with it.
Thus, the surface states for an exchange field in arbitrary direction
can be found from the zero energy surface states of $H_0$
using the method from appendix II. 

We first determine the zero energy surface states of $H_0$ by
noting that $H_0$ commutes with $\Gamma_z$ and anticommutes
with the operator $\mathbb{I}_{2 \times 2}\otimes \tau_z$.
The common eigenstates of these two operators are
just $(1,0,0,0)^T$, $(0,1,0,0)^T$,
$(0,0,1,0)^T$, and $(0,0,0,1)^T$. 
We try the following two ans\"atze (the other two eigenstates 
leading to exponentially increasing functions again):
\begin{eqnarray}
\label{eq:sur_s_B_z}
\psi_{1,\mathbf k}(y)=(1,0,0,0)^Tf_{\mathbf k}(z),\\
\psi_{2,\mathbf k}(y)=(0,1,0,0)^Tf_{\mathbf k}(z),
\label{eq:sur_s_B_z_2}
\end{eqnarray}
where $f_{\mathbf k}(z)$ is solution of the equation
\begin{equation}
\label{eq:solu_f_B_z}
[\tilde m_0(\mathbf k)+B_1\partial_z^2+2A_1\partial_z]f_{\mathbf k}(z)=0,
\end{equation}
Solving the differential equation (\ref{eq:solu_f_B_z}) we
find that $f_{\mathbf k}(z)$ is given by
\begin{equation}
\label{eq:solu_f_B_z_2}
f_{\mathbf k}(z)=e^{-\frac{A_1}{B_1} z}\sinh\left(\sqrt{\frac{A_1^2}{B_1^2}-\frac{\tilde m_0(\mathbf k)}{B_1}} \;
z\right).
\end{equation}
This solution can only fulfil the boundary condition for $z \rightarrow
\infty$, if $\tilde m_0(\mathbf k) > 0$. For those $\mathbf k$ values
where this condition is not fulfilled anymore, a surface state
does not exist. The form of the solutions Eq.~(\ref{eq:sur_s_B_z})
and (\ref{eq:sur_s_B_z_2}) means that the surface state only occupies
orbital 1, leaving orbital 2 empty. Correspondingly, the surface state at the opposite
surface of the system is found to occupy orbital 2 only.

To obtain the surface states of $H$ from the ones of $H_0$
we have to determine those linear combinations of
$\psi_{1,\mathbf k}$ and $\psi_{2,\mathbf k}$ that diagonalize
$H'$, i.e.
\begin{eqnarray}
\label{eq:zero_mode_sol_B_z_3}
\Psi_{1,\mathbf k}&=a_1(\mathbf k)\psi_{1,\mathbf k}(z)+b_1(\mathbf k)\psi_{2,\mathbf k}(z),\\
\Psi_{2,\mathbf k}&=a_2(\mathbf k)\psi_{1,\mathbf k}(z)+b_2(\mathbf k)\psi_{2,\mathbf k}(z).
\end{eqnarray}
If we write $\xi_i(\mathbf k)=(a_i(\mathbf k),b_i(\mathbf k))^T$ the
coefficients are determined from the equation
\begin{equation}
\label{eq:zero_mode_equ}
[(V_x+m_1(\mathbf k))\sigma_x+(V_y+m_2(\mathbf k))\sigma_y+V_z\sigma_z]\xi_1(\mathbf k)=E_1(\mathbf k)\xi_1(\mathbf k).
\end{equation}
The eigenenergies are given by
\begin{eqnarray}
\label{eq:s_s_ener_B_z}
E_{1,\pm}(\mathbf k)=\pm\sqrt{V_z^2+(2 A_2 \sin k_x + V_x)^2+(2 A_2 \sin k_y +V_y)^2}.
\end{eqnarray}
This expression tells us that the components of the exchange field parallel to
the surface ($V_x$ and $V_y$) for small momenta just shift and split the surface Dirac cone
without opening a gap and without changing the group velocity. The component $V_z$
perpendicular to the surface opens a gap, however. This behavior is
in agreement with previous work \cite{RLChu}. A flat band does not
appear in this geometry. Even though the system is still a Weyl semimetal
in the bulk, a surface Fermi arc does not appear, because the Weyl nodes
all sit on the $k_z$-axis. Thus, their projection onto the surface
Brillouin zone is the single point $k_x=k_y=0$ and the Fermi arc
is not present.

The full surface state wave functions can be written in the form
\begin{eqnarray}
\label{eq:sur_s_B_z_4}
\Psi_{+,\mathbf k}(y)&=\left( e^{-i\phi_{\mathbf k}} \cos
\frac{\theta_{\mathbf k}}{2},
\sin \frac{\theta_{\mathbf k}}{2},0,0 \right)^T f_{\mathbf k}(z) \\
\Psi_{-,\mathbf k}(y)&=\left( -e^{-i\phi_{\mathbf k}} \sin
\frac{\theta_{\mathbf k}}{2},
\cos \frac{\theta_{\mathbf k}}{2},0,0 \right)^T f_{\mathbf k}(z)
\end{eqnarray} 
Here, we have introduced two spherical angles $\phi$ and $\theta$
that define the direction of the vector $(m_1+V_x,m_2+V_y,V_z)$:
\begin{eqnarray}
\nonumber
\cos \phi_{\mathbf k} \sin \theta_{\mathbf k} &= \frac{2 A_2 \sin k_x+V_x}{
\sqrt{V_z^2+(2 A_2 \sin k_x + V_x)^2+(2 A_2 \sin k_y +V_y)^2}} \\
\nonumber
\sin \phi_{\mathbf k} \sin \theta_{\mathbf k} &= \frac{2 A_2 \sin k_y+V_y}{
\sqrt{V_z^2+(2 A_2 \sin k_x + V_x)^2+(2 A_2 \sin k_y +V_y)^2}} \\
\cos \theta_{\mathbf k} &= \frac{V_z}{
\sqrt{V_z^2+(2 A_2 \sin k_x + V_x)^2+(2 A_2 \sin k_y +V_y)^2}}
\nonumber
\end{eqnarray} 
For the spin texture of these surface states we find
\begin{eqnarray}
\nonumber
\left\langle \Psi_{\pm,\mathbf k} \left| \hat{s}_{1,x} \right|
\Psi_{\pm,\mathbf k} \right\rangle &= \pm \cos \phi_{\mathbf k} \sin \theta_{\mathbf k} \\
\nonumber
\left\langle \Psi_{\pm,\mathbf k} \left| \hat{s}_{1,y} \right|
\Psi_{\pm,\mathbf k} \right\rangle &= \pm \sin \phi_{\mathbf k} \sin \theta_{\mathbf k} \\
\nonumber
\left\langle \Psi_{\pm,\mathbf k} \left| \hat{s}_{1,z} \right|
\Psi_{\pm,\mathbf k} \right\rangle &= \pm \cos \theta_{\mathbf k} 
\end{eqnarray}
The spin in orbital 2 vanishes. This means that the spin is directed
along the vector $(m_1+V_x,m_2+V_y,V_z)$. For $V_z=0$ the spin-$z$ component
vanishes and the spin is oriented within the plane of the surface,
in contrast to the cases with the $y$-boundary.

\begin{figure}
\begin{minipage}{0.495\textwidth}
\centering
\includegraphics[width=\textwidth]{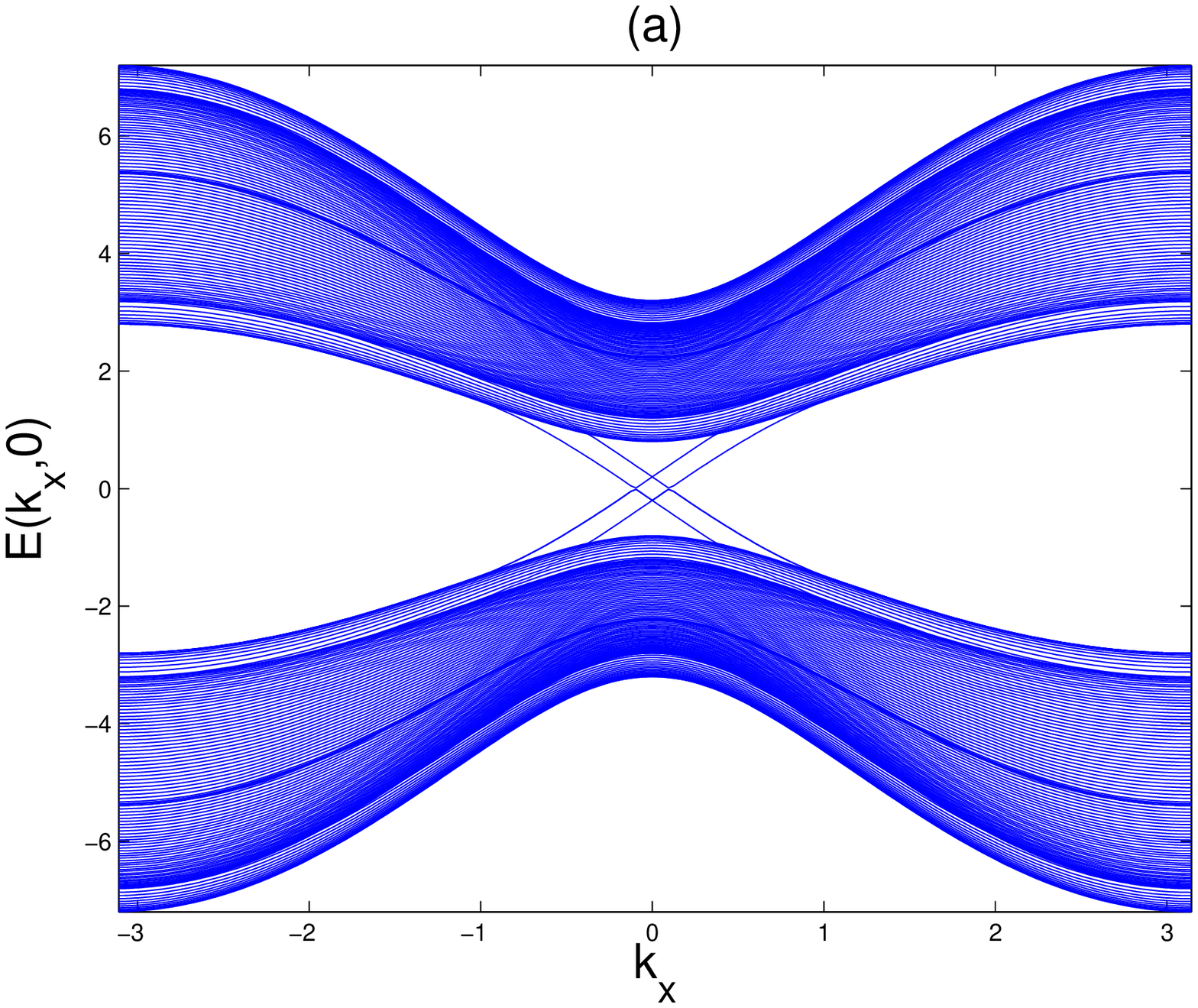}
\end{minipage}
\begin{minipage}{0.495\textwidth}
\centering
\includegraphics[width=\textwidth]{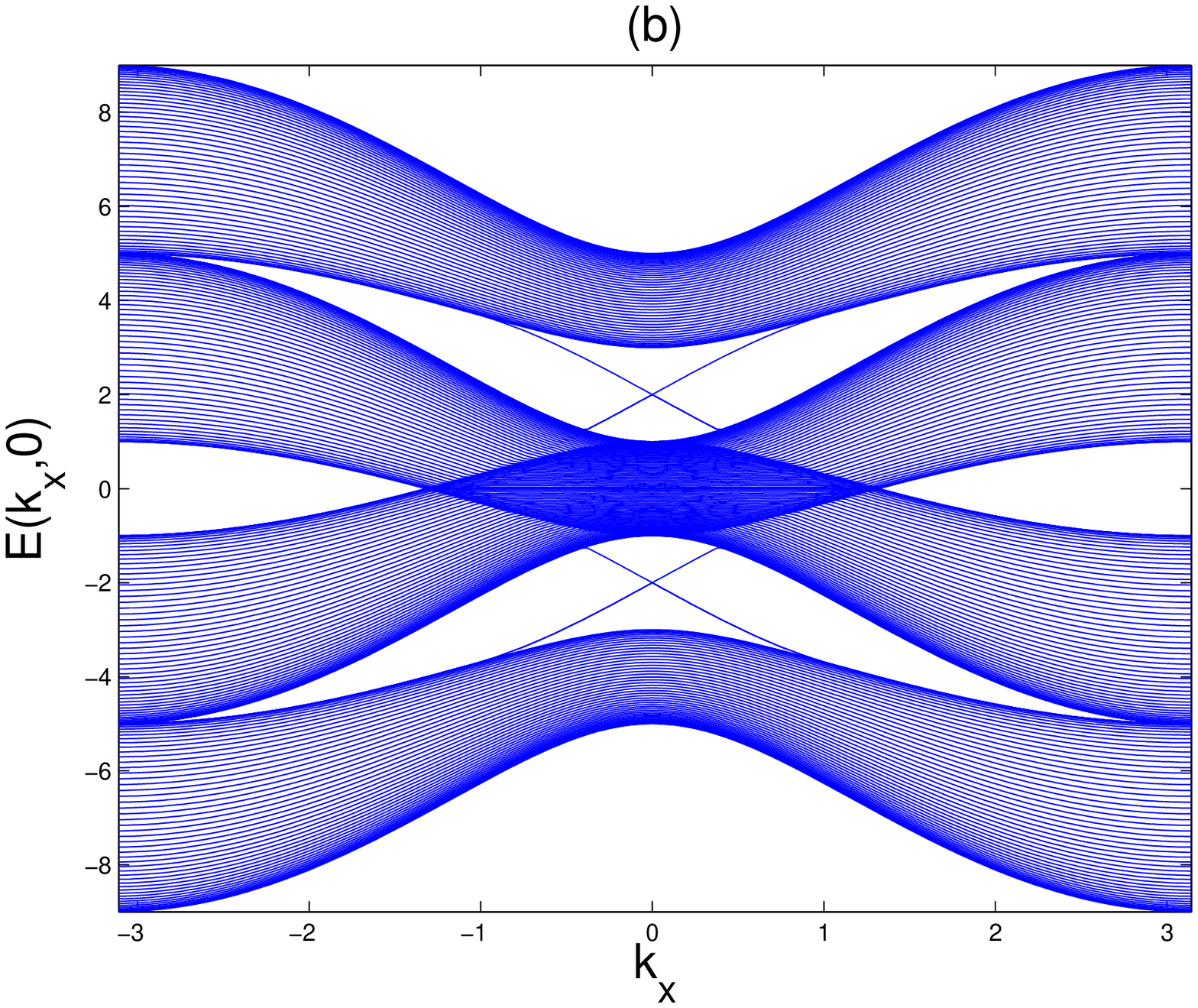}
\end{minipage}
\begin{minipage}{0.495\textwidth}
\centering
\includegraphics[width=\textwidth]{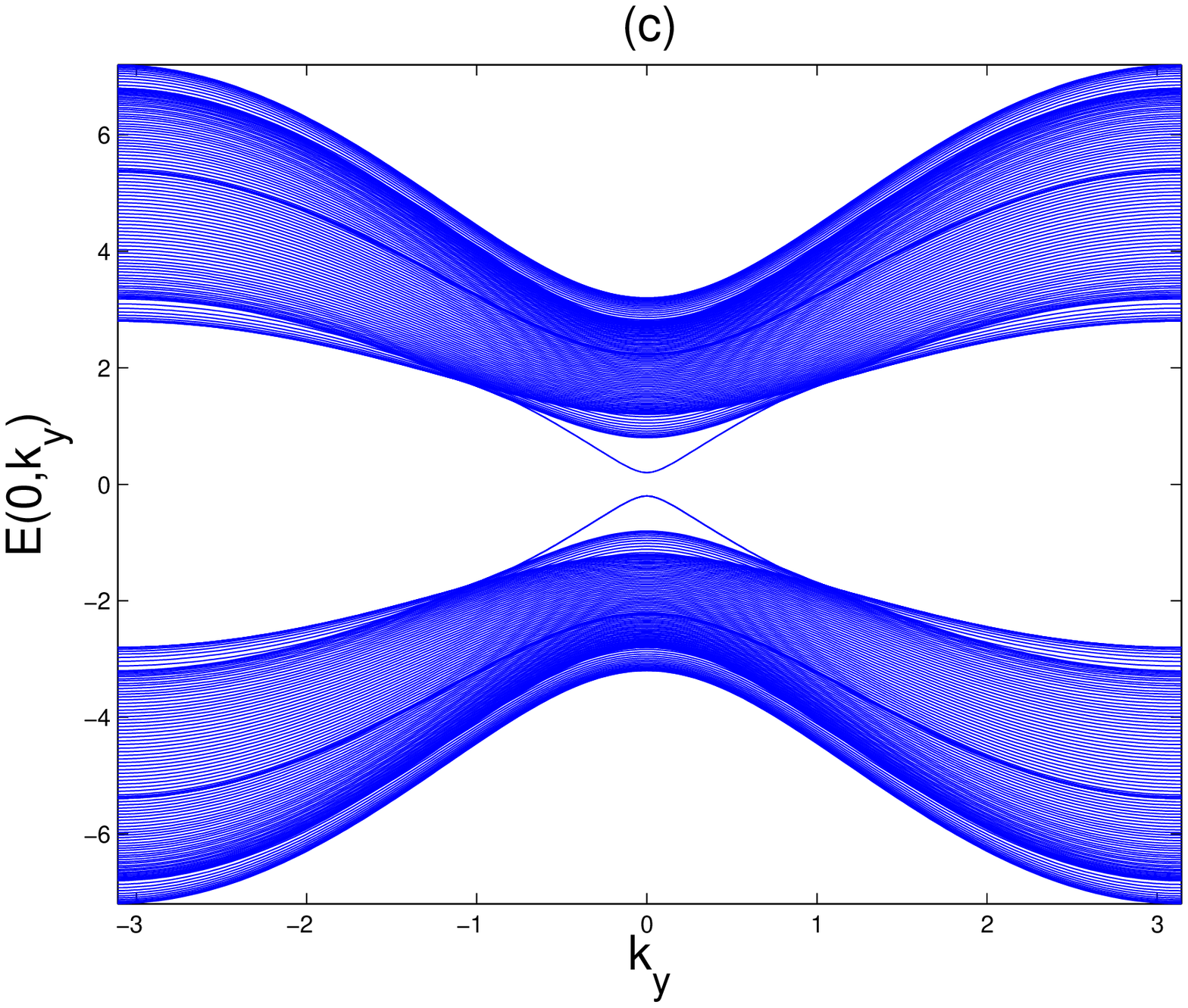}
\end{minipage}
\begin{minipage}{0.495\textwidth}
\centering
\includegraphics[width=\textwidth]{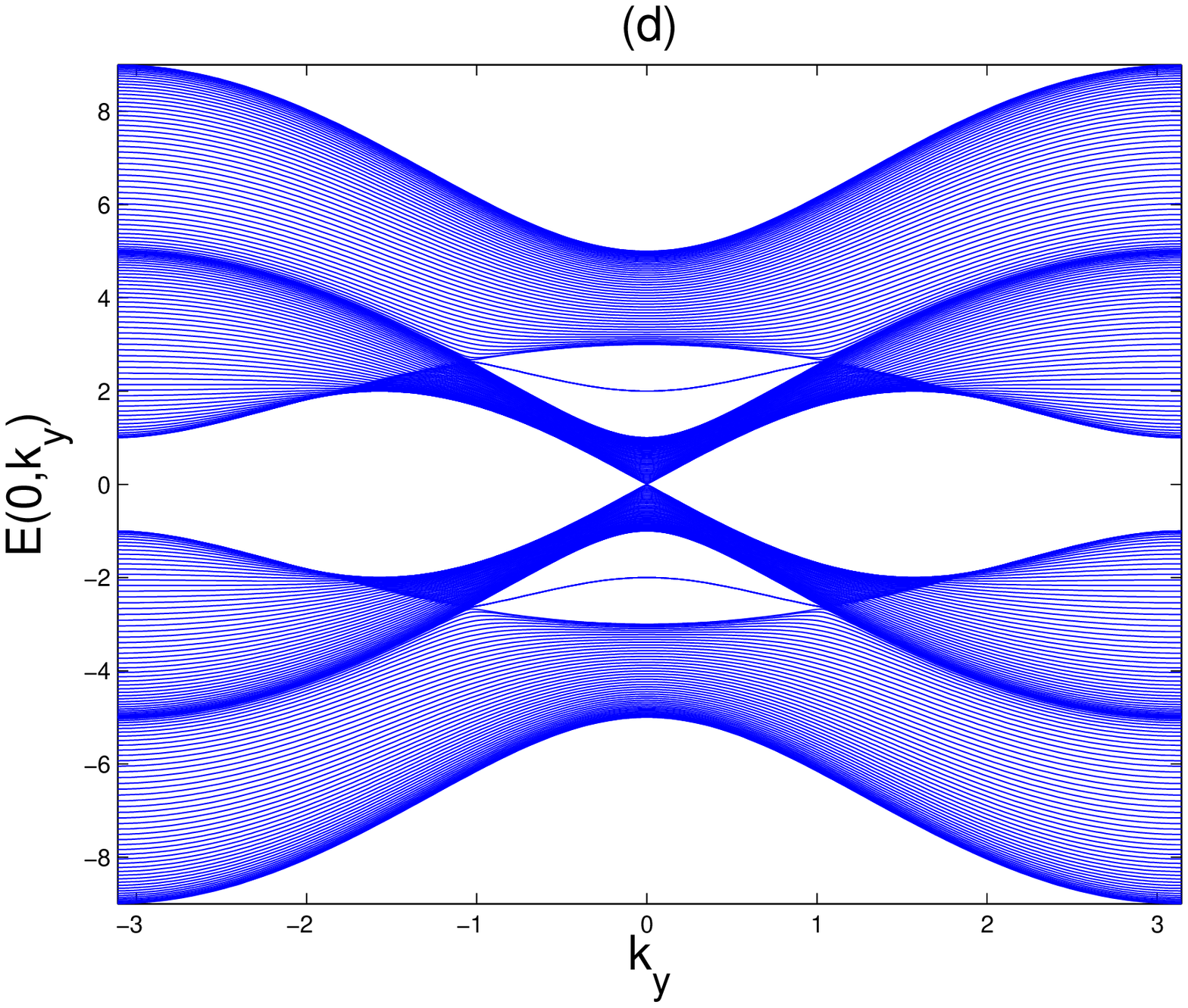}
\end{minipage}
\caption{\label{Fig9}
Numerical dispersions of bulk and surface states for model I with boundary in $z$-direction for 
$V_x/M=0.2$ in (a) and (c) and $2.0$ in (b) and (d).
In (a) and (b) $k_y=0$, and in (c) and (d) $k_x=0$.
The other parameters are same as in figure~\ref{Fig1}
}
\end{figure}

\begin{figure}
\begin{minipage}{0.495\textwidth}
\centering
\includegraphics[width=\textwidth]{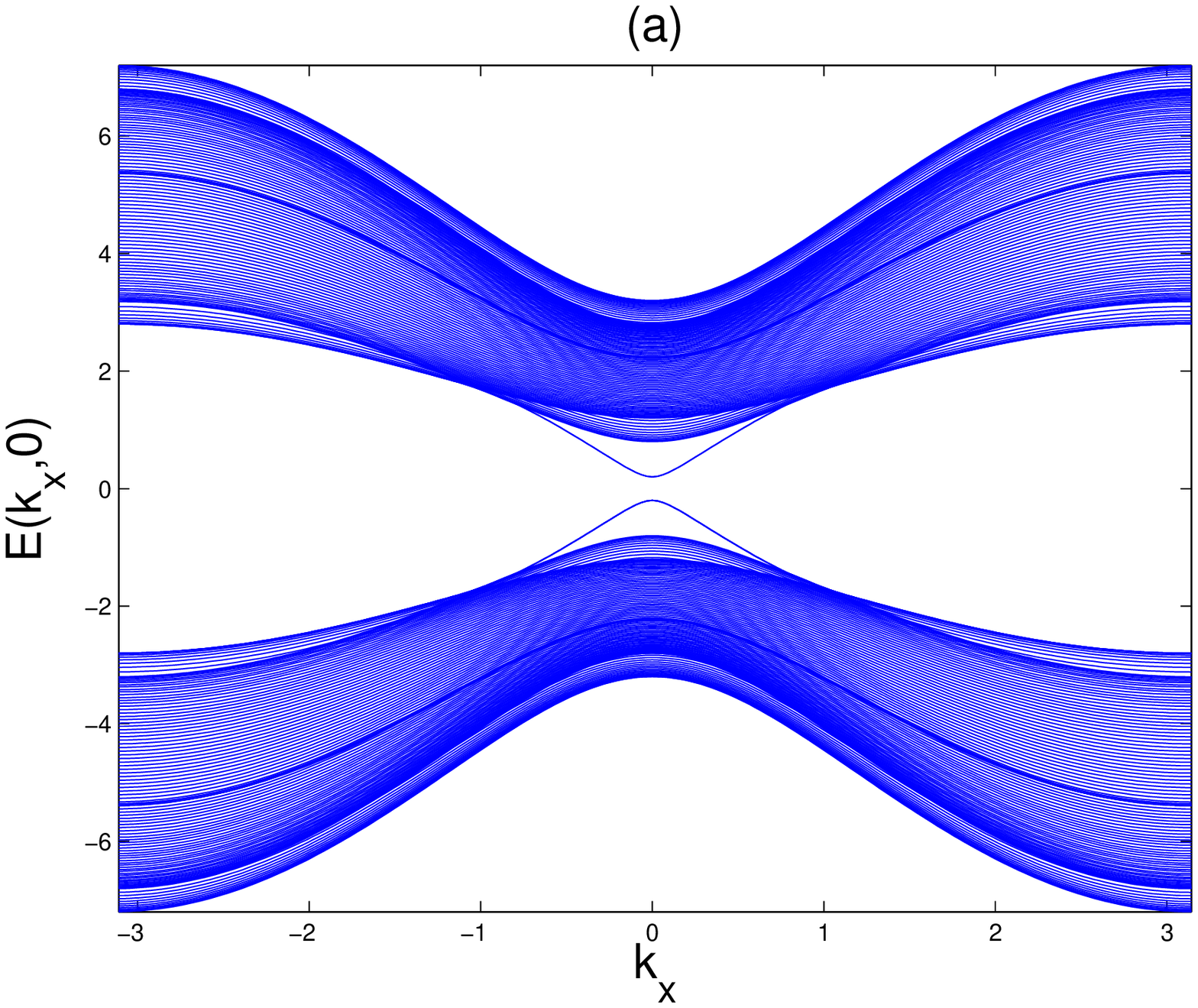}
\end{minipage}
\begin{minipage}{0.495\textwidth}
\centering
\includegraphics[width=\textwidth]{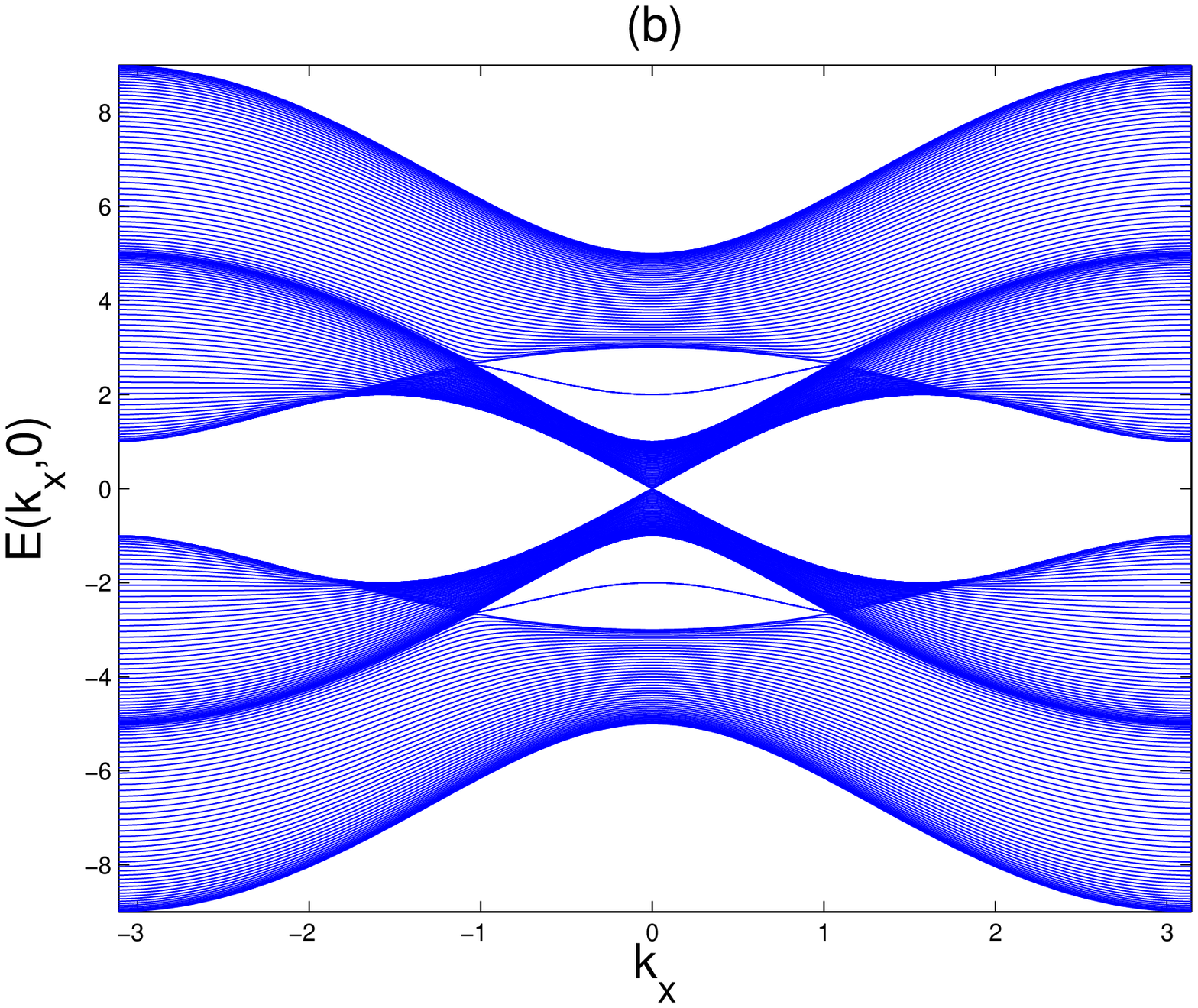}
\end{minipage}
\begin{minipage}{0.495\textwidth}
\centering
\includegraphics[width=\textwidth]{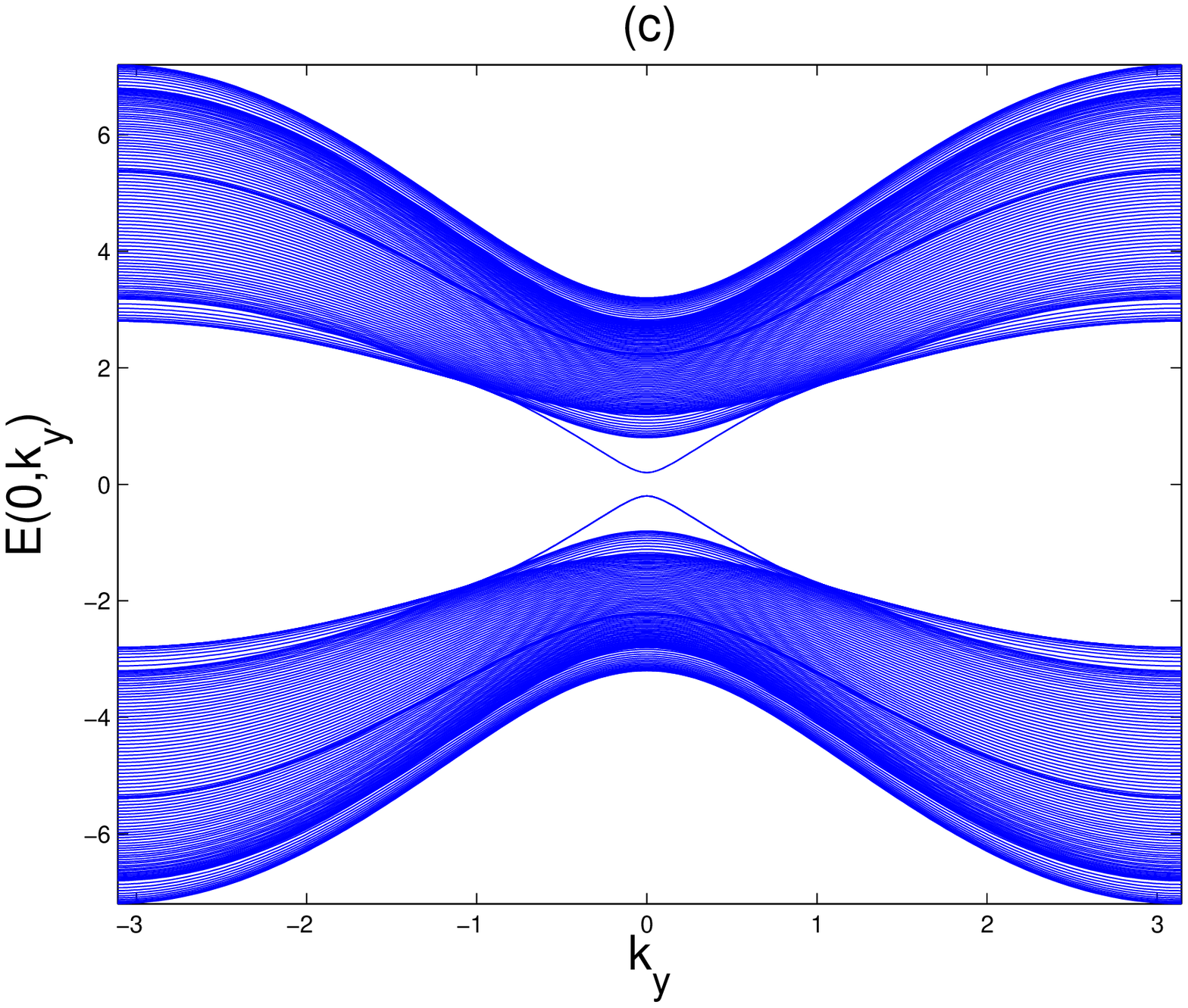}
\end{minipage}
\begin{minipage}{0.495\textwidth}
\centering
\includegraphics[width=\textwidth]{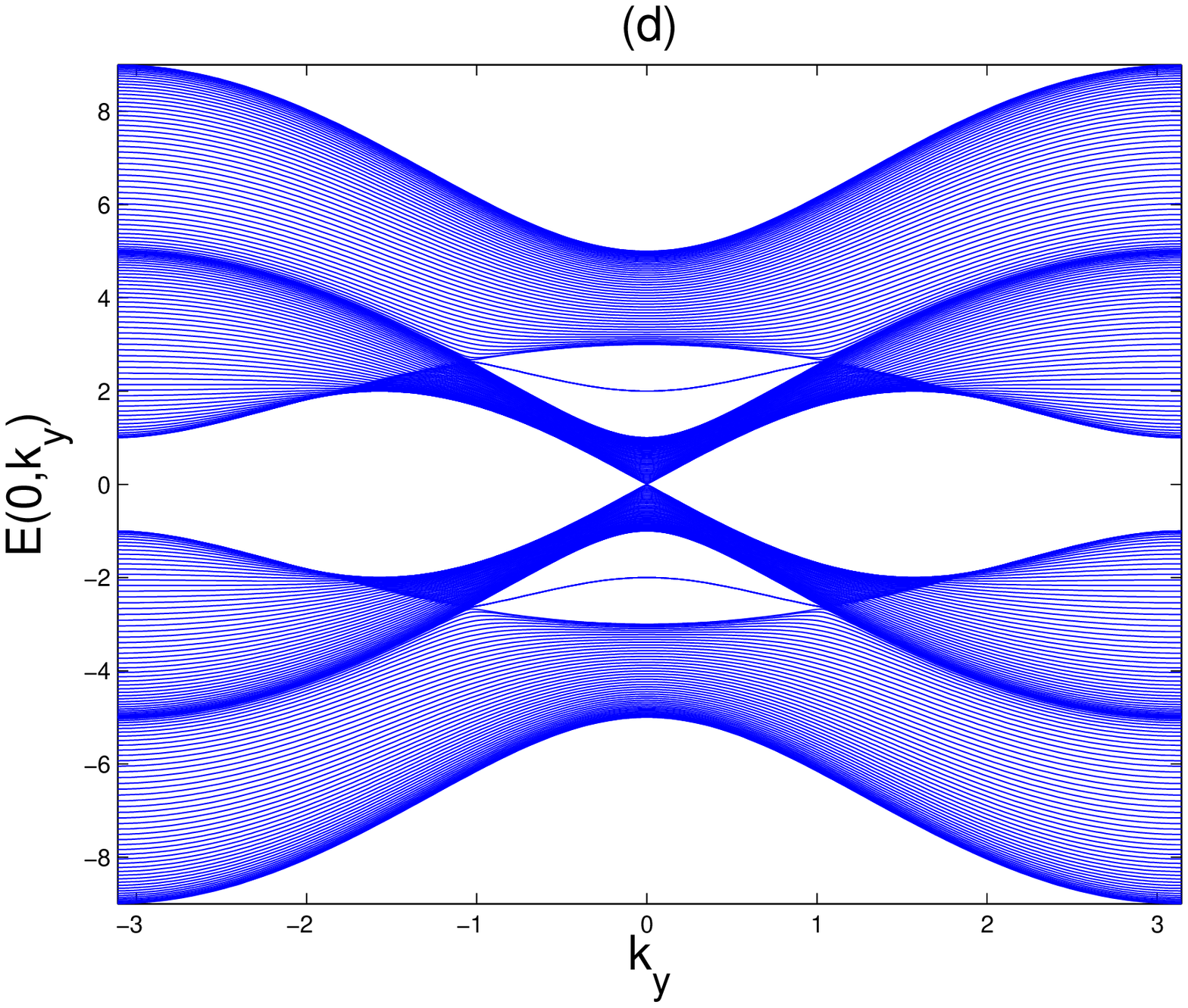}
\end{minipage}
\caption{\label{Fig10}
Numerical dispersions of bulk and surface states for model I with boundary in $z$-direction for 
$V_z/M=0.2$ in (a) and (c) and $2.0$ in (b) and (d).
In (a) and (b) $k_y=0$, and in (c) and (d) $k_x=0$.
The other parameters are same as in figure~\ref{Fig1}
}
\end{figure}

Figure~\ref{Fig9} shows numerical dispersions from a finite size system with
finite $V_x$ in agreement with Eq.~(\ref{eq:s_s_ener_B_z}). For larger values
of the exchange field $V_x>M$ the bulk gap has closed. The surface states
then exist within the bulk bands.

Figure~\ref{Fig10} shows numerical dispersions with finite $V_z$
confirming the opening of a gap for small values of $V_z$. 
Again, for larger values of the exchange field $V_z>M$ the
surface states only exist within the bulk bands and no flat band appears.

\section{Model II}

In this section we discuss the two nonequivalent cases for the
particle-hole symmetric model II. Model II is distinguished from model I
only in the coupling in $k_z$-direction by the matrix
$\Gamma_{II}^3=\sigma_z \otimes \tau_z$ instead of $\Gamma_I^3$.
The other $\Gamma$-matrices are the same as in model I.
The matrix $\Gamma_{II}^3$ has different commutation and
anti-commutation relations than $\Gamma_I^3$.
Model II is more symmetric in the sense, that the three spatial
directions possess equivalent couplings. Therefore, it does not
matter which direction of the boundary we consider.
For convenience we choose a boundary in $y$-direction, because this allows us
to build on the results from model I found in the previous
section.

\subsection{Finite $V_x$ and $V_y$}

\label{subsec2vxy}
The case with both $V_x$ and $V_y$ nonzero, but $V_z=0$ can be treated
along the same lines as has been discussed for model I in
section~\ref{subsec1vxy}.
We can again go over to polar coordinates in this
case and write $V_x=V_0 \cos \vartheta$ and $V_y=V_0 \sin \vartheta$.
The bulk energy bands Eq.~(\ref{eq:model2bulk})
can then be brought into the form:
\begin{eqnarray} 
\nonumber
	E_i^{II}(\mathbf k)&=\pm \Big\{ \left(m_1 \sin \vartheta - m_2 \cos
          \vartheta \right)^2+m_3^2 + \\ \label{eq:model2bulkb}
&+ \left( V_0 \pm \sqrt{m_0^2 + \left(m_1 \cos \vartheta + m_2 \sin
          \vartheta \right)^2} \right)^2 \Big\}^{1/2}
\end{eqnarray}
Again, the system is
insulating in the absence of an exchange field and the gap closes,
when $V_0$ reaches a critical value $V_{cr}\sim M$. In contrast to model I,
in model II $V_{cr}$ depends on the direction of the
exchange field as detailed in appendix I. In the semimetallic
state the Fermi surface is defined by three equations
$m_1 \sin \vartheta - m_2 \cos \vartheta = 0$, $m_3=0$, and
$V_0^2 = m_0^2 + \left(m_1 \cos \vartheta + m_2 \sin
          \vartheta \right)^2 $ now.
Therefore, the Fermi surface is pointlike, i.e. zero-dimensional.
Similarly as in section~\ref{Weyl} it can be shown by linearization
around the point nodes that the system is a Weyl semimetal in this case, too.

\begin{figure}[t]
\begin{minipage}{0.495\textwidth}
\centering
\includegraphics[width=\textwidth]{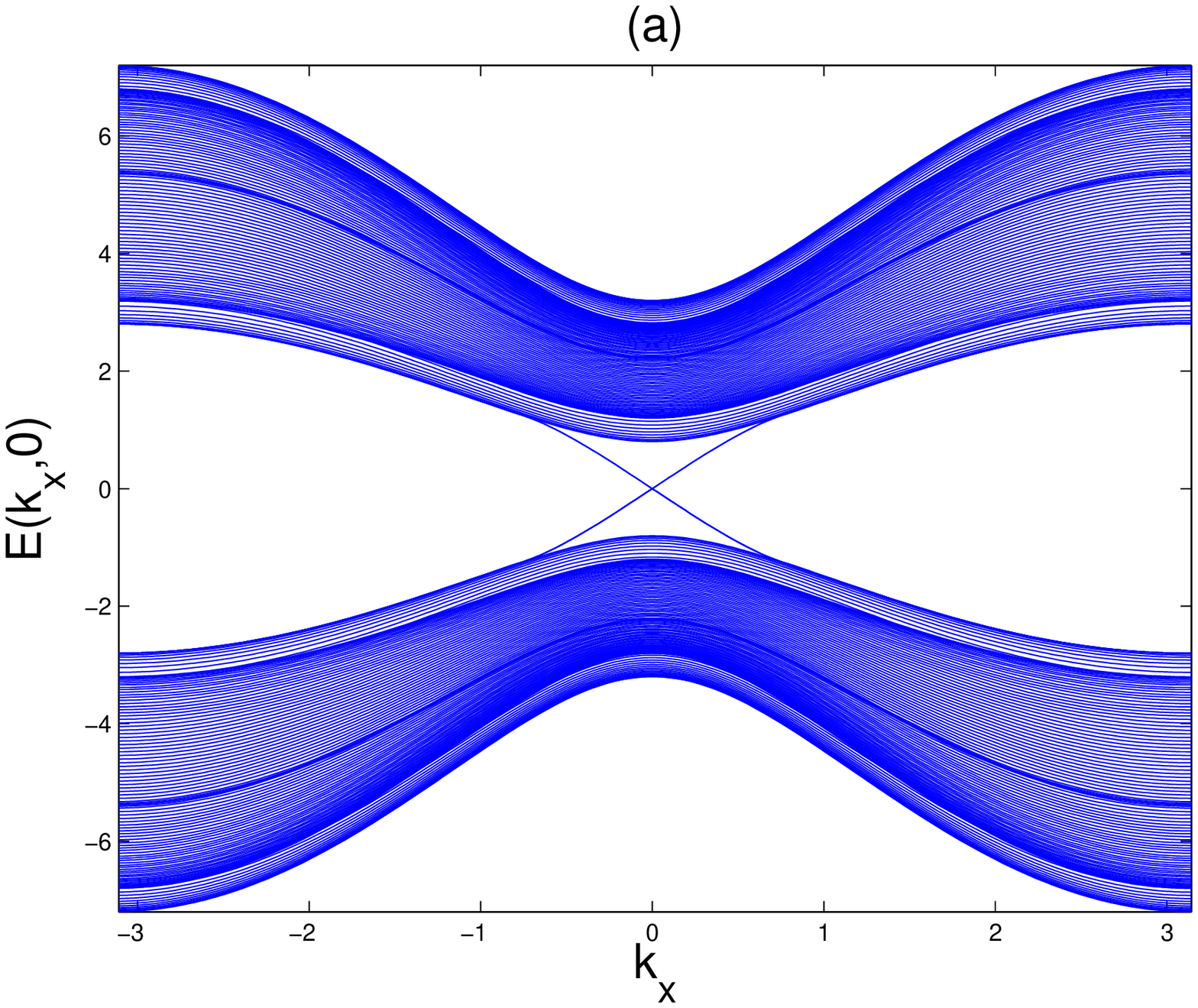}
\end{minipage}
\begin{minipage}{0.495\textwidth}
\centering
\includegraphics[width=\textwidth]{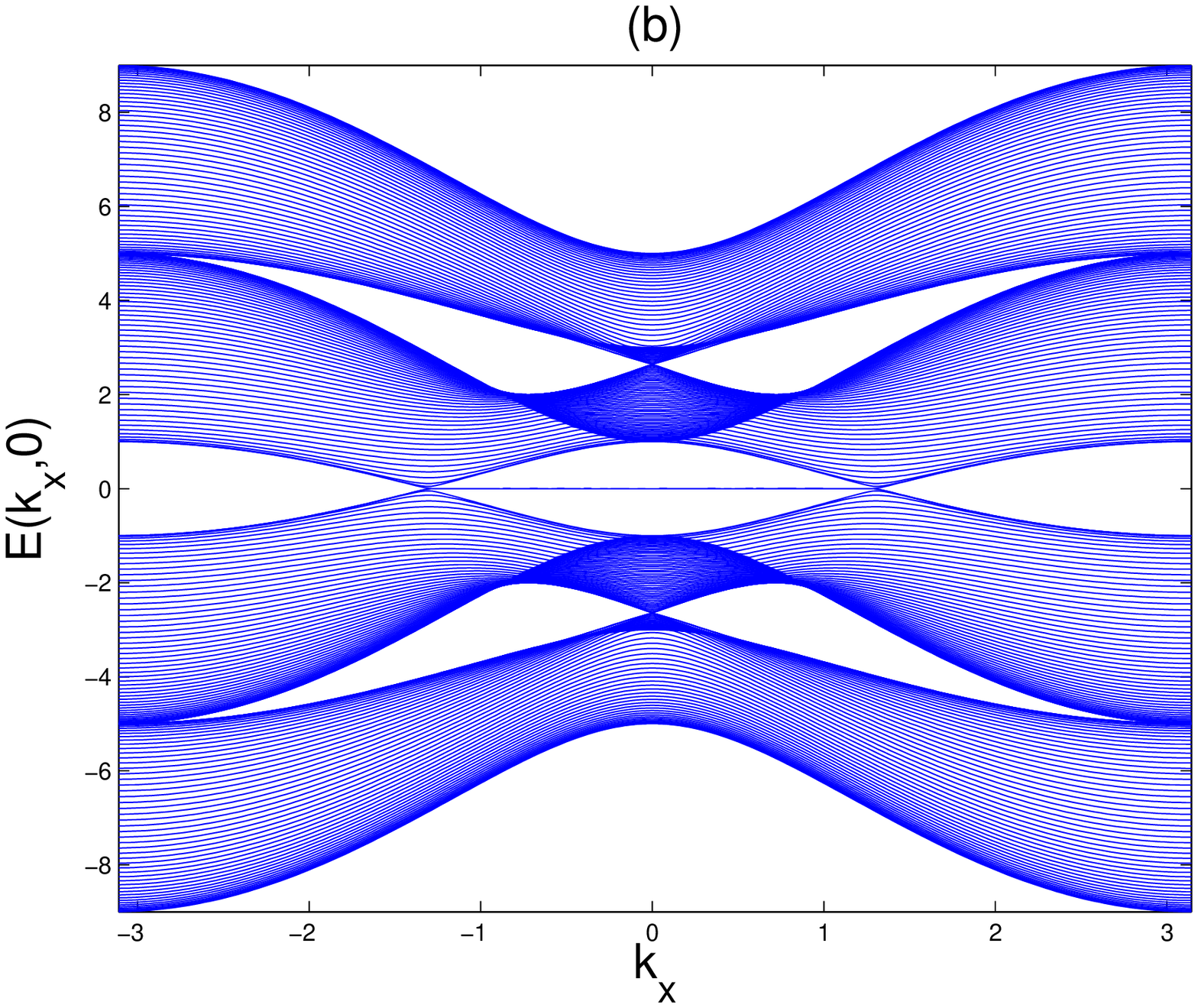}
\end{minipage}
\begin{minipage}{0.495\textwidth}
\centering
\includegraphics[width=\textwidth]{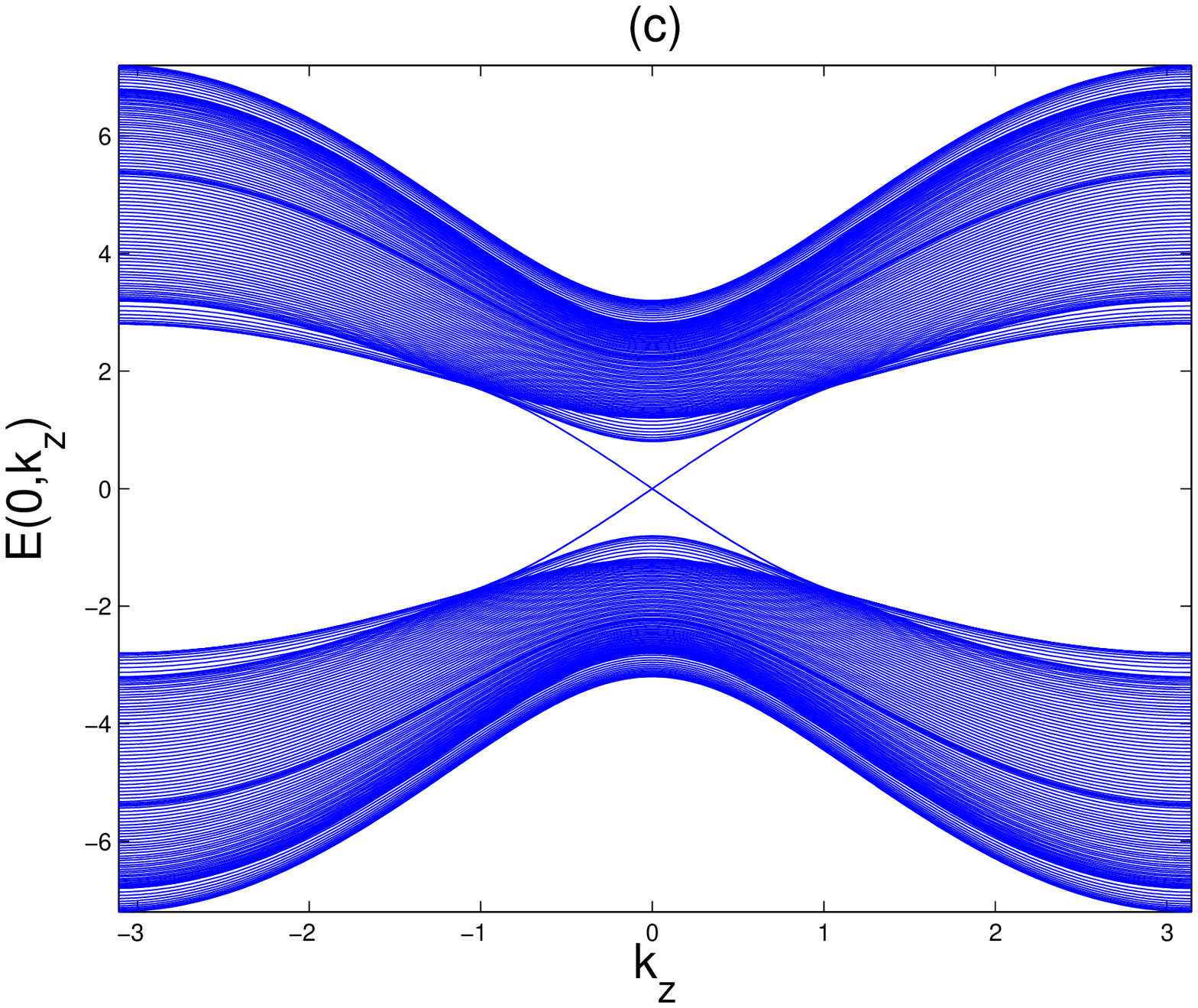}
\end{minipage}
\begin{minipage}{0.495\textwidth}
\centering
\includegraphics[width=\textwidth]{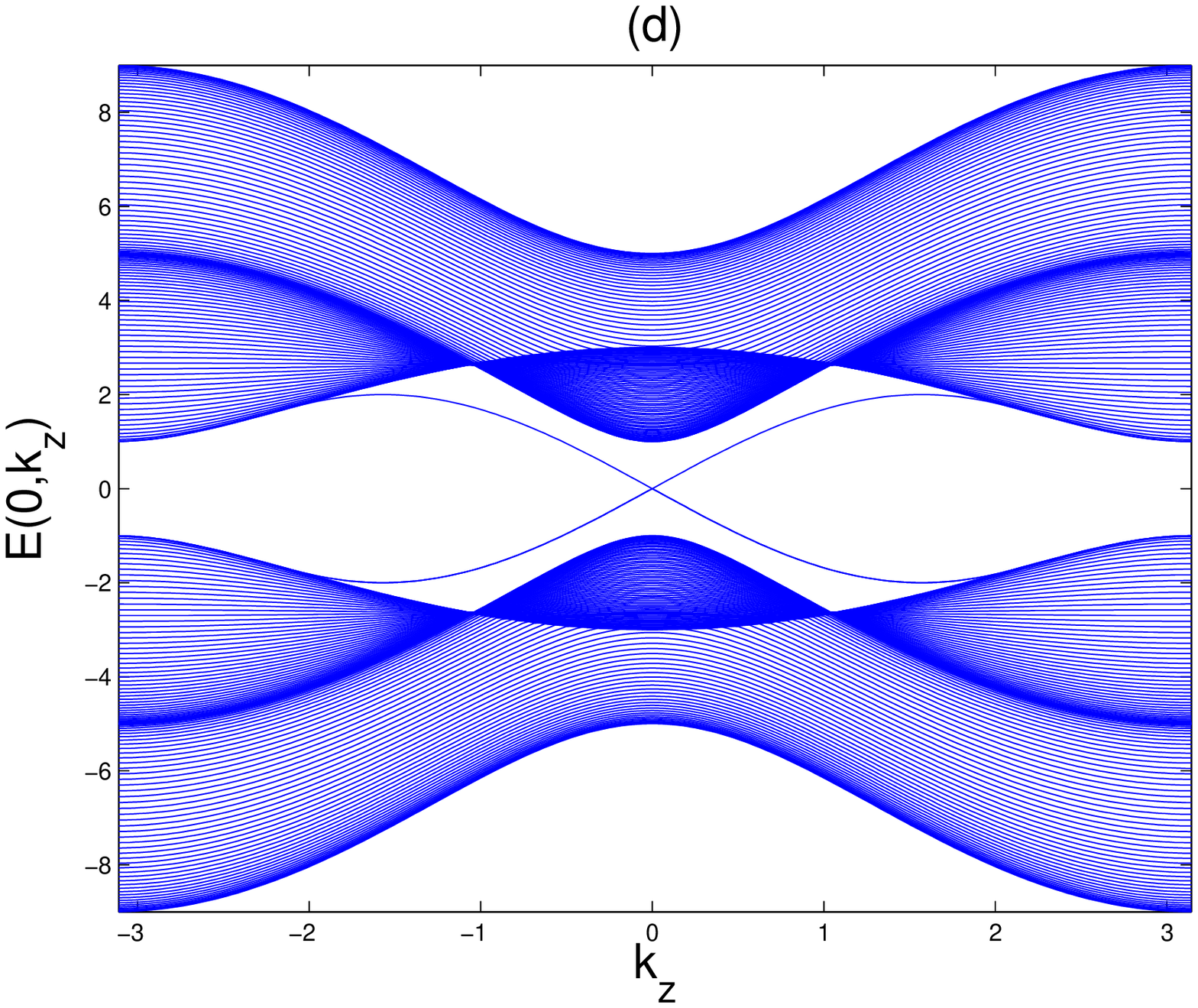}
\end{minipage}
\caption{\label{Fig11}
Numerical dispersions of bulk and surface states for model II with
$V_x/M=0.2$ in (a) and (c) and $2.0$ in (b) and (d).
In (a) and (b) $k_z=0$, and in (c) and (d) $k_x=0$.
The other parameters are same as in figure~\ref{Fig1}
}
\end{figure}

\begin{figure}[t]
\begin{minipage}{0.495\textwidth}
\centering
\includegraphics[width=\textwidth]{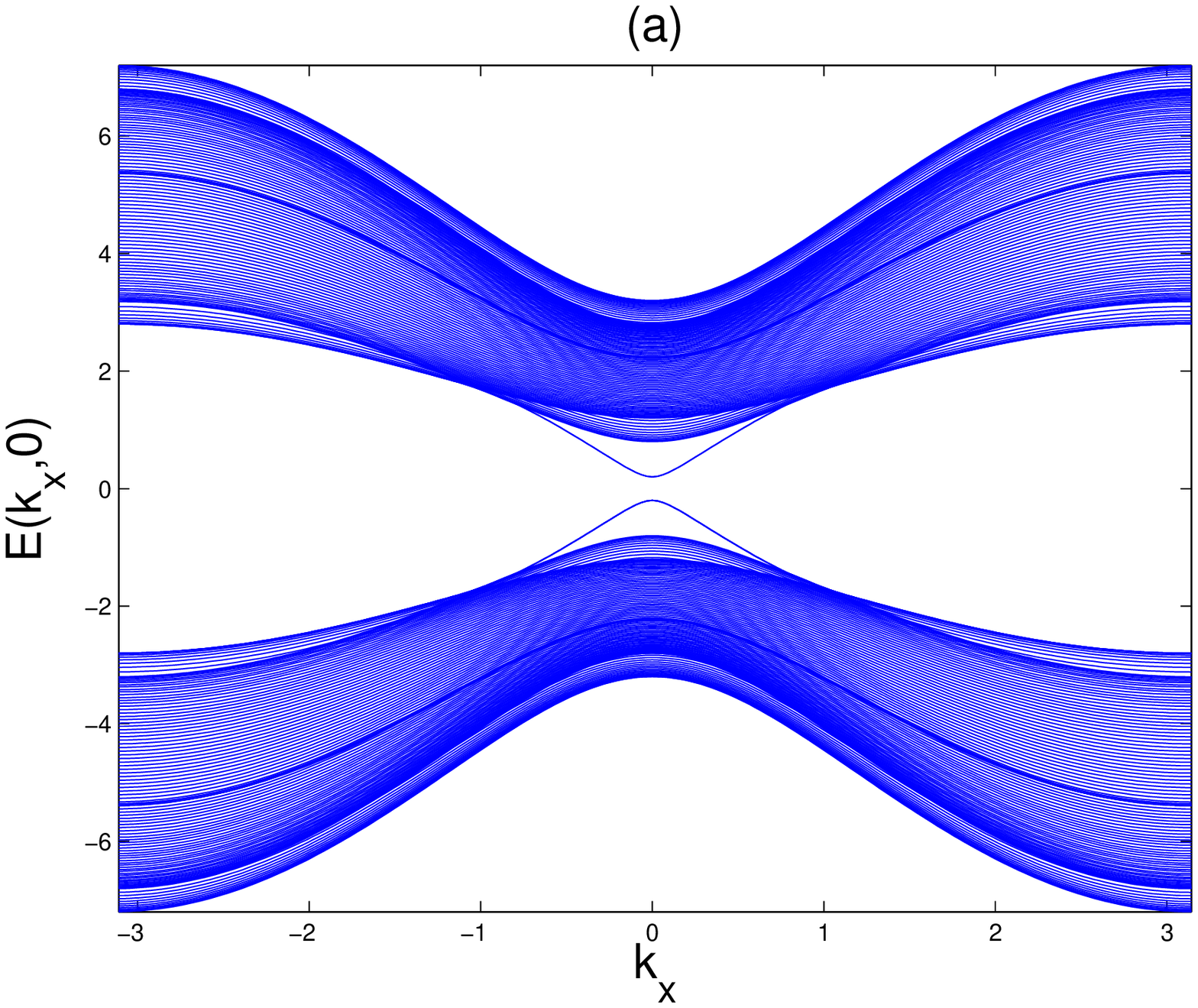}
\end{minipage}
\begin{minipage}{0.495\textwidth}
\centering
\includegraphics[width=\textwidth]{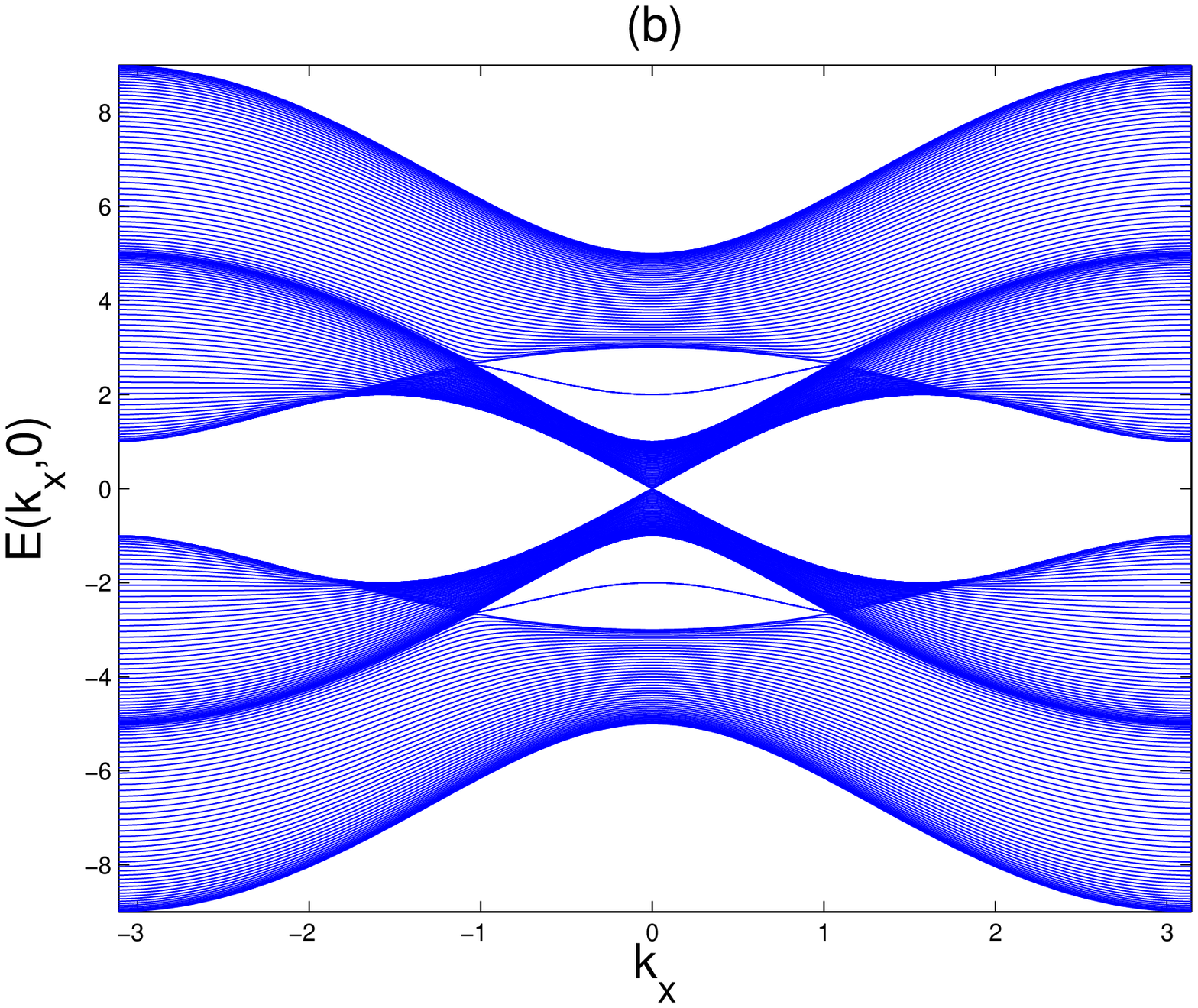}
\end{minipage}
\begin{minipage}{0.495\textwidth}
\centering
\includegraphics[width=\textwidth]{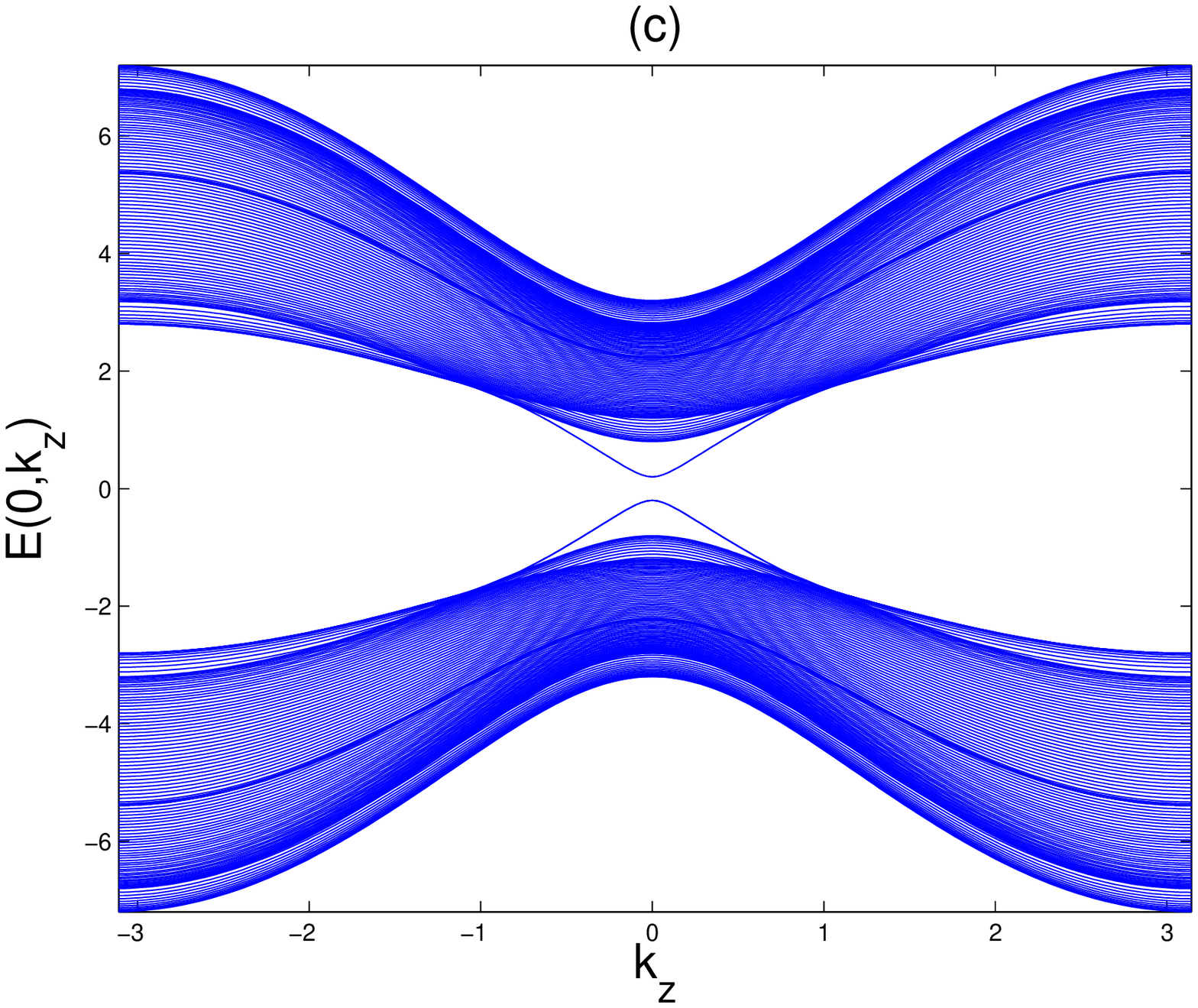}
\end{minipage}
\begin{minipage}{0.495\textwidth}
\centering
\includegraphics[width=\textwidth]{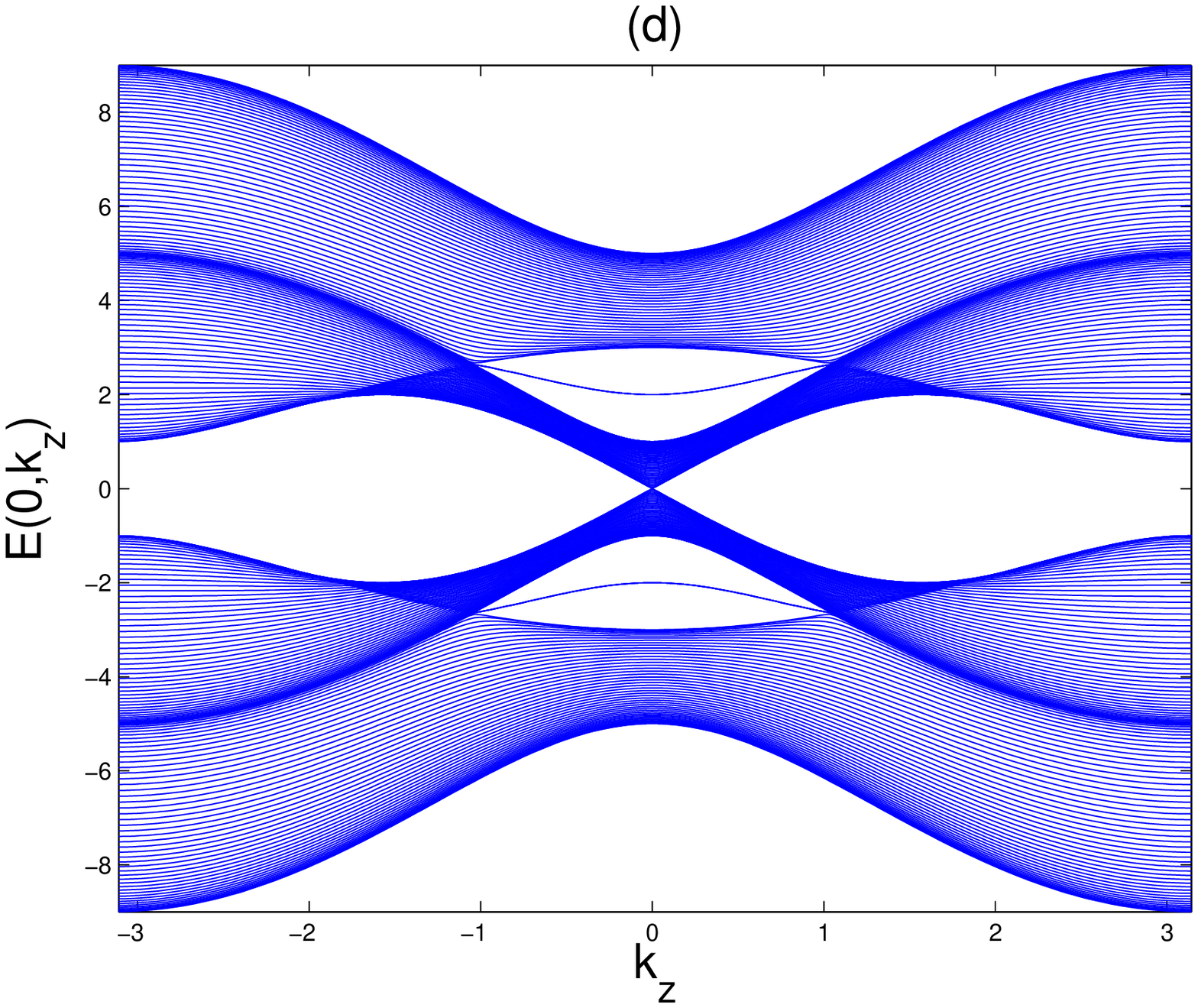}
\end{minipage}
\caption{\label{Fig12}
Numerical dispersions of bulk and surface states for model II with 
$V_y/M=0.2$ in (a) and (c) and $2.0$ in (b) and (d).
In (a) and (b) $k_z=0$, and in (c) and (d) $k_x=0$.
The other parameters are same as in figure~\ref{Fig1}
}
\end{figure}

To determine the surface states, we can start from the same Hamiltonian
$H_0'$ in Eq.~(\ref{eq:H'_0_V_x}) and its zero energy surface
states in Eq.~(\ref{eq:surf_solu_B_y_V_x_k_0}) and (\ref{eq:surf_solu_B_y_V_x_k_0_2}).
We can then treat the terms
\begin{equation}
H'(\mathbf k)=m_1(\mathbf k)\Gamma^1+m_3(\mathbf k)\Gamma_{II}^3+V_y\Gamma_y.
\label{eq:H'_B_y_V_y_model2}
\end{equation}
using degenerate perturbation theory again. In contrast to
the matrix $\Gamma_I^3$ the matrix $\Gamma_{II}^3$ leads to
a diagonal coupling of the two surface states instead of
an off-diagonal coupling. As a result the dispersion of the
surface states in $k_z$-direction remains unaffected by
the spatial overlap $\beta(V_x,\mathbf k)$ Eq.~(\ref{eq:bertu_etu_k_B_y_V_y}).
Consequently, we find the following surface state
dispersions for the full Hamiltonian within
perturbation theory:
\begin{equation}
\label{eq:ener_bertu_B_y_V_x_M_II}
E_{\pm}(\mathbf k)=\pm \sqrt{\beta(V_x,\mathbf k)^2\left( V_y^2+4 A_2^2 \sin^2k_x \right)+4A_1^2\sin^2k_z}.
\end{equation}
This dispersion shows that the $V_y$ component of the exchange field,
which is perpendicular to the surface, opens a gap in the
surface state spectrum for small values of $V_y$.
The component $V_x$ parallel to the surface leads to an anisotropic reduction of
the dispersion in $k_x$-direction, but not in $k_z$-direction.
In Figures~\ref{Fig11} and \ref{Fig12} we show the corresponding
numerical results on a finite lattice for $V_x$ nonzero and $V_y$ nonzero,
respectively, which agree with this behavior.

The surface state wave functions for this case can be written in the form
\begin{eqnarray}
\label{eq:sur_s_B_z_4_model2}
\Psi_{+,\mathbf k}(y)&=e^{-i\phi_{\mathbf k}} \cos
\frac{\theta_{\mathbf k}}{2} \psi_{1,\mathbf k}(y) +
\sin \frac{\theta_{\mathbf k}}{2} \psi_{2,\mathbf k}(y) \\
\Psi_{-,\mathbf k}(y)&=-e^{-i\phi_{\mathbf k}} \sin
\frac{\theta_{\mathbf k}}{2} \psi_{1,\mathbf k}(y) +
\cos \frac{\theta_{\mathbf k}}{2} \psi_{2,\mathbf k}(y)
\end{eqnarray} 
Here, we have introduced two spherical angles $\phi_{\mathbf k}$ and $\theta_{\mathbf k}$
that define the direction of the vector $(\beta m_1, \beta V_y, m_3)$:
\begin{eqnarray}
\nonumber
\cos \phi_{\mathbf k} \sin \theta_{\mathbf k} &= \frac{2 \beta A_2 \sin k_x}{
\sqrt{\beta^2 V_y^2+4 \beta^2 A_2^2 \sin^2 k_x +4 A_1^2 \sin k_z }} \\
\nonumber
\sin \phi_{\mathbf k} \sin \theta_{\mathbf k} &= \frac{\beta V_y}{
\sqrt{\beta^2 V_y^2+4 \beta^2 A_2^2 \sin^2 k_x +4 A_1^2 \sin k_z }} \\
\cos \theta_{\mathbf k} &= \frac{2 A_1 \sin k_z}{
\sqrt{\beta^2 V_y^2+4 \beta^2 A_2^2 \sin^2 k_x +4 A_1^2 \sin k_z }}
\nonumber
\end{eqnarray} 
For the spin texture of these surface states we find in orbital 1:
\begin{eqnarray}
\nonumber
\left\langle \Psi_{\pm,\mathbf k} \left| \hat{s}_{1,x} \right|
\Psi_{\pm,\mathbf k} \right\rangle &= \pm \frac{\beta}{2} \cos \phi_{\mathbf k} \sin \theta_{\mathbf k} \\
\nonumber
\left\langle \Psi_{\pm,\mathbf k} \left| \hat{s}_{1,y} \right|
\Psi_{\pm,\mathbf k} \right\rangle &= \pm \frac{\beta}{2} \sin \phi_{\mathbf k} \sin \theta_{\mathbf k} \\
\nonumber
\left\langle \Psi_{\pm,\mathbf k} \left| \hat{s}_{1,z} \right|
\Psi_{\pm,\mathbf k} \right\rangle &= \pm \frac{1}{2} \cos \theta_{\mathbf k} 
\end{eqnarray}
In orbital 2 the $x$- and $z$-components of the spin turn out to
be inverse:
\begin{eqnarray}
\nonumber
\left\langle \Psi_{\pm,\mathbf k} \left| \hat{s}_{2,x} \right|
\Psi_{\pm,\mathbf k} \right\rangle &= \mp \frac{\beta}{2} \cos \phi_{\mathbf k} \sin \theta_{\mathbf k} \\
\nonumber
\left\langle \Psi_{\pm,\mathbf k} \left| \hat{s}_{2,y} \right|
\Psi_{\pm,\mathbf k} \right\rangle &= \pm \frac{\beta}{2} \sin \phi_{\mathbf k} \sin \theta_{\mathbf k} \\
\nonumber
\left\langle \Psi_{\pm,\mathbf k} \left| \hat{s}_{2,z} \right|
\Psi_{\pm,\mathbf k} \right\rangle &= \mp \frac{1}{2} \cos \theta_{\mathbf k} 
\end{eqnarray}
Here, we see that in contrast to the corresponding case for model I
the spin possesses components in all three spatial directions.
In the limit $V_y \rightarrow 0$, the angle $\phi_{\mathbf k}$ goes to zero and
the spin-$y$ component vanishes. The spatial overlap factor $\beta$
is seen to suppress only the $x$- and $y$-components of the spin.
For the total spin we find:
\begin{eqnarray}
\nonumber
\left\langle \Psi_{\pm,\mathbf k} \left| \Gamma_x \right|
\Psi_{\pm,\mathbf k} \right\rangle &= 0 \\
\nonumber
\left\langle \Psi_{\pm,\mathbf k} \left| \Gamma_y \right|
\Psi_{\pm,\mathbf k} \right\rangle &= \pm \beta \sin \phi_{\mathbf k} \sin \theta_{\mathbf k} \\
\nonumber
\left\langle \Psi_{\pm,\mathbf k} \left| \Gamma_z \right|
\Psi_{\pm,\mathbf k} \right\rangle &= 0
\end{eqnarray}
The total spin is directed perpendicular to the surface.
It vanishes in the limit $V_y \rightarrow 0$.

From the numerical results in Fig.~\ref{Fig11}~(b) and (d) 
we see that a one-dimensional flat band
appears for nonzero $V_x>M$. Again, the appearence of this flat band
can be understood by a topological winding number. In the present case
the system obeys the chiral symmetry $\Theta_3$ for $k_z=0$, as was
discussed in section~\ref{Secsymmetries}. The Fermi surface
is zero dimensional, so we may expect a one-dimensional flat band.
As the Hamiltonians for model I and model II are identical for $k_z=0$,
the off-diagonal block form of the Hamiltonian is the same
as in Eq.~(\ref{eq:Schnyder_form_2}), i.e.
\begin{equation}
\label{eq:Schnyder_form_model2}
\fl D(\mathbf k)=\left(\begin{array}{cc}
m_0(\mathbf k) & -m_1(\mathbf k)+im_2(\mathbf k)+V_x-iV_y \\
m_1(\mathbf k)+im_2(\mathbf k)+V_x+iV_y & m_0(\mathbf k)
\end{array}\right).
\end{equation}
The corresponding winding number is then given by
\begin{equation}
\label{eq:WN_schnyder_model2}
w(k_x)=\frac{1}{2\pi}\textrm{Im}\int_{-\pi}^{\pi}\, dk_y\, \partial_{k_y} \ln \textrm{det} D(\mathbf k).
\end{equation}
The one-dimensional flat band in the present case thus
corresponds to a $k_z=0$ cut of the flat band area
shown in Fig.~\ref{Fig5}~(a). We can find the full
dispersion of this flat band also for finite $k_z$
by noting that the matrix $\Gamma_{II}^3$ anticommutes
with the Hamiltonian for $k_z=0$. As a result the
zero energy states for $k_z=0$ are eigenstates of
$\Gamma_{II}^3$, too. Therefore, the dispersion
of the surface flat band for finite $k_z$ is
given by
\begin{equation}
\label{eq:ener_B_y_V_x_M_II_flat}
E_{\pm}(\mathbf k)=\pm 2A_1\sin k_z.
\end{equation}
Like for model I the zero energy surface of this 
one-dimensional flat band is a Fermi arc, whose end 
points are the projections
of the Weyl nodes onto the surface Brillouin zone.

\subsection{Finite $V_z$ and $V_y$}

\label{subsec2vzy}
The case with both $V_z$ and $V_y$ nonzero, but $V_x=0$ can be treated
like the corresponding case for model I in section~\ref{subsec1vz}.
The Fermi surface turns out to be zero-dimensional and the
system possesses the chiral symmetry $\Theta_1$ for $k_x=0$.
For determination of the surface states we can start from
the zero energy surface states Eqs.~(\ref{eq:surf_solu_B_y_V_z_k_0}) and 
(\ref{eq:surf_solu_B_y_V_z_k_0_2}) of the Hamiltonian Eq.~(\ref{eq:model1hpp})
and treat $H'=m_1 \Gamma^1 + m_3 \Gamma_{II}^3 + V_y \Gamma_y$
as a perturbation. The energies of the surface
states are then found to be
\begin{equation}
\label{eq:ener_bertu_B_y_V_z_M_II}
E_{\pm}(\mathbf k)=\pm \sqrt{\beta(V_z,\mathbf k)^2\left( V_y^2+4 A_1^2 \sin^2k_z \right)+4A_2^2\sin^2k_x}.
\end{equation}
This is the same kind of dispersion as in
Eq.~(\ref{eq:ener_bertu_B_y_V_x_M_II})
with the roles of the $x$- and $z$-coordinates interchanged.
The spin texture of the surface states in orbital 1 is found to be:
\begin{eqnarray}
\nonumber
\left\langle \Psi_{\pm,\mathbf k} \left| \hat{s}_{1,x} \right|
\Psi_{\pm,\mathbf k} \right\rangle &= \pm \frac{1}{2} \frac{m_1}{\sqrt{m_1^2 + \beta^2
    \left( m_3^2 + V_y^2\right)}} \\
\nonumber
\left\langle \Psi_{\pm,\mathbf k} \left| \hat{s}_{1,y} \right|
\Psi_{\pm,\mathbf k} \right\rangle &= \pm \frac{1}{2} \frac{\beta V_y}{\sqrt{m_1^2 + \beta^2
    \left( m_3^2 + V_y^2\right)}} \\
\nonumber
\left\langle \Psi_{\pm,\mathbf k} \left| \hat{s}_{1,z} \right|
\Psi_{\pm,\mathbf k} \right\rangle &= \pm \frac{1}{2} \frac{\beta m_3}{\sqrt{m_1^2 + \beta^2
    \left( m_3^2 + V_y^2\right)}}
\end{eqnarray}
and in orbital 2:
\begin{eqnarray}
\nonumber
\left\langle \Psi_{\pm,\mathbf k} \left| \hat{s}_{2,x} \right|
\Psi_{\pm,\mathbf k} \right\rangle &= \mp \frac{1}{2} \frac{m_1}{\sqrt{m_1^2 + \beta^2
    \left( m_3^2 + V_y^2\right)}} \\
\nonumber
\left\langle \Psi_{\pm,\mathbf k} \left| \hat{s}_{2,y} \right|
\Psi_{\pm,\mathbf k} \right\rangle &= \pm \frac{1}{2} \frac{\beta V_y}{\sqrt{m_1^2 + \beta^2
    \left( m_3^2 + V_y^2\right)}} \\
\nonumber
\left\langle \Psi_{\pm,\mathbf k} \left| \hat{s}_{2,z} \right|
\Psi_{\pm,\mathbf k} \right\rangle &= \mp \frac{1}{2} \frac{\beta m_3}{\sqrt{m_1^2 + \beta^2
    \left( m_3^2 + V_y^2\right)}}
\end{eqnarray}
For the total spin we find:
\begin{eqnarray}
\nonumber
\left\langle \Psi_{\pm,\mathbf k} \left| \Gamma_x \right|
\Psi_{\pm,\mathbf k} \right\rangle &= 0 \\
\nonumber
\left\langle \Psi_{\pm,\mathbf k} \left| \Gamma_y \right|
\Psi_{\pm,\mathbf k} \right\rangle &= \pm \frac{\beta V_y}{\sqrt{m_1^2 + \beta^2
    \left( m_3^2 + V_y^2\right)}} \\
\nonumber
\left\langle \Psi_{\pm,\mathbf k} \left| \Gamma_z \right|
\Psi_{\pm,\mathbf k} \right\rangle &= 0
\end{eqnarray}
The total spin is again directed perpendicular to the surface and
vanishes in the limit $V_y \rightarrow 0$.

We find a one-dimensional flat band for finite $V_z>M$,
that can be understood by a topological winding number using the
chiral symmetry $\Theta_1$ for $k_x=0$, analogously to the previous case.
As the matrix $\Gamma_1$ anticommutes with the Hamiltonian for $k_x=0$
we eventually find the dispersion of the surface flat band in this
case as
\begin{equation}
\label{eq:ener_B_y_V_x_M_II_flat_2}
E_{\pm}(\mathbf k)=\pm 2A_2\sin k_x.
\end{equation}
Again, the zero energy surface of this 
one-dimensional flat band is a Fermi arc, whose end 
points are the projections
of the Weyl nodes onto the surface Brillouin zone.

\section{Effect of broken particle-hole symmetry}
\label{Secbrokenph}

So far we have studied the case $C=D_1=D_2=0$. When these parameters become
nonzero, the particle-hole symmetry of the system is broken. Also, the chiral
symmetries discussed in section~\ref{Secsymmetries} are not obeyed anymore.
For this reason the topological winding numbers that we used in the previous
sections to determine the presence or absence of a surface flat band
cannot be used anymore. However, this does not mean that the surface bands
completely disappear. In the following we will demonstrate by both numerical
and analytical calculation that surface bands, which are energetically well 
separated from the bulk bands still exist in the broken particle-hole case.
We will see that the surface bands become dispersive now with the dispersion
increasing proportional to $D_1$ and $D_2$. Similar behavior has been noted
in other systems before as well \cite{Matsuura,PALee,Lau}.
For the numerical calculations we use parameters that are realistic
for Bi$_2$Se$_3$ and have been given in Ref.~\cite{Zhang:NPhys09}.
The Hamiltonian is now
\begin{equation} \label{eq:hamiltonian_real_gen}
	H(\mathbf k)=\epsilon_0(\mathbf k) \mathbb{I}_{4 \times 4} +
\sum_{i=0}^3 m_i(\mathbf k) \Gamma^i + \sum_{\alpha \in \{x,y,z \}} V_{\alpha}\Gamma_{\alpha}
\end{equation}
with $\epsilon_0(\mathbf k) = C + 2 D_2 ( 1- \cos k_x )+ 2 D_2 ( 1- \cos k_y
)+ 2 D_1 ( 1- \cos k_z )$,
$m_0(\mathbf k) = M - 2 B_2 ( 1- \cos k_x ) - 2 B_2 ( 1- \cos k_y )- 2 B_1 ( 1- \cos k_z )$,
$m_1(\mathbf k) = 2 A_2 \sin k_x$, $m_2(\mathbf k) = 2 A_2 \sin k_y$, and
$m_3(\mathbf k) = 2 A_1 \sin k_z$.
For Bi$_2$Se$_3$ parameters derived from Ref.~\cite{Zhang:NPhys09} are: 
$A_1=0.575\, \textrm{eV}$,  $A_2=0.495\, \textrm{eV}$, $B_1=2.74\, \textrm{eV}$,
$B_2=3.30\, \textrm{eV}$, $C=-0.0068\, \textrm{eV}$,
$D_1=0.36\, \textrm{eV}$, $D_2=1.14\, \textrm{eV}$, and $M=0.28 \,
\textrm{eV}$.
Here, we have used the lattice constants $a=4.14$~\AA \; and $15 c=28.64$~\AA \;
for conversion into our lattice model. In the following,
we use the same parameters for both model I and II.

It is clear that the bulk energy bands are just shifted by $\epsilon_0(\mathbf
k)$ with respect to the particle-hole symmetric cases studied in the previous 
sections. The former zero energy points, at which two bulk bands
touch each other, will then have energy 
$\epsilon_0(\mathbf k)$. For model I with $V_z=0$ and $V>V_{cr}$ we had a 
one-dimensional Fermi surface, whose degeneracy will now be lifted.
Thus, these nodes will generally overlap with the two touching bulk bands.
However, for the particle-hole symmetric model I with $V_z \neq 0$ and 
$V>M$ we have a Weyl semimetal
with just two Fermi points. As these two Weyl nodes sit at symmetry related
momentum points, they are shifted by the same amount $\epsilon_0(\mathbf k)$.
Therefore, the particle-hole broken system remains a Weyl semimetal unless
the dispersion $\epsilon_0(\mathbf k)$ becomes so large that the Weyl nodes
start to overlap with the bulk bands. This same argument also holds for
model II with $V>M$. Thus, the Weyl semimetallic phases in model I and II are preserved
under not too large particle-hole symmetry breaking and we may still
expect the presence of surface Fermi arcs, as we will show explicitly
from our numerical calculations below. In the following we restrict the
discussion to the geometries in which we found surface flat bands for
the particle-hole symmetric cases.

\subsection{Model I with boundary perpendicular to the $y$-direction}

In the absence of an exchange field the surface state dispersions can be found
by replacing the momentum $k_y$ by the momentum operator $-i \partial_y$
and solving the $4 \times 4$ matrix Schr\"odinger equation directly
with an exponential ansatz following Ref.~\cite{Shan}. This way one finds
\begin{eqnarray} 
\nonumber
	E_{\pm}(\mathbf k)&=C+Mt_2+2(D_1-B_1t_2)(1-\cos k_z)
          \\ \label{eq:ener_ex_realistic_I_y}
&\pm\sqrt{1-t_2^2}\sqrt{4 A_2^2\sin^2 k_x+4 A_1^2\sin^2 k_z},
\end{eqnarray}
where $t_2=D_2/B_2\approx 0.35$. If $t_2>1$ surface states do not
exist.

\begin{figure}[t]
\begin{minipage}{0.495\textwidth}
\centering
\includegraphics[width=\textwidth]{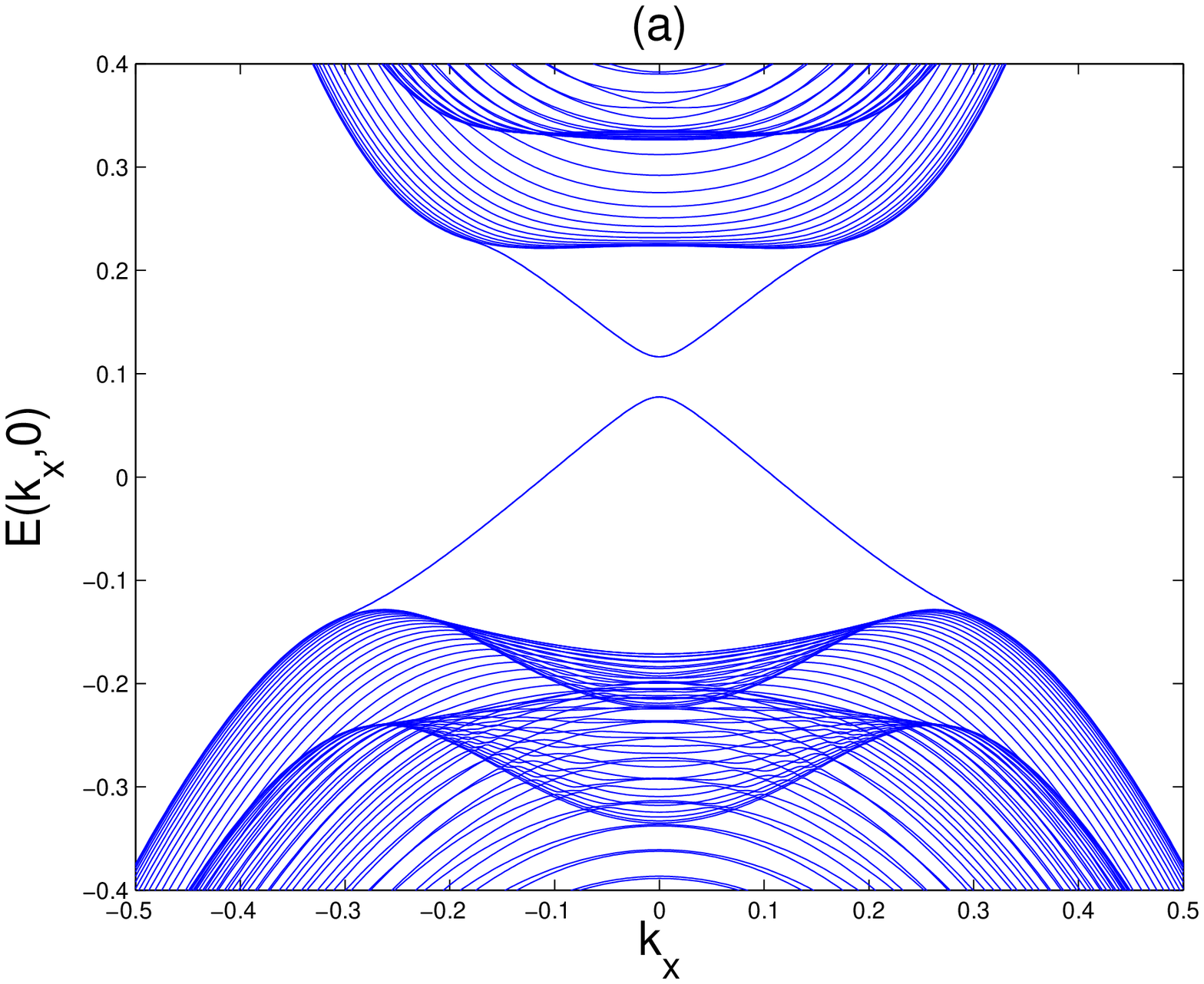}
\end{minipage}
\begin{minipage}{0.495\textwidth}
\centering
\includegraphics[width=\textwidth]{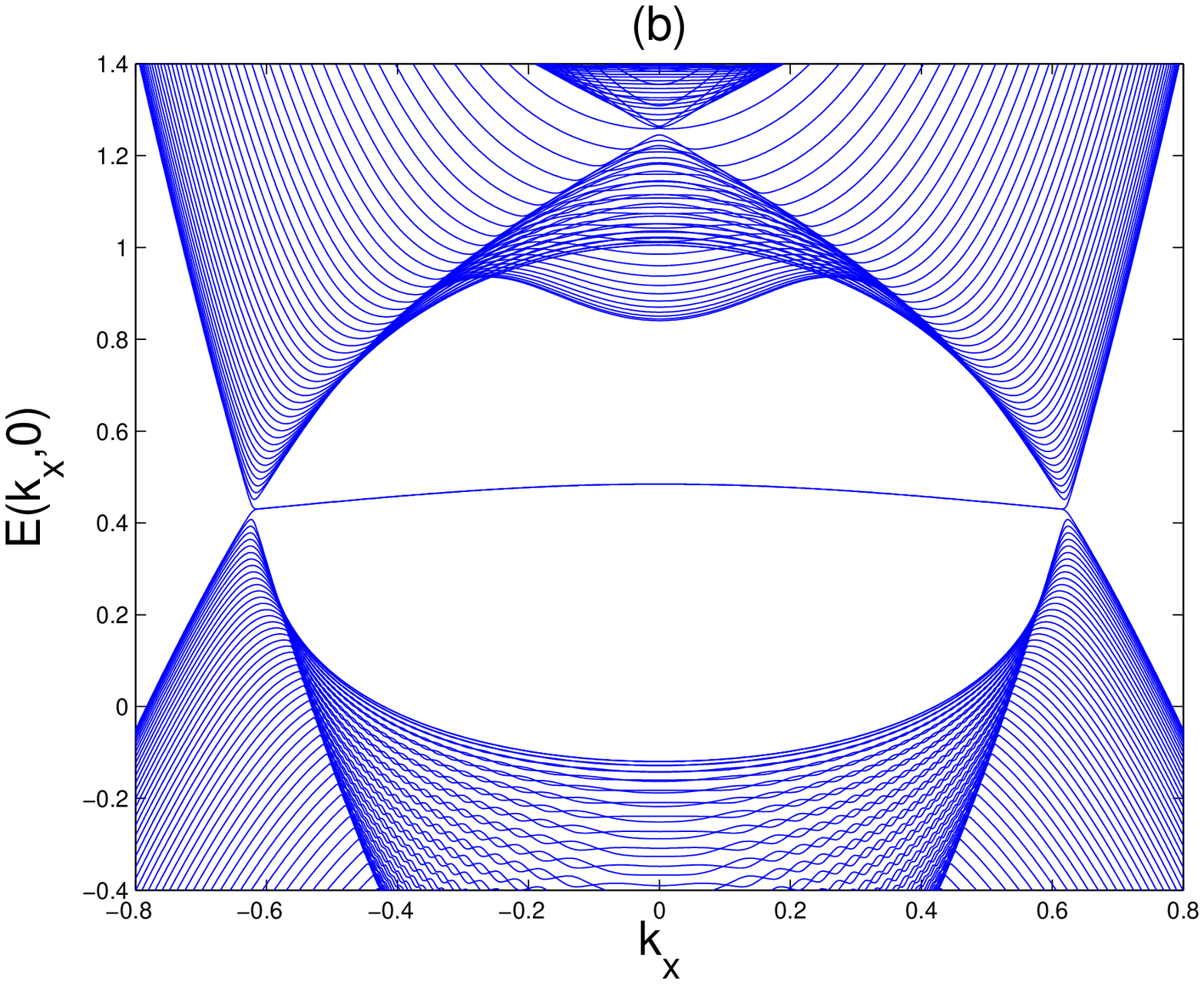}
\end{minipage}
\begin{minipage}{0.495\textwidth}
\centering
\includegraphics[width=\textwidth]{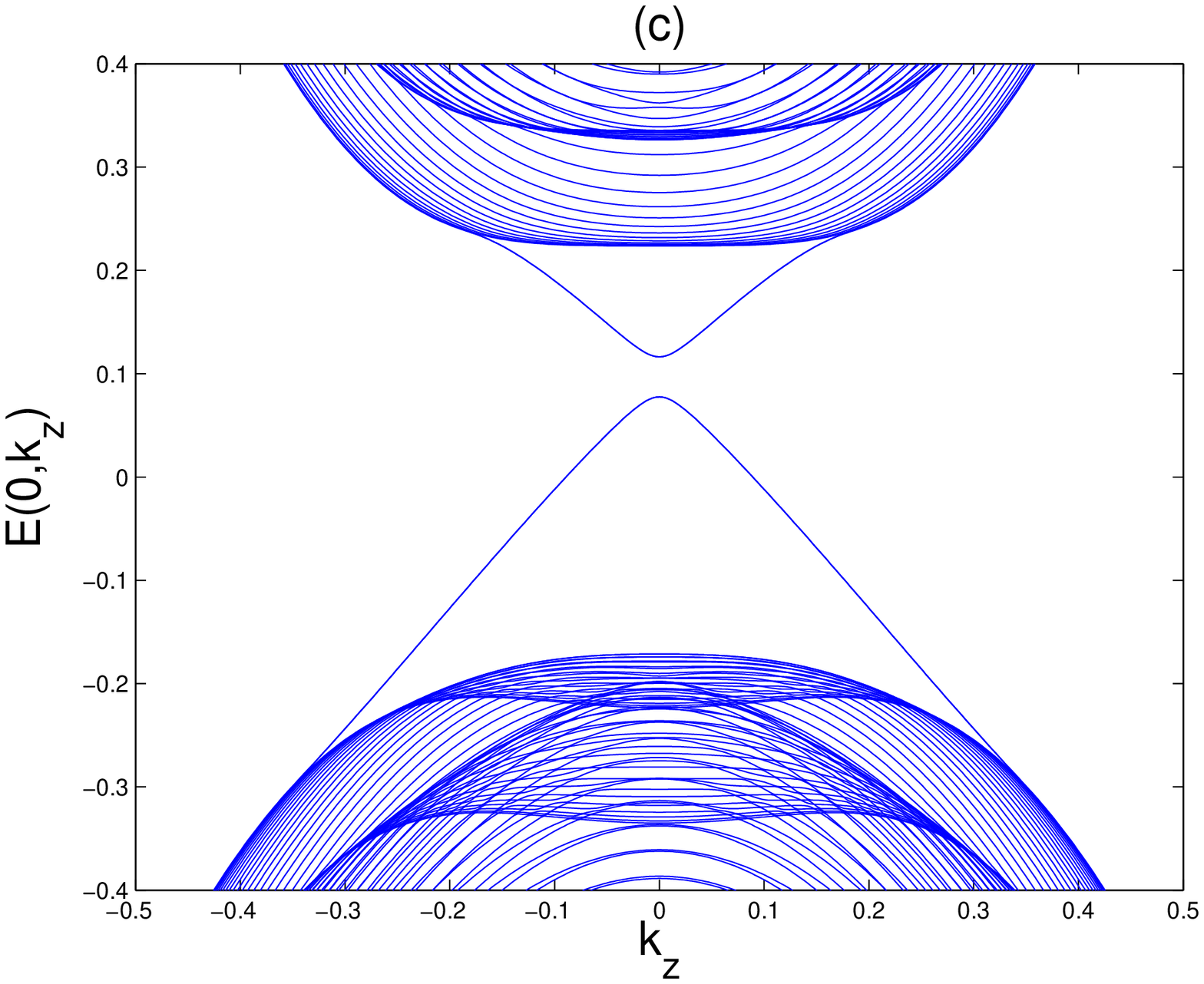}
\end{minipage}
\begin{minipage}{0.495\textwidth}
\centering
\includegraphics[width=\textwidth]{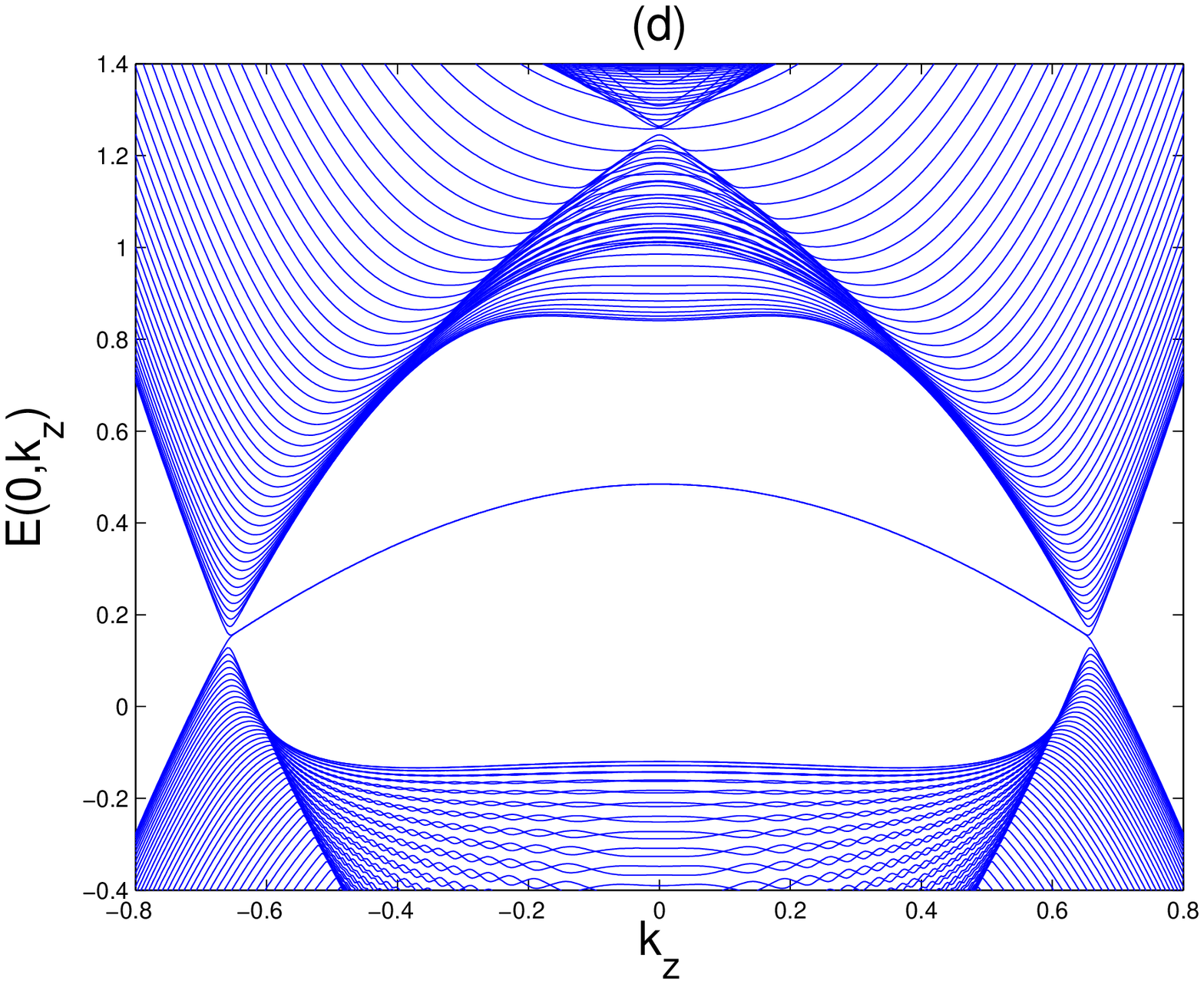}
\end{minipage}
\caption{\label{Fig13}
Numerical dispersions of bulk and surface states for model I with
$V_x/M=0.2$ in (a) and (c) and $4.0$ in (b) and (d).
In (a) and (b) $k_z=0$, and in (c) and (d) $k_x=0$.
The other parameters are the ones for Bi$_2$Se$_3$ as described in the text.
}
\end{figure}

When an exchange field is turned on, the Schr\"odinger equation
leads to 8th order polynomials, whose zeroes cannot be given
in closed form. However, analytical results can be obtained
for fields in high symmetry directions and small values
of momentum $k_x$ and $k_z$ by expansion.

When the exchange field points into $y$-direction we find
the following dispersions
\begin{eqnarray}
\label{eq:ener_ex_realistic_I_y_V_y}
\nonumber E_{\pm}(\mathbf k)&=C+Mt_2+2(D_1-B_1t_2)(1-\cos k_z)\\
&\pm\sqrt{\beta^2_3(k_z)A_2^2\sin^2 k_x+(V_y-\sqrt{1-t_2^2}A_1\sin k_z)^2}.
\end{eqnarray}
Here, $\beta_3$ is a spatial overlap factor of the form
\begin{equation} \label{eq:beta3}
\beta_3(k_z) = \frac{2A_2^2 \sqrt{\left[M-2B_1 \left( 1-\cos k_z \right) \right]^2
- 4 \frac{t_2^2}{1-t_2^2} A_1^2 \sin^2 k_z}}{2A_2^2\left[M-2B_1 \left( 1-\cos
    k_z \right) \right] +t_2^2 B_2 A_1^2 \sin^2 k_z}
\end{equation}
Eq.~(\ref{eq:ener_ex_realistic_I_y_V_y}) tells us that the Dirac cone remains ungapped for
an exchange field in $y$-direction and is shifted in $k_z$ direction,
like in the particle-hole symmetric case in section~\ref{sec:mIyVy}.

\begin{figure}[t]
\centering
\includegraphics[width=0.7\textwidth,angle=270]{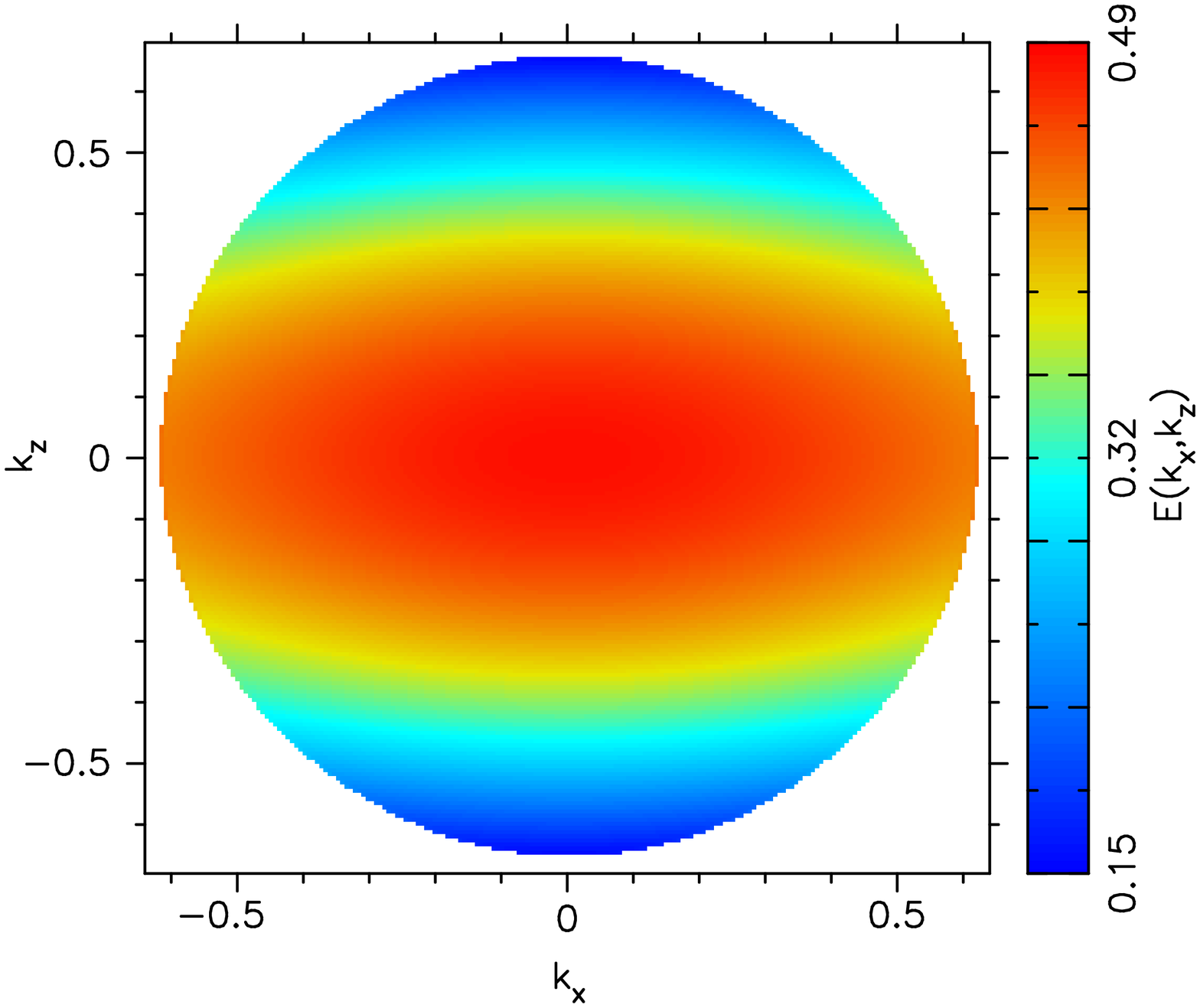}
\caption{\label{Fig14}
Dispersion of the surface band for model I
with $V_x/M=4.0$. Red color indicates high energy values and blue color low values.
The other parameters are the same as in figure~\ref{Fig13}
}
\end{figure}

With exchange field in $x$-direction the following low field dispersions are found
\begin{eqnarray}
\label{eq:ener_ex_realistic_I_y_V_x}
\nonumber E_{\pm}(\mathbf k)&=C+Mt_2+2(D_1-B_1t_2)(1-\cos k_z )\\
&\pm\sqrt{t_2^2V_x^2+\left(1-t_2^2\right) \beta^2_4(V_x) 
\left( 4 A_2^2\sin^2 k_x+ 4 A_1^2\sin^2 k_z \right) },
\end{eqnarray}
where the spatial overlap
\begin{equation} \label{eq:beta4}
\beta_4(V_x) = \frac{4A_2^2 \sqrt{M^2-V_x^2}}{4A_2^2 M +\left(1-t_2^2\right) B_2 V_x^2}
\end{equation}
In contrast to Eq.~(\ref{eq:ener_bertu_B_y_V_x}) we now find the opening of a 
gap in the dispersion, which is of the order of $t_2 V_x$.
Corresponding numerical results on a finite lattice for $V_x=0.2 M$ are shown
in Fig.~\ref{Fig13}~(a) and (c). If we increase $V_x$ beyond $M$, we enter
a state that corresponds to the two-dimensional flat band state discussed
in section~\ref{subsecflat1}. As one can see in Fig.~\ref{Fig13}~(b) and (d)
the surface band becomes dispersive now. However, it still exists and remains
well separated from the two bulk bands. The dispersion of the surface band over the
surface Brillouin zone is shown in color coded scale in Fig.~\ref{Fig14}.
The dispersion is much smaller along $k_x$-direction than in $k_z$-direction.
An approximate analytical expression valid for small $k_x$ and $k_z$ can
be obtained from Eq.~(\ref{eq:ener_ex_realistic_I_y_V_x}), if one sets $\beta_4=0$.
In the total density of states of
this system the surface band appears as a peak, similar to what was found for
the edge states in a two-dimensional system in our previous work
(see Fig.~5 in Ref.~\cite{PD}).

\begin{figure}[t]
\begin{minipage}{0.495\textwidth}
\centering
\includegraphics[width=\textwidth]{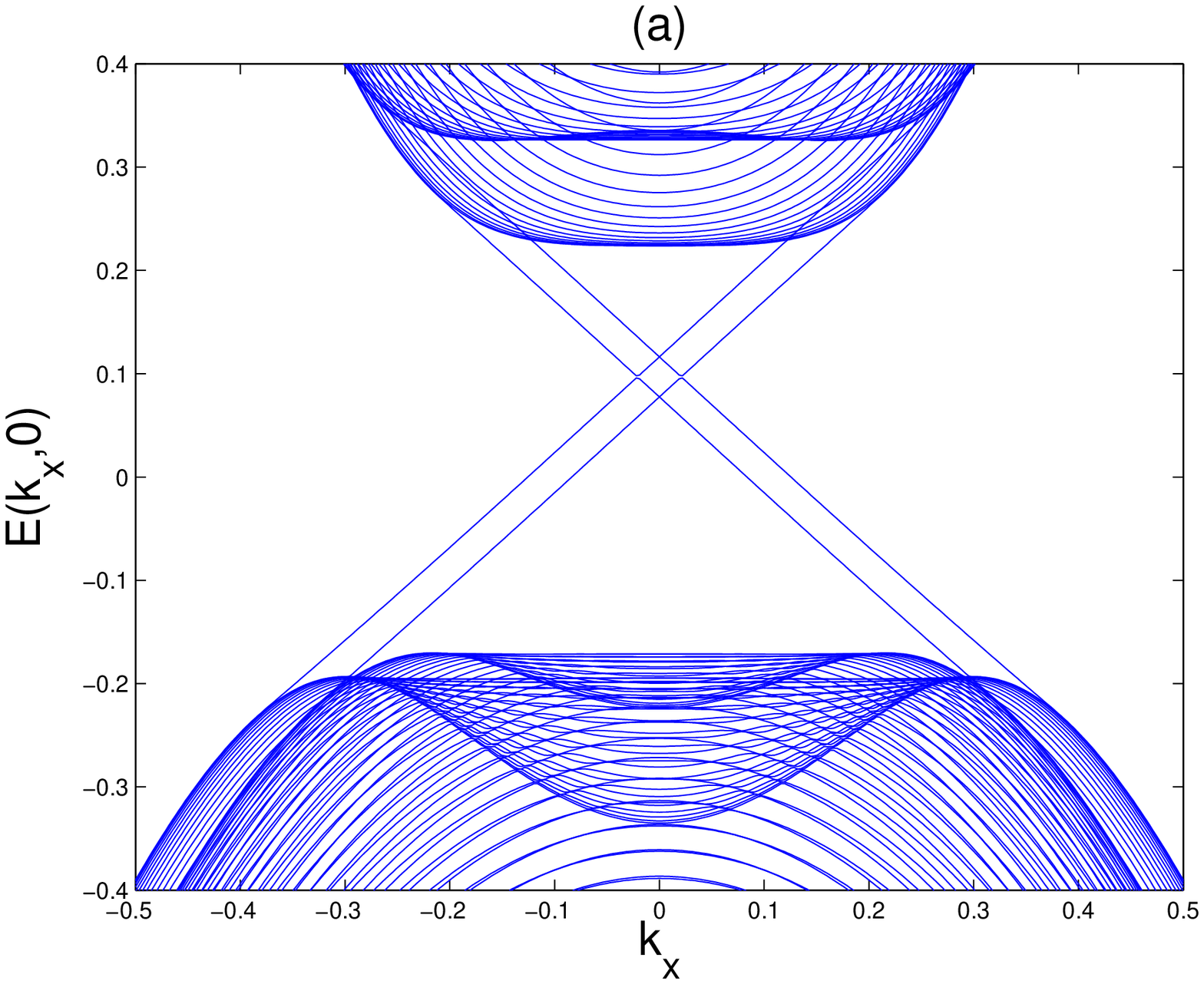}
\end{minipage}
\begin{minipage}{0.495\textwidth}
\centering
\includegraphics[width=\textwidth]{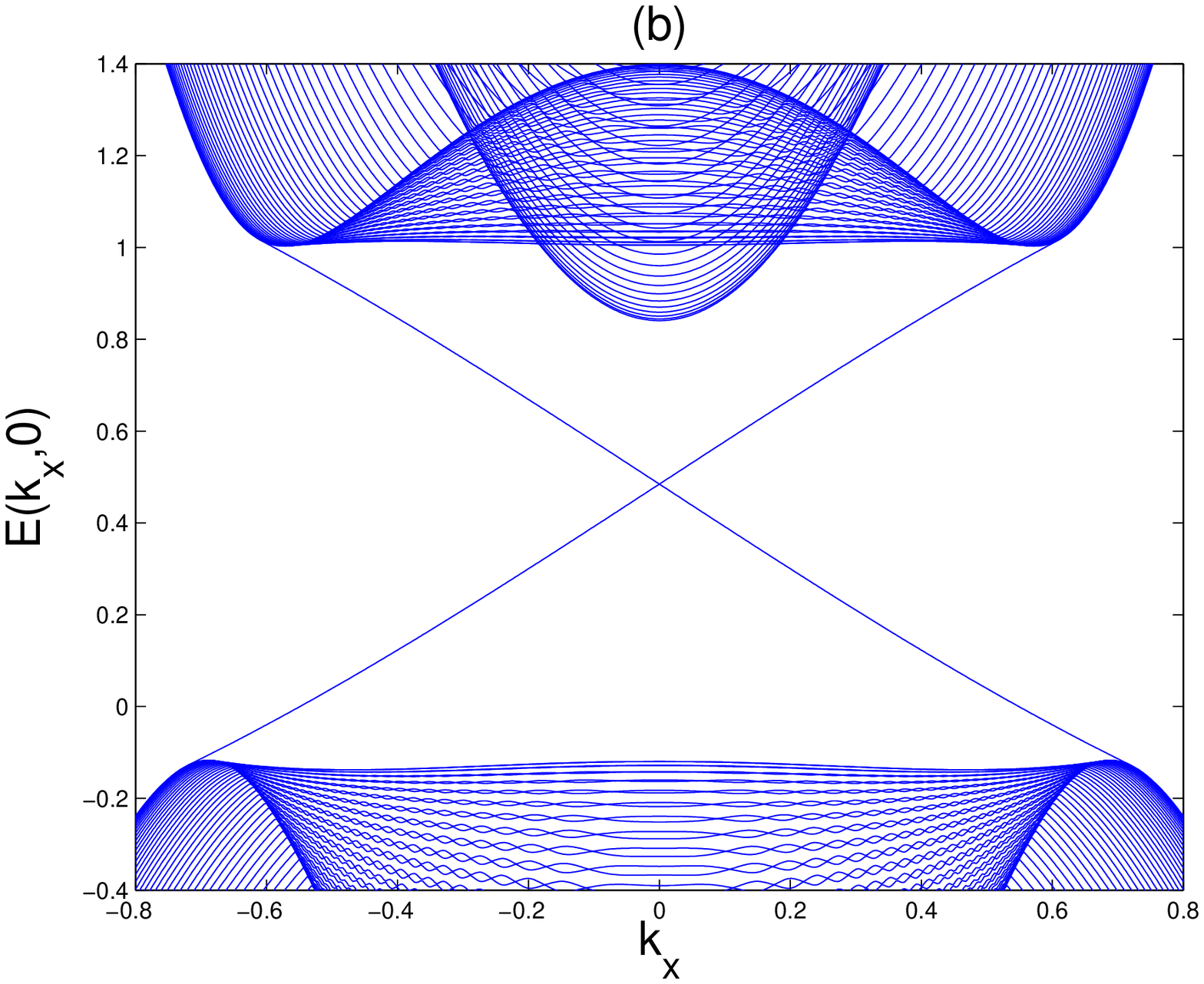}
\end{minipage}
\begin{minipage}{0.495\textwidth}
\centering
\includegraphics[width=\textwidth]{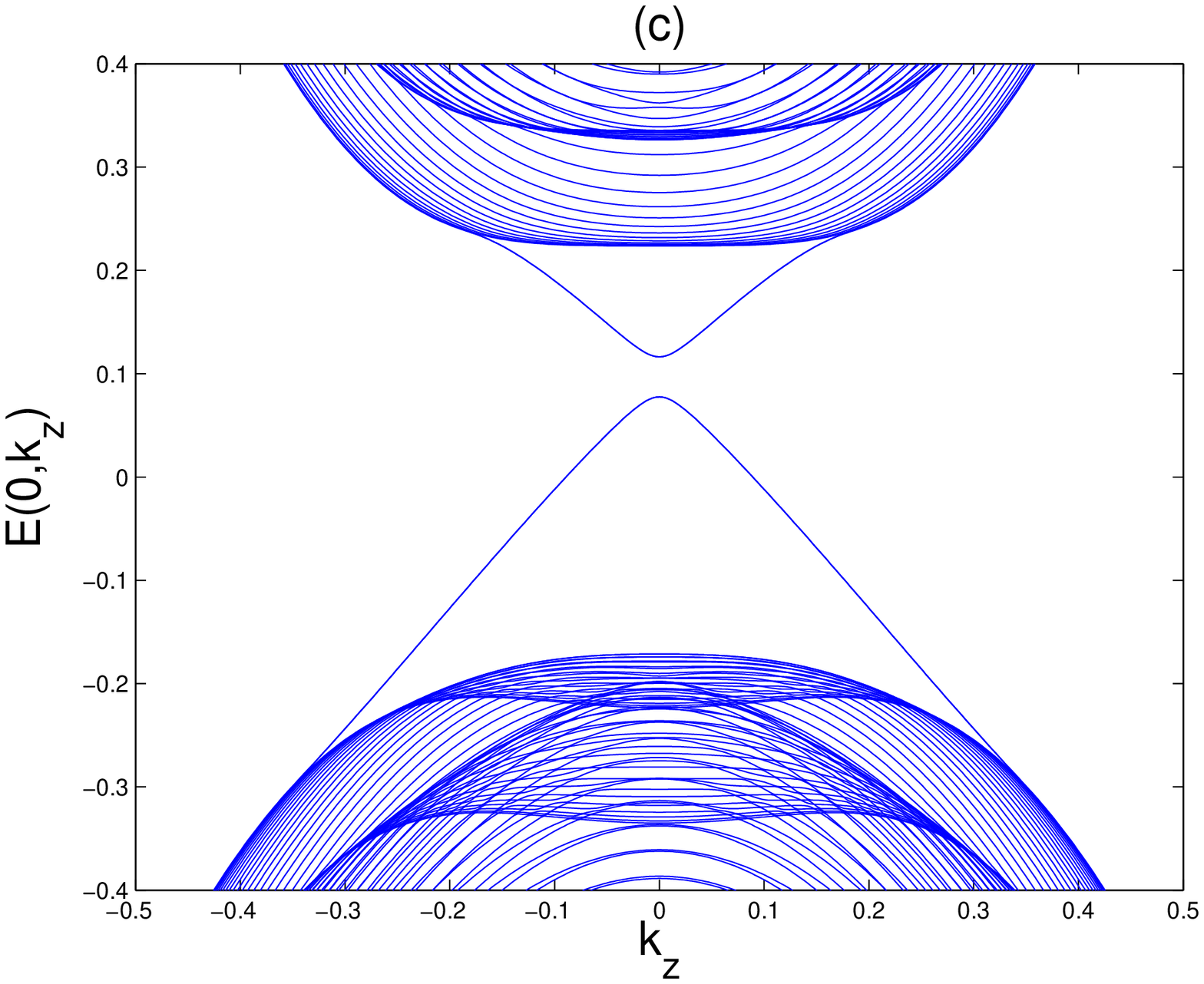}
\end{minipage}
\begin{minipage}{0.495\textwidth}
\centering
\includegraphics[width=\textwidth]{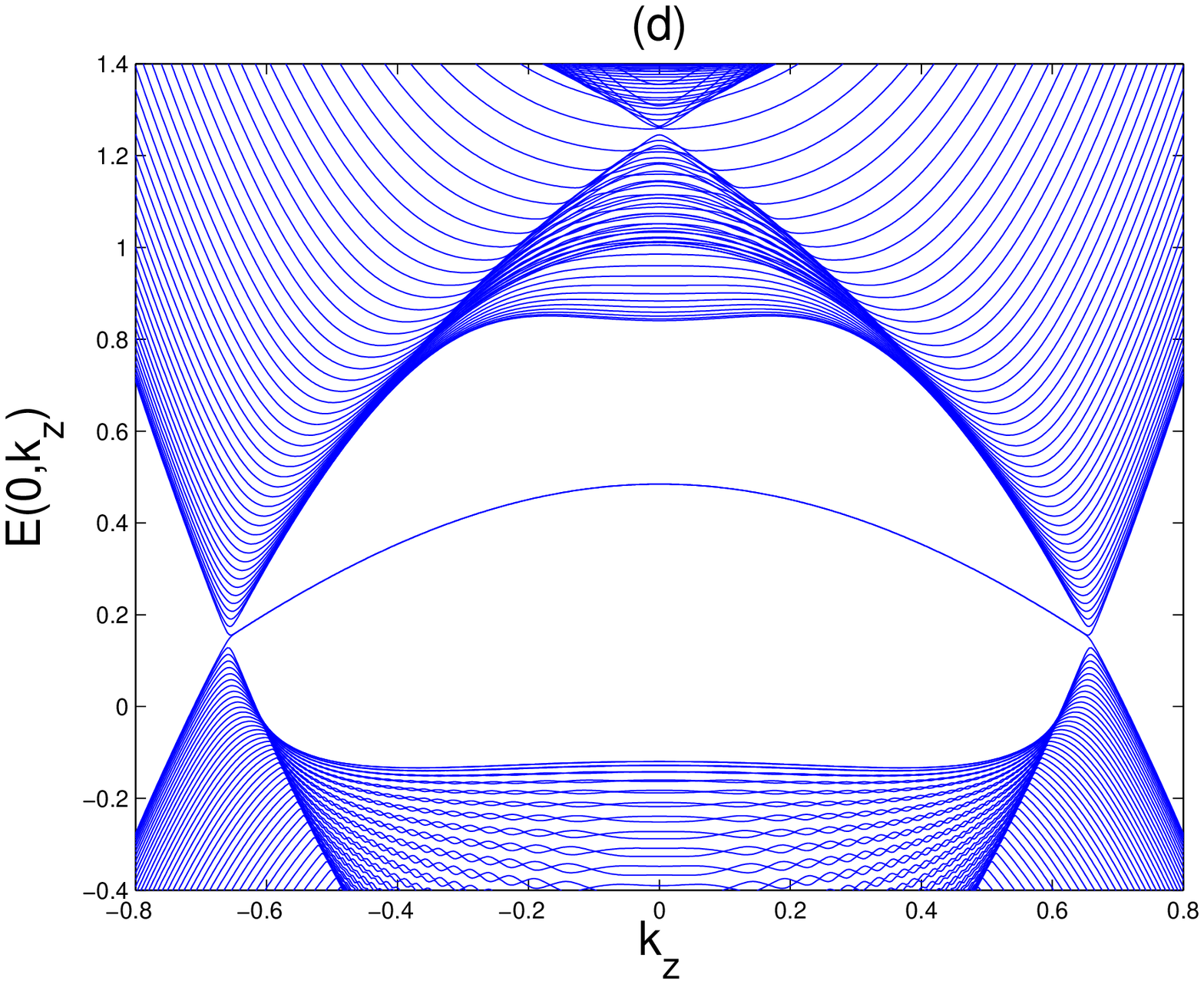}
\end{minipage}
\caption{\label{Fig15}
Numerical dispersions of bulk and surface states for  model I with
$V_z/M=0.2$ in (a) and (c) and $4.0$ in (b) and (d).
In (a) and (b) $k_z=0$, and in (c) and (d) $k_x=0$.
The other parameters are same as in figure~\ref{Fig13}
}
\end{figure}

\begin{figure}[t]
\centering
\includegraphics[width=0.7\textwidth,angle=0]{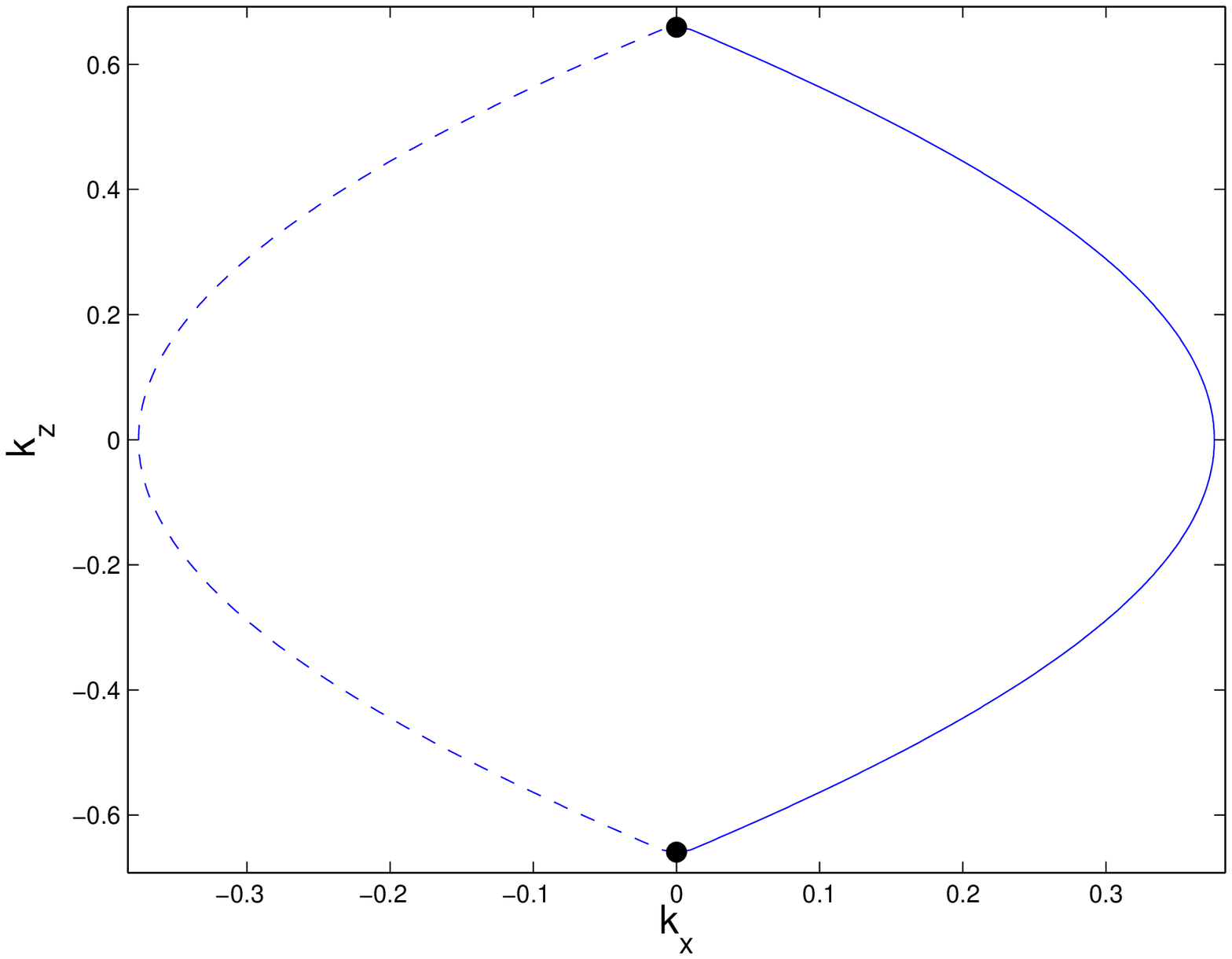}
\caption{\label{Fig16}
Surface Fermi arc 
for the set of parameters shown in figure~\ref{Fig15}~(b) and (d). 
The solid line shows the line of constant energy $E=0.144$~eV for the states
at the $y=0$ surface, the dashed line for the opposite surface at $y=L$.
The black dots at $(k_x,k_z)=(0,\pm 0.66)$ are the projections of
the Weyl nodes onto the surface Brillouin zone. The Fermi arcs end there.
}
\end{figure}

With exchange field in $z$-direction we find the dispersions
\begin{eqnarray}
\label{eq:ener_ex_realistic_I_y_V_z}
\fl E_{\pm}(\mathbf k) =C+M t_2+2(D_1-B_1t_2)(1-\cos k_z) & \\ \nonumber 
\pm\sqrt{4 \beta^2_5(V_z)\left(1-t_2^2\right) A_1^2\sin^2 k_z+(t_2 V_z-2 \sqrt{1-t_2^2}A_2\sin k_x)^2}.
\end{eqnarray}
where $\beta_5$ is given by
\begin{equation} \label{eq:beta5}
\fl \beta_5(V_z) = \frac{4A_2^2 \sqrt{\tilde m_0^2 - V_z^2
- \frac{1}{1-t_2^2} \left( 4 t_2^2 A_2^2 \sin^2 k_x + 4 V_z t_2 A_2 \sqrt{1-t_2^2}
\sin k_x \right)}}{ 4A_2^2 \tilde m_0 + 4 B_2 V_z t_2 A_2 \sqrt{1-t_2^2} \sin
  k_x + B_2 V_z^2 + t_2^2 B_2 \left( 4 A_2^2 \sin^2 k_x - V_z^2 \right)}
\end{equation}
with $\tilde m_0=M-2B_2 \left( 1-\cos k_x \right)-2B_1 \left( 1-\cos k_z \right)$.
From this dispersion we see that the Dirac cone remains ungapped for
an exchange field in $z$-direction and is shifted in $k_x$ direction.
The numerical results shown in Fig.~\ref{Fig15}~(a) and (c) 
confirm this behavior. In Fig.~\ref{Fig15}~(b) and (d) results are shown
for $V_z>M$. From these figures we see that also the one-dimensional
flat band discussed in section~\ref{subsecflat2} becomes dispersive now
due to the broken particle-hole symmetry. However, the system remains
a Weyl semimetal, as was pointed out above. The projections of the
two Weyl nodes onto the surface Brillouin zone for $V_z=4 M$ are found at 
$(k_x,k_z)=(0,\pm 0.66)$ in Fig.~\ref{Fig15}~(d). These are the
points, where the surface band ends. In Fig.~\ref{Fig16}
we show the lines of constant energy $E=0.144$~eV for the
states on the $y=0$ surface (solid line) and on the $y=L$
surface (dashed line). Here, one can see that the line
of constant energy at the energy of the Weyl nodes becomes a
curved and open Fermi arc, as expected in a Weyl semimetal \cite{Wan,Balents}.
In contrast, the Fermi arc was a straight line in the 
particle-hole symmetric case discussed in section~\ref{Weyl}.

\subsection{Model I with boundary perpendicular to the $z$-direction}

In this case we find approximate surface state dispersions 
for small momenta $k_x$ and $k_y$ of the form
\begin{eqnarray}
\label{eq:ener_ex_realistic_I_z_A_V}
\fl E_{\pm}(\mathbf k)=C+Mt_1+2 (D_2-B_2t_1)(2-\cos k_x -\cos k_y) &\\ \nonumber 
\pm\sqrt{V_z^2+(V_x+2\sqrt{1-t_1^2} A_2\sin k_x )^2+(V_y+
2\sqrt{1-t_1^2} A_2\sin^2 k_y )^2},
\end{eqnarray}
where $t_1=D_1/B_1$. Here, the $z$-component of the exchange field
opens a gap in the Dirac dispersion, while both $x$- and $y$-components
lead to a shift of the Dirac cone, leaving it intact.
In this geometry there appear no Fermi arcs at the surface,
as in the corresponding particle-hole symmetric case.
This can be understood from the fact that both Weyl nodes sit on the
$k_z$-axis. For that reason their projections onto a $k_z=$~const.
plane fall on top of each other and the Fermi arc shrinks to zero.
It is interesting to see that the physics of the Weyl semimetal state
cannot be observed on a $z$-surface due to the structure of the Weyl state
here. Experimentally one needs to look at the side surfaces or at a
surface having a finite angle with the $z$-surface to
observe the Fermi arc. 

\begin{figure}
\begin{minipage}{0.495\textwidth}
\centering
\includegraphics[width=\textwidth]{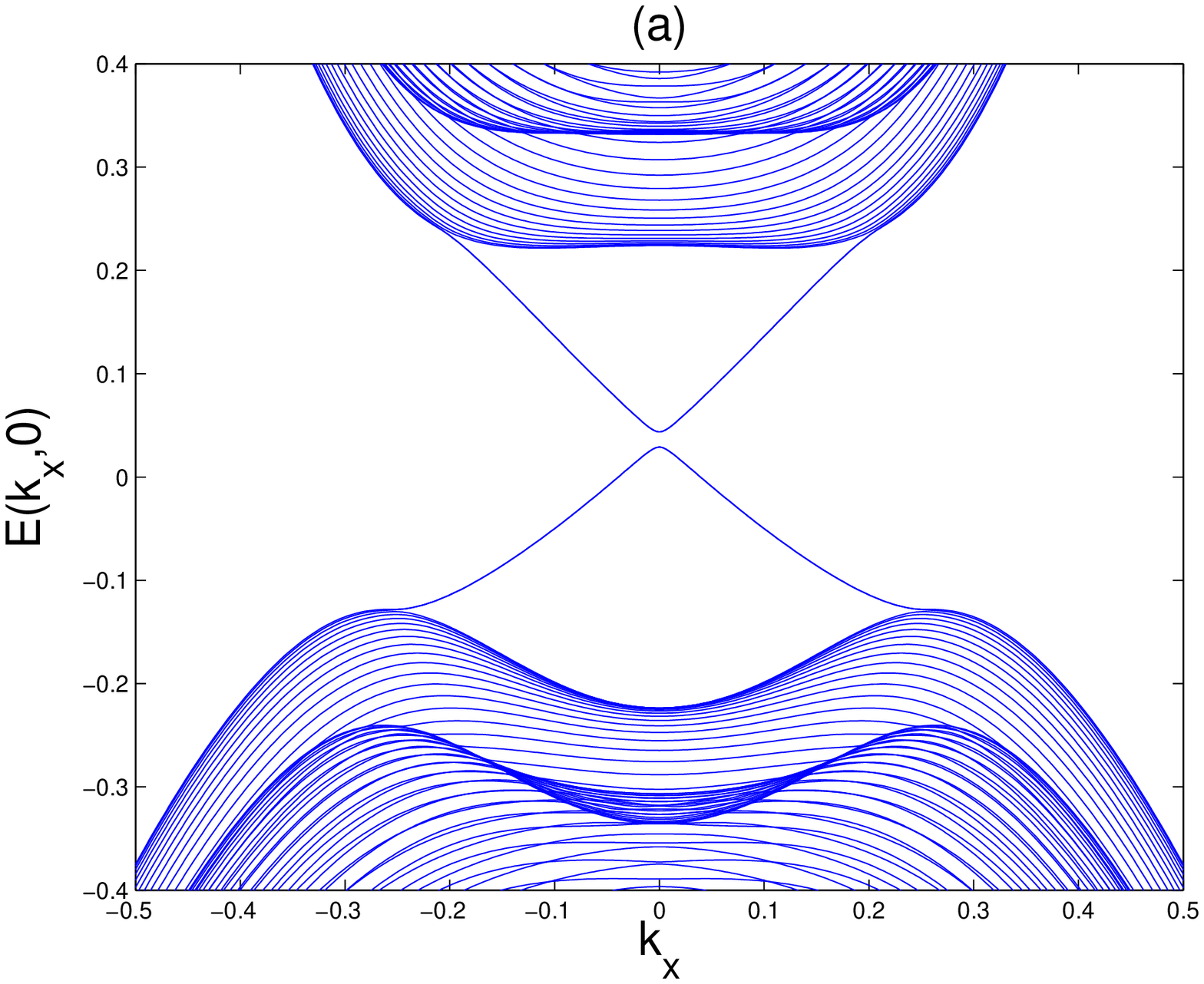}
\end{minipage}
\begin{minipage}{0.495\textwidth}
\centering
\includegraphics[width=\textwidth]{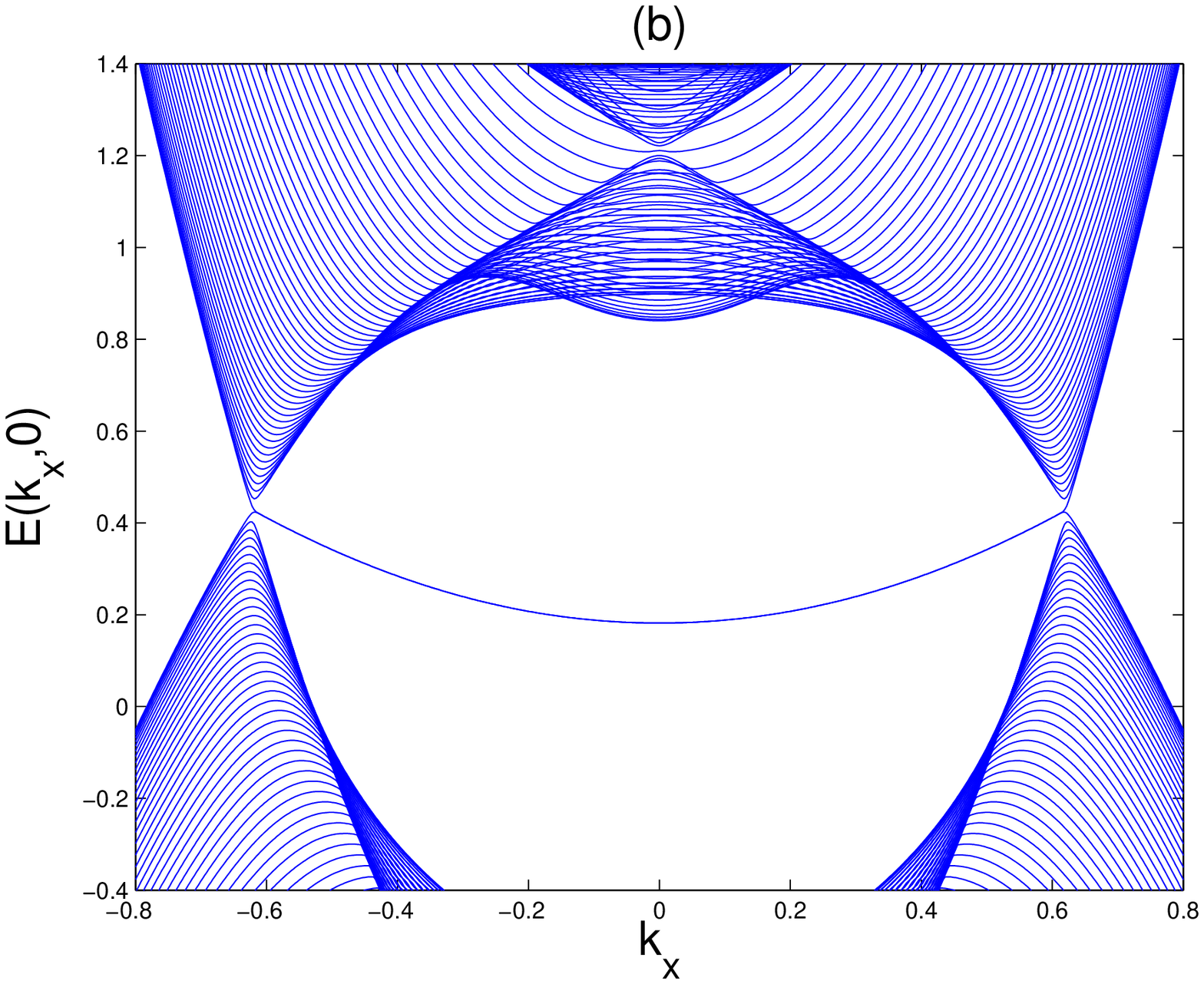}
\end{minipage}
\begin{minipage}{0.495\textwidth}
\centering
\includegraphics[width=\textwidth]{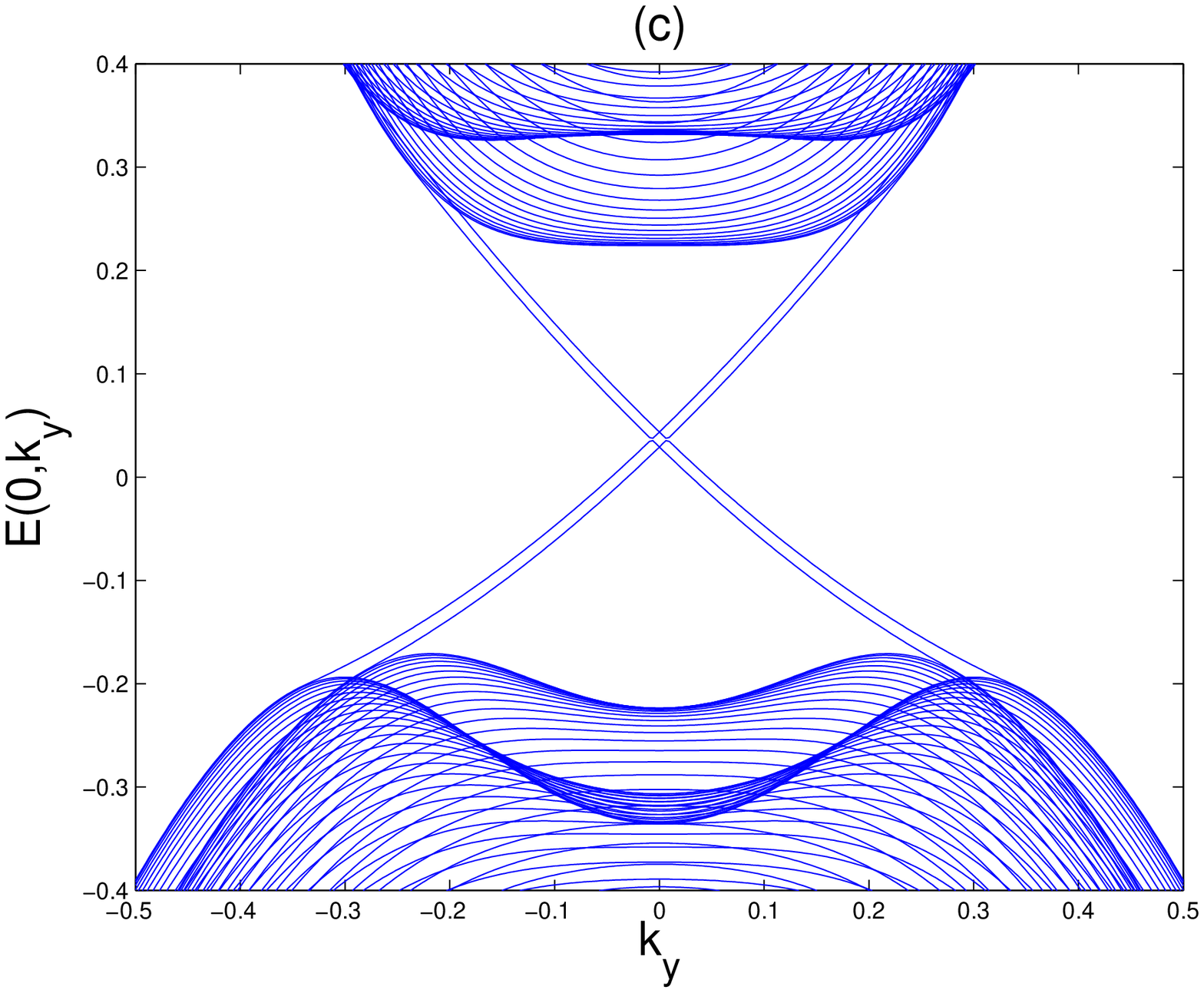}
\end{minipage}
\begin{minipage}{0.495\textwidth}
\centering
\includegraphics[width=\textwidth]{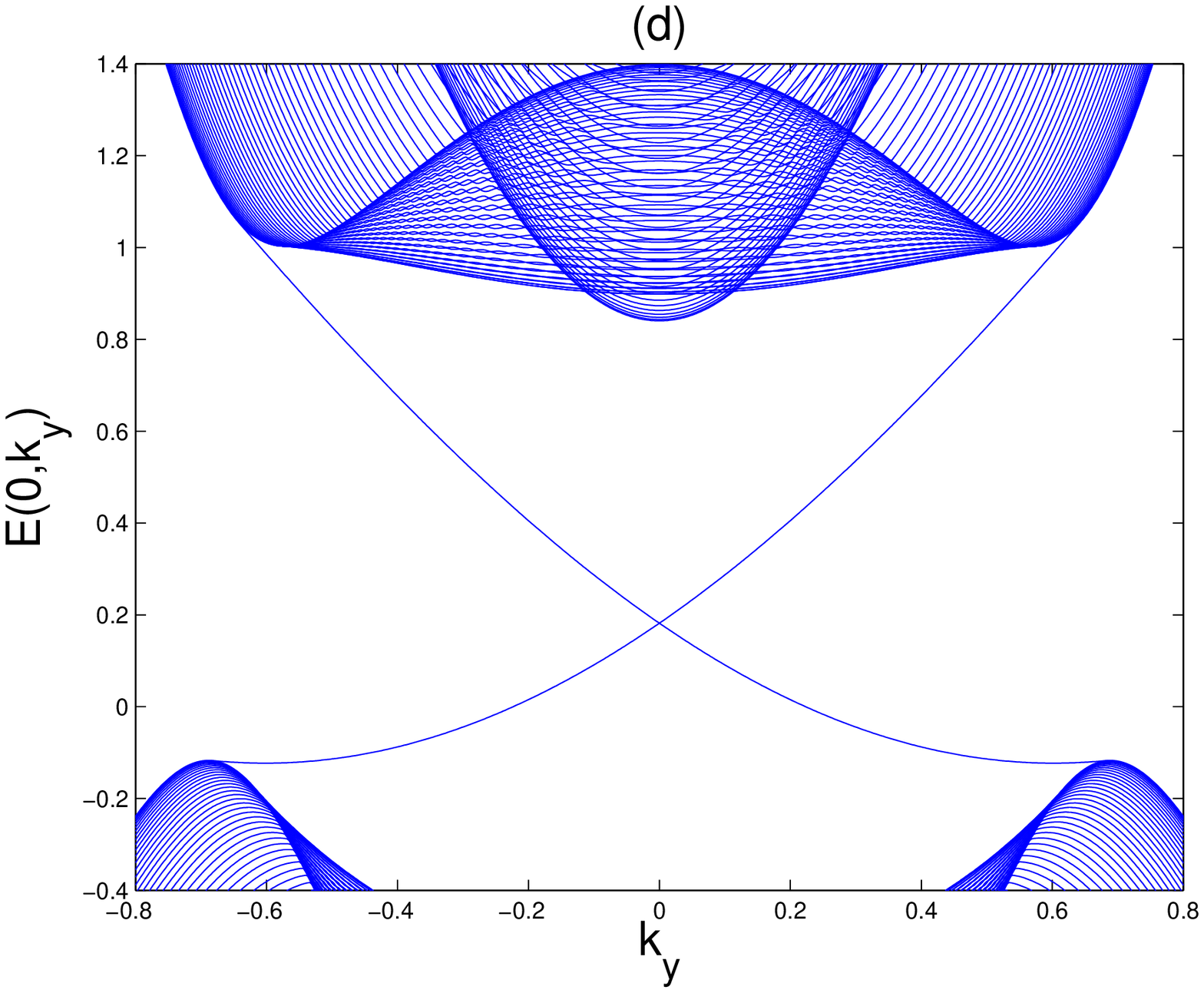}
\end{minipage}
\caption{\label{Fig17}
Numerical dispersions of bulk and surface states for model II with
$V_x/M=0.2$ for (a) and (c) and $4.0$ for (b) and (d).
In (a) and (b) $k_y=0$, and in (c) and (d) $k_x=0$.
The other parameters are same as in figure~\ref{Fig13}
}
\end{figure}

\subsection{Model II with boundary perpendicular to the $z$-direction}

In the absence of an exchange field the surface state dispersions are
given by
\begin{eqnarray} 
\nonumber
	E_{\pm}(\mathbf k)&=C+Mt_1+2(D_2-B_2t_1)(2-\cos k_x-\cos k_y)
          \\ \label{eq:ener_ex_realistic_II_z}
&\pm\sqrt{1-t_1^2}2 A_2 \sqrt{\sin^2 k_x+\sin^2 k_y},
\end{eqnarray}
where $t_1=D_1/B_1\approx 0.13$. If $t_1>1$ surface states do not
exist.

When the exchange field points into $x$-direction we find
the following surface state dispersions
\begin{eqnarray}
\label{eq:ener_ex_realistic_II_z_V_x}
E_{\pm}(\mathbf k)&=C+M t_1+(D_2-B_2t_1) (2-\cos k_x -\cos k_y )\\
\nonumber 
&\pm\sqrt{4 \beta^2_6(V_x)\left(1-t_1^2\right) A_2^2\sin^2 k_x+
(t_1 V_x-2 \sqrt{1-t_2^2}A_2\sin k_y)^2}.
\end{eqnarray}
where $\beta_6$ is given by
\begin{equation} \label{eq:beta6}
\fl \beta_6(V_x) = \frac{4A_2^2 \sqrt{\tilde m_0^2 - V_x^2
- \frac{1}{1-t_1^2} \left( t_1^2 A_2^2 \sin^2 k_y + 4 V_x t_1 A_2 \sqrt{1-t_1^2}
\sin k_x \right)}}{ 4A_2^2 \tilde m_0^2 + 4 B_2 V_x t_1 A_2 \sqrt{1-t_1^2} \sin
  k_y + B_2 V_x^2 + t_1^2 B_2 \left( 4 A_2^2 \sin^2 k_y - V_x^2 \right)}
\end{equation}
with $\tilde m_0=M-2B_2 \left( 1-\cos k_x \right)-2B_2 \left( 1-\cos k_y \right)$.
From this dispersion we see that the Dirac cone remains ungapped for
an exchange field in $x$-direction and is shifted in $k_y$ direction.
Corresponding numerical results are shown in Fig.~\ref{Fig17}~(a) and (c) 
confirming this behavior. In Fig.~\ref{Fig17}~(b) and (d) we show results
for $V_x>M$. Here, we see that again the corresponding one-dimensional
flat band discussed in section~\ref{subsec2vxy} becomes dispersive now
due to the broken particle-hole symmetry. Again, this system remains
a Weyl semimetal, however. As pointed out in appendix I, in model II the Weyl 
nodes follow the direction of the exchange field and sit on the $k_x$-axis
in the present case. The projections of the
two Weyl nodes onto the surface Brillouin zone for $V_z=4 M$ are found at 
$(k_x,k_y)=(\pm 0.623,0)$ in Fig.~\ref{Fig17}~(d). These are the
points, where the surface band ends. Similarly as in Fig.~\ref{Fig16}
the lines of constant energy $E=0.421$~eV become curved and open
Fermi arcs (not shown).

When the exchange field points into $z$-direction we find
the following surface state dispersions
\begin{eqnarray}
\label{eq:ener_ex_realistic_II_z_V_z}
\nonumber E_{\pm}(\mathbf k)&=C+Mt_1+2(D_2-B_2t_1) (2- \cos k_x - \cos k_y )\\
&\pm\sqrt{V_z^2+(1-t_1^2)4 A_2^2 \left( \sin^2(k_x)+ \sin^2(k_y) \right)}.
\end{eqnarray}
Thus, the application of an exchange field perpendicular to the surface
leads to a gap in the Dirac dispersion. When $V_z$ is increased
beyond $V_{cr}$ no flat band appears. This can be understood
from the fact that in model II the Weyl nodes follow the
field direction. Thus, for fields perpendicular to the
surface their projections onto the surface Brillouin zone fall
on top of each other and the Fermi arc disappears.

\section{Summary and Conclusions}

We have studied the modification of the surface states of a three dimensional
topological insulator by a ferromagnetic exchange field using the
two models that are presently discussed for topological insulators.

For model I, which is appropriate for the Bi$_2$Se$_3$ class of layered materials
the surface states on a side surface behave qualitatively different than
the ones on the top or bottom surface. For exchange fields smaller than the
bulk gap the velocity of the Dirac cone can be tuned down to smaller values
in an anisotropic way depending on the direction of the exchange field.
For exchange fields larger than a critical value of the order of the bulk gap 
the system becomes a
topologically nontrivial semimetal. We have shown that in a particle-hole
symmetric system the Fermi surface of this semimetal is a line in
momentum space, if the exchange field is directed within the $xy$-plane.
In this case a two-dimensional flat band appears at a side surface.
If the exchange field possesses a finite component in $z$-direction,
there exist only single points in the Brillouin zone, where the
bulk gap vanishes. We have shown that in this general case the system
becomes a Weyl semimetal. Associated with this peculiar state of matter
we find Fermi arcs at the side surfaces and one-dimensional flat bands.

If particle-hole symmetry is not obeyed, the flat bands become
dispersive, but remain well separated from the bulk bands. The
Weyl semimetallic phase and the existence of a surface Fermi arc 
is preserved also under broken particle-hole symmetry.

In model II, which is more isotropic than model I, the behavior
on the top and side surfaces is qualitatively the same.
For small exchange fields, it is again possible to tune
the velocity of the Dirac cone by the exchange field anisotropically.
For exchange fields larger than a critical value of the order of 
the bulk gap model II always enters a Weyl semimetal state.
Associated with this we also find Fermi arcs and one-dimensional 
surface flat bands in this case. However, in model II there
is no case in which a two-dimensional surface flat band appears.

If particle-hole symmetry is not obeyed, the flat bands become
again dispersive. The Weyl semimetallic phase and the existence 
of a surface Fermi arc again
is preserved under broken particle-hole symmetry.

The Weyl nodes in model I sit on the $k_z$ axis, while they move
with the direction of the exchange field in model II. As a consequence, 
in model I the Fermi arcs and flat bands do not appear on the top
and bottom surfaces, while in model II this is possible.

We have shown that the appearence of flat bands in our particle-hole
symmetric cases could always be classified by topological
invariants that have recently been given by Matsuura et al \cite{Matsuura}.
The invariants are related to different chiral symmetries in the
different cases.

Surface flat bands have been proposed in systems like graphene, 
superfluid $^3$He, or unconventional superconductors before.
However, in these systems cryogenic temperatures are required
to observe the flat bands. In the materials discussed here,
the energy scale is set by the bulk gap, which is of the
order of 0.3~eV in Bi$_2$Se$_3$. Thus, flat bands could be
observed already at room temperature. The fact that
one can turn on or off a surface flat band by rotation
of the magnetization of a ferromagnet makes the present
system particularly interesting for applications in
spintronics.
 
\section*{Acknowledgments}

We would like to thank A.~P.~Schnyder, A.~M.~Lunde, 
G.~Reiss, C.~Timm, and A.~Altland for valuable discussions.

\section*{Appendix I}

In this appendix we derive the ranges of the exchange field under which
model I and II become semimetallic, i.e. the gap vanishes. It is only
in the semimetallic phase that the system possesses surface flat bands.
In particular it is of interest to know the critical value of the
exchange field that is necessary to drive the system from the insulating
into the semimetallic phase.

For model I we start from the bulk bandstructure of the four bands given
in Eq.~(\ref{eq:model1bulk}). We omit the diagonal term $\epsilon_0(\mathbf k)
\mathbb{I}_{4 \times 4}$ in the Hamiltonian as it just shifts the four bands by
$\epsilon_0(\mathbf k)$ and thus does not affect the direct gap of the system.
For large values of $D_1$ and $D_2$ this term can lead
to an indirect gap, however. 

As is clear from the symmetric spectrum Eq.~(\ref{eq:model1bulk}) the
gap closes when there exist momenta ${\mathbf k}$ for which
$E_i^I({\mathbf k})$ becomes zero. To determine those momenta it is
beneficial to write the exchange field in spherical coordinates,
i.e. $V_x=V \cos \varphi \cos \vartheta$, $V_y=V \sin \varphi \cos
\vartheta$, and $V_z=V \sin \vartheta$. If one exploits the fact that
\begin{samepage}
\begin{eqnarray*}
\left( m_1 \sin \varphi \cos \vartheta - m_2 \cos \varphi \cos
\vartheta\right)^2 +
\left( m_1 \cos \varphi \cos \vartheta + m_2 \sin \varphi \cos
\vartheta\right)^2 \\  
= \left( m_1^2 + m_2^2 \right) \left( 1 - \sin^2 \vartheta
\right) \, ,
\end{eqnarray*}
\end{samepage}
Eq.~(\ref{eq:model1bulk}) can be written in the form
\begin{eqnarray} 
\nonumber
\fl	E_i^{I}(\mathbf k)&=\pm \Big\{ \left(m_1 \sin \varphi \cos \vartheta - m_2 \cos
          \varphi \cos \vartheta \right)^2+ \sin^2 \vartheta \left( m_1^2 +
          m_2^2 \right) + \\ \label{eq:model1bulkd}
\fl &+ \left( V \pm \sqrt{m_0^2 + m_3^2 + \left(m_1 \cos \varphi \cos \vartheta
            + m_2 \sin \varphi \cos \vartheta \right)^2} \right)^2 \Big\}^{1/2}
\end{eqnarray}
This expression can become zero only, if all squared expressions below
the square-root become zero simultaneously. For $V_z \neq 0$
this means $m_1=m_2=0$ and $V^2= m_0^2
+ m_3^2$. For a given value of $V$ the gap will close only in a single
pair of bands. From $m_1=m_2=0$ it follows that $k_x=0$ or $\pi$ and
$k_y=0$ or $\pi$. The equation $V^2= m_0^2+ m_3^2$ can then only be
fulfilled for selected values of $k_z$. Thus the system will have
single Fermi points, when a solution exists. For the special case
$\vartheta=0$, i.e. when the exchange field lies within the $x$-$y$-plane,
there are just two equations to be fulfilled, i.e.
$m_1 \sin \varphi = m_2 \cos \varphi $
and $V^2=m_0^2 + m_3^2 + \left(m_1 \cos \varphi + m_2 \sin \varphi
\right)^2$. 
In this case the Fermi surface will be a line in momentum space.

To determine the ranges of the exchange field for which these solutions
exist, we determine the minimum and maximum value of $m_0^2+ m_3^2$.
Let us set $c=\cos k_z$. Then we have to minimize the function
\begin{equation}
\label{eq:fc}
f(c) = \left[ M' - 2B_1 \left( 1-c \right) \right]^2 + 4 A_1^2 \left( 1- c^2 \right)
\end{equation}
in the interval $c \in \left[ -1,1 \right]$.
Here, we have set
\begin{eqnarray}
\nonumber
M' &= \left\{ \begin{array}{cl} 
M & \mbox{for} \, k_x=k_y=0 \\
M - 4 B_2 & \mbox{for} \, (k_x=0 \, \mbox{and} \, k_y=\pi) \, \,
\mbox{or} \, \, (k_x=\pi \, \mbox{and} \, k_y=0)  \\
M - 8 B_2 & \mbox{for} \, k_x=k_y=\pi 
\end{array} \right.
\end{eqnarray}
The function $f(c)$ is quadratic in $c$, so there can only be a single
extremum within $c \in \left[ -1,1 \right]$ or the extremum will
be on the boundaries $c=\pm 1$. On the boundaries we have
\begin{eqnarray}
\nonumber
f(c=1) &= {M'}^2 \quad \mbox{and} \quad f(c=-1) = \left( 4 B_1 - M' \right)^2
\end{eqnarray}
As we assume $B_1>M>0$ and $B_2>M>0$, we have $f(c=-1) > f(c=1)$.
To determine a possible extremum inside the interval $c \in \left[ -1,1
  \right]$
we take the derivative of $f(c)$ yielding
\begin{eqnarray}
\nonumber
\frac{df}{dc} &= 4 M' B_1 - 8 B_1^2 + \left( 8 B_1^2 -8 A_1^2 \right) c
\end{eqnarray}
This becomes zero, if
\begin{equation}
\label{eq:cmin}
c = c_{min} = \frac{B_1 \left( 2 B_1 - M' \right)}{2 \left( B_1^2 - A_1^2 \right)}
\end{equation}
The second derivative of $f(c)$ is
\begin{eqnarray}
\nonumber
\frac{d^2f}{dc^2} = 8\left( B_1^2 - A_1^2 \right) \,.
\end{eqnarray}
As we assume $B_1 > A_1$, this is positive. We thus find that there
exists a minimum inside the interval $c \in \left[ -1,1 \right]$,
if
\begin{eqnarray}
\nonumber
\left| \frac{B_1 \left( 2 B_1 - M' \right)}{2 \left( B_1^2 - A_1^2 \right)}
\right| \le 1
\end{eqnarray}
which is equivalent to the condition
\begin{eqnarray}
\nonumber
M' B_1 \ge 2 A_1^2
\end{eqnarray}
This condition can only be fulfilled for positive $M'$, i.e. only
for $k_x=k_y=0$. Thus, in the case $M B_1 \ge 2 A_1^2$, the minimum
for the exchange field can be found by introducing Eq.~(\ref{eq:cmin})
into Eq.~(\ref{eq:fc}). After some algebra one finds
\begin{eqnarray}
\nonumber
f(c_{min})= \frac{A_1^2 \left( 4 B_1 M - M^2 - 4 A_1^2\right)}{B_1^2-A_1^2}
\end{eqnarray}
Thus we find that the minimum critical value $V_{cr}$ for the magnitude
of the exchange field, which is necessary to bring the system
into the semimetallic phase, is given by
\begin{equation}
\label{eq:Vcr}
V_{cr} = \left\{ \begin{array}{cl} 
M & \mbox{for} \, M B_1 < 2 A_1^2 \\
A_1 \sqrt{\frac{4 B_1 M - M^2 - 4 A_1^2}{B_1^2-A_1^2}}
& \mbox{for} \, M B_1 \ge 2 A_1^2
\end{array} \right.
\end{equation}
In total we find the following three ranges for the magnitude
of the exchange field, in which the system becomes semimetallic:
$\left[ V_{cr}, 4B_1-M \right]$, $\left[ 4B_2-M, 4B_1+4B_2-M \right]$,
and $\left[ 8B_2-M, 4B_1+8B_2-M \right]$. For $B_2 \ge B_1$ these ranges
overlap, while for $B_2 < B_1$ they are separate. As
$B_1$ and $B_2$ are usually of the order of one to several eV, we
do not expect that large enough exchange fields can be applied in practice
to actually observe these different ranges. However, the minimum
critical field $V_{cr}$, which is of the order of $M$ or less,
is within experimental reach.

The position of the Fermi points is found from the quadratic equation
\begin{eqnarray}
\label{eq:fcv0}
V^2= f(c)
\end{eqnarray}
For $V \in \left[ V_{cr}, 4B_1-M \right]$ the Fermi points sit on
the $k_z$-axis ($k_x=k_y=0$). Their positions are given by
\begin{eqnarray}
\label{eq:c12model1}
c_{1/2} = c_{min} \pm \frac{\sqrt{A_1^2 \left( M^2 - 4 M B_1 + 4 A_1^2 \right) 
+ V^2 \left( B_1^2 - A_1^2 \right)}}{2 \left( B_1^2 - A_1^2 \right) }
\end{eqnarray}
For the case $M B_1 \ge 2 A_1^2$ and $V < M$ both solutions are
within the interval $c \in \left[ -1,1 \right]$ and we thus have
a total of four Fermi points at the positions
$k_{z,1/2}=\pm \arccos c_1$ and $k_{z,3/4}=\pm \arccos c_2$.
In the other case $M B_1 < 2 A_1^2$ only solution $c_2$
with the minus sign in Eq.(\ref{eq:c12model1}) is within
$\left[ -1,1 \right]$ and we have just two Fermi points at
$k_{z,1/2}=\pm \arccos c_2$.

For the special case $V_z=0$ the minimum value of the exchange field
to bring the system into the semimetallic state remains the same.
In this case the function
\begin{eqnarray}
\nonumber
V^2= m_0^2 + m_3^2 + \left( m_1 \cos \varphi + m_2 \sin \varphi \right)^2
\end{eqnarray}
has to be minimized. However, as the minimum of $m_0^2 + m_3^2$ occurs
at $k_x=k_y=0$ and $m_1=m_2=0$ there, the minimum of $V$ remains the
same value $V_{cr}$ Eq.~(\ref{eq:Vcr}). Also, for $V>V_{cr}$ the two
(or four) points determined above still lie on the Fermi surface.
However, as pointed out above, the Fermi surface becomes a line
(or two lines) now, which approximately run within the plane
perpendicular to the exchange field (The precise condition is that
the vector $(m_1,m_2)^T=2 A_2(\sin k_x,\sin k_y)^T$ should be 
perpendicular to the exchange field $(V_x,V_y)$.)

Let us next look at model II. We again write
the exchange field in spherical coordinates,
i.e. $V_x=V \cos \varphi \cos \vartheta$, $V_y=V \sin \varphi \cos
\vartheta$, and $V_z=V \sin \vartheta$. 
This time Eq.~(\ref{eq:model2bulk}) can be written in the form
\begin{eqnarray} 
\nonumber
\fl	E_i^{II}(\mathbf k)&=\pm \Big\{ \left(m_1 \sin \varphi - m_2 \cos
          \varphi \right)^2+ \left( m_1 \cos \varphi \sin \vartheta +
          m_2 \sin \varphi \sin \vartheta - m_3 \cos \vartheta \right)^2 + \\ \label{eq:model2bulkd}
\fl &+ \left( V \pm \sqrt{m_0^2 + \left(m_1 \cos \varphi \cos \vartheta
            + m_2 \sin \varphi \cos \vartheta + m_3 \sin \vartheta \right)^2} \right)^2 \Big\}^{1/2}
\end{eqnarray}
This expression can become zero only, if all three squared expressions below
the square-root vanish simultaneously. This means that the system will have
Fermi points, when a zero energy solution exists.
The first two expression become zero, if
the vector $(m_1,m_2,m_3)$
is parallel to $(\cos \varphi \cos \vartheta, \sin \varphi \cos \vartheta, \sin
\vartheta)$, i.e. parallel to the exchange field. The third equation can
then be written as
\begin{eqnarray}
\nonumber
V^2 &= m_0^2 + m_1^1 + m_2^2 + m_3^2
\end{eqnarray}
It is clear that the minimum critical value $V_{cr}$ for the magnitude
of the exchange field, which is necessary to bring the system
into the semimetallic phase fulfils $V_{cr} \le M$, because a zero energy 
state can always be found for $k_x=k_y=k_z=0$ and $V=M$. 
A general expression for $V_{cr}$ can in principle be obtained
analytically for exchange field in any direction. However,
the expressions become quite complicated. Therefore, here
we focus on the two cases that the exchange field points
either in $x$-direction or in $z$-direction.

For exchange field in $x$-direction we have $m_2=m_3=0$.
The minimum can thus be found on the $k_x$-axis by
minimizing the function $f(c)$ Eq.~(\ref{eq:fc}) with
$c=\cos k_x$ and $B_1$ and $A_1$ being replaced by
$B_2$ and $A_2$. Following the same calculation as for model I
we then find for the critical value $V_{cr,x}$ in this case:
\begin{equation}
\label{eq:Vcrx}
V_{cr,x} = \left\{ \begin{array}{cl} 
M & \mbox{for} \, M B_2 < 2 A_2^2 \\
A_2 \sqrt{\frac{4 B_2 M - M^2 - 4 A_2^2}{B_2^2-A_2^2}}
& \mbox{for} \, M B_2 \ge 2 A_2^2
\end{array} \right.
\end{equation}
Analogously to model I, for $V>V_{cr,x}$ we have either
two or four Fermi points lying on the $k_x$-axis.
Their positions are found from
\begin{eqnarray}
\label{eq:c12model2x}
c_{1/2} = c_{min} \pm \frac{\sqrt{A_2^2 \left( M^2 - 4 M B_2 + 4 A_2^2 \right) 
+ V^2 \left( B_2^2 - A_2^2 \right)}}{2 \left( B_2^2 - A_2^2 \right) }
\end{eqnarray}

For exchange field in $z$-direction we have $m_1=m_2=0$.
The minimum can thus be found on the $k_z$-axis by
minimizing the same function $f(c)$ Eq.~(\ref{eq:fc}) 
as for model I. We thus find for the critical value $V_{cr,z}$ in this case:
\begin{equation}
\label{eq:Vcrz}
V_{cr,z} = \left\{ \begin{array}{cl} 
M & \mbox{for} \, M B_1 < 2 A_1^2 \\
A_1 \sqrt{\frac{4 B_1 M - M^2 - 4 A_1^2}{B_1^2-A_1^2}}
& \mbox{for} \, M B_1 \ge 2 A_1^2
\end{array} \right.
\end{equation}
From this expression we see that for model II in general the critical
value $V_{cr}$ depends on the direction of the exchange field, in
contrast to model I, where it was isotropic.
The position of the Fermi points in this case is given by
Eq.~(\ref{eq:c12model1}) for $V>V_{cr,z}$. Again, one
sees that the position of the Fermi points in model II
varies with the direction of the exchange field, in contrast
to model I.

\section*{Appendix II}

During the course of this work we repeatedly encounter the situation
that the Hamiltonian can be written in the following form:
\begin{eqnarray}
\nonumber
H({\bf k}) &= H_0({\bf k}) + H'({\bf k}) \\
H'({\bf k}) &= f({\bf k}) A + g({\bf k}) B
\nonumber
\end{eqnarray}
where $f(\bf k)$ and $g(\bf k)$ are complex functions. 
$A$ is a momentum independent operator that commutes with $H_0(\bf k)$,
while $B$ anticommutes with $H_0(\bf k)$. We would like to
determine the eigenstates and eigenvalues of $H(\bf k)$ from the known
eigenstates and eigenvalues of $H_0(\bf k)$. Due to the symmetries
and the $4 \times 4$ structure for each momentum $\bf k$ $H_0(\bf k)$
possesses at most two degenerate states with energy $E_0(\bf k)$ and two 
degenerate states with energy $-E_0(\bf k)$.

Let $\left| 1 \right\rangle$ be an eigenstate of $H_0$ with energy $E$.
Then  $A \left| 1 \right\rangle$ will also be an eigenstate of
$H_0$ with energy $E$ and $B \left| 1 \right\rangle$ 
will be an eigenstate of $H_0$ with energy $-E$. Thus,
\begin{equation}
A \left| 1 \right\rangle = {\sum_j}^+ a_j \left| j \right\rangle
\end{equation}
where the sum goes over all the eigenstates of $H_0$ with energy E,
and 
\begin{equation}
B \left| 1 \right\rangle = {\sum_n}^- b_n \left| n \right\rangle
\end{equation}
where the sum goes over all the eigenstates of $H_0$ with energy -E.
In total we have
\begin{equation}
H' \left| 1 \right\rangle = f({\bf k}) {\sum_j}^+ a_j \left| j \right\rangle + 
g({\bf k}) {\sum_n}^- b_n \left| n \right\rangle
\end{equation}
Thus, $H'$ can only couple the eigenstates of $H_0$ with energies $E$ and
$-E$. This means that the total Hamiltonian $H$ in the basis of the
eigenstates of $H_0$ is reduced to (at most) $4 \times 4$ blocks.
The problem of finding the eigenstates and eigenvalues of $H(\bf k)$ 
thus reduces to $4 \times 4$ matrices within the $\pm E$ spaces of $H_0$.

In the cases discussed in this manuscript we are interested in
surface states localized on a single side of the system. We construct
$H_0$ in such a way that its surface states have zero energy.
We then have to diagonalize $H'$ in the $E=0$ subspace of $H_0$.
The eigenvalues and eigenstates of $H'$ in this subspace are then 
the surface states of $H$ and their energies that we look for.
As the number of localized states at one surface of the system is
equal to the number of localized states at the oppsite surface
due to parity symmetry, we are only left to solve a $2 \times 2$ problem.
So the general procedure, which we take in this manuscript, is to
first determine the zero energy eigenstates of $H_0$ (if they exist)
and then determine the eigenvalues and eigenstates of $H'$
in this subspace.

\section*{Appendix III}

In this appendix we derive an analytical condition
for the existence of the two-dimensional flat band based
on the winding number 
\begin{equation}
\label{eq:WN_schnyder_4}
w(k_x,k_z)=\frac{1}{2\pi}\textrm{Im}\int_{-\pi}^{\pi}\, dk_y\, \left( \textrm{det} D(\mathbf k) \right)^{-1} \partial_{k_y} \textrm{det} D(\mathbf k).
\end{equation}
To evaluate this expression analytically it is useful to
employ the residue theorem in the following way \cite{Gerber}:
for given $k_x$ and $k_z$ the quantity
\begin{equation}
\label{eq:path}
\gamma(\mathbf k)=\textrm{det} D(\mathbf k)
= m_0^2 + m_1^2 + m_2^2 + m_3^2 -V_x^2 -V_y^2 + 2 i m_1 V_y -2 i m_2 V_x
\end{equation}
for $k_y \in [-\pi,\pi]$ defines a closed path in the complex plane.
Therefore, by transforming $z=\textrm{det} D(\mathbf k)$ 
the winding number can be expressed as the integral
\begin{equation}
\label{eq:WN_schnyder_5}
w(k_x,k_z)=\frac{1}{2\pi}\textrm{Im}\oint_\gamma\, \frac{dz}{z} 
= I(\gamma,0).
\end{equation}
where $I(\gamma,0)$ is the winding number of the curve $\gamma$ around
the origin. Thus, if the curve does not enclose the origin, the
winding number becomes zero and no zero energy surface state exists. 
Now, the path $\gamma$ crosses the
real axis, when $\textrm{Im} \; \textrm{det} D(\mathbf k)=0$. This
happens, when
\begin{equation}
\label{eq:crossing_cond}
m_1 V_y = m_2 V_x \qquad \Longleftrightarrow \qquad \sin k_y = \frac{V_y}{V_x} \sin k_x
\end{equation}
For a given $k_x$ this equation has two solutions $k_{y,1}$ and $k_{y,2}=\pi
-k_{y,1}$, if $\frac{V_y}{V_x} \sin k_x \in (-1,1)$. The path encloses
the origin only, if these two crossings possess different sign
of the real part, i.e. the criterion for the existence of the
flat band becomes
\begin{equation}
\label{eq:flatband_cond}
0 > \gamma(k_x,k_{y,1},k_z) \; \gamma(k_x,k_{y,2},k_z) .
\end{equation}
The boundary of the flat band in $k_x$-$k_z$-space is then given
by the criterion
\begin{equation}
\label{eq:flatband_bondary_cond}
0 = \gamma(k_x,k_{y,1},k_z) \; \gamma(k_x,k_{y,2},k_z) .
\end{equation}
Now, as on the Fermi surface we have $\textrm{det} D(\mathbf k)=0$,
it follows that the boundary of the flat band is just the projection
of the (one-dimensional) Fermi surface onto the $k_x$-$k_z$-plane.

In the general case the analytical expression Eq.~(\ref{eq:flatband_cond})
becomes quite complicated. However, the expression can be simplified
in the case $V_y=0$, as then $k_{y,1}=0$ and $k_{y,2}=\pi$.
Then we have
\begin{eqnarray}
\gamma(k_x,0,k_z) &= \tilde m_0(\mathbf k)^2 + 4 A_2^2 \sin^2 k_x + 4 A_1^2 \sin^2 k_z - V_x^2\\
\gamma(k_x,\pi,k_z) &= \left[ \tilde m_0(\mathbf k) - 4 B_2 \right]^2 
 + 4 A_2^2 \sin^2 k_x + 4 A_1^2 \sin^2 k_z - V_x^2 \nonumber
\end{eqnarray}
where $\tilde m_0(\mathbf k)=M - 2 B_2 ( 1- \cos k_x ) - 2 B_1 ( 1- \cos k_z
)$. If we look at $k_x=k_z=0$, we see that $\gamma(0,0,0)$ becomes
negative, when $|V_x|>M$. $\gamma(0,\pi,0)$ becomes negative, when
$|V_x|>4B_2-M$. As we assume $B_2>M$ here, $\gamma(0,\pi,0)$ will
change sign only at much larger fields $|V_x|$. Therefore, a flat
band will be present in the vicinity of $k_x=k_z=0$ for
$4B_2-M>|V_x|>M$.


\section*{References}
\begin{thebibliography}{99}

\bibitem{Bernevig}
B.~A.~Bernevig, T.~L.~Hughes, and S.-C.~Zhang, \href{http://dx.doi.org/10.1126/science.1133734}{Science {\bf 314}, 1757 (2006)}.

\bibitem{Fu}
L.~Fu, C.~L.~Kane and E.~J.~Mele, \href{http://link.aps.org/doi/10.1103/PhysRevLett.98.106804}{Phys. Rev. Lett. {\bf 98}, 106803 (2007)}.

\bibitem{Koenig}
M.~K\"onig, S.~Wiedmann, C.~Br\"une, A.~Roth, H.~Buhmann,
L.W.~Molenkamp, X.-L.~Qi, and S.-C.~Zhang, \href{http://dx.doi.org/10.1126/science.1148047}{Science {\bf 318}, 766 (2007)}.

\bibitem{Hsieh1}
D.~Hsieh, D.~Qian, L.~Wray, Y.~Xia, Y.~Hor, R.~J.~Cava, and M.~Z.~Hasan,
\href{http://dx.doi.org/10.1038/nature06843}{Nature (London) {\bf 452}, 970 (2008)}.

\bibitem{Chen}
Y.~L.~Chen, J~G.~Analytis, J.-H.~Chu, Z.~K.~Liu, S.-K.~Mo, X.~L.~Qi,
H.~J.~Zhang, D.~H.~Lu, X.~Dai, Z.~Fang, S.~C.~Zhang, I.~R.~Fisher, Z.~Hussain,
and Z.-X.~Shen, \href{http://dx.doi.org/10.1126/science.1173034}{Science {\bf 325}, 178 (2009)}.

\bibitem{Xia}
Y.~Xia, D.~Qian, D.~Hsieh, L. Wray, A.~Pal, H.~Lin, A.~Bansil, D.~Grauer,
Y.~S.~Hor, R.~J.~Cava, and M.~Z.~Hasan,  \href{http://dx.doi.org/10.1038/nphys1274}{Nature Phys. {\bf 5}, 398 (2009)}.

\bibitem{Hsieh2}
D.~Hsieh, Y.~Xia, D.~Qian, L.~Wray, F.~Meier, J.~H.~Dil, J.~Osterwalder,
L.~Patthey, A.~V.~Fedorov, H.~Lin, A.~Bansil, D.~Grauer, Y.~S.~Hor,
R.~J.~Cava, and M.~Z.~Hasan, \href{http://link.aps.org/doi/10.1103/PhysRevLett.103.146401}{Phys. Rev. Lett. {\bf 103}, 146401 (2009)}.

\bibitem{Kuroda}
K.~Kuroda et al, \href{http://link.aps.org/doi/10.1103/PhysRevLett.108.206803}{Phys. Rev. Lett. {\bf 108}, 206803 (2012)}. 

\bibitem{Chadov}
S.~Chadov, X.~Qi, J.~K\"ubler, G.~H.~Fecher, C.~Felser, and
S.~C.~Zhang, \href{http://dx.doi.org/10.1038/nmat2770}{Nature Mat. {\bf 9}, 541 (2010)}. 

\bibitem{Lin}
H.~Lin, L.~A.~Wray, Y.~Xia, S.~Xu, S.~Jia, R.~J.~Cava, A.~Bansil, and
M.~Z.~Hasan, \href{http://dx.doi.org/10.1038/nmat2771}{Nature Mat. {\bf 9}, 546 (2010)}. 

\bibitem{Hasan}
M.~Z.~Hasan and C.~L.~Kane, \href{http://link.aps.org/doi/10.1103/RevModPhys.82.3045}{Rev. Mod. Phys. {\bf 82}, 3045 (2010)}.

\bibitem{Ando}
Y.~Ando, \href{http://dx.doi.org/10.1143/JPSJ.82.102001}{J. Phys. Soc. Japan {\bf 82}, 102001 (2013)}.

\bibitem{Garate}
I. Garate and M. Franz, \href{http://link.aps.org/doi/10.1103/PhysRevLett.104.146802}{Phys. Rev. Lett. {\bf 104}, 146802 (2010)}. 

\bibitem{Yokoyama}
T.~Yokoyama , Y.~Tanaka, and N.~Nagaosa, \href{http://link.aps.org/doi/10.1103/PhysRevB.81.121401}{Phys. Rev. B {\bf 81}, 121401(R) (2010)}.
 
\bibitem{Yu:Science10}
R.~Yu, W.~Zhang, H.-J.~Zhang, S.-C.~Zhang, X.~Dai, and Z.~Fang, 
\href{http://dx.doi.org/10.1126/science.1187485}{Science {\bf 329},  61  (2010)}.

\bibitem{Chen2}
Y.~L.~Chen et al,
\href{http://dx.doi.org/10.1126/science.1189924}{Science {\bf 329},  659  (2010)}.

\bibitem{Hor}
Y.~S.~Hor, et al, \href{http://link.aps.org/doi/10.1103/PhysRevB.81.195203}{Phys. Rev. B {\bf 81}, 195203 (2010)}.

\bibitem{Moodera}
P.~Wei, F.~Katmis, B.~A.~Assaf, H.~Steinberg, P.~Jarillo-Herrero, D.~Heiman,
and J.~S.~Moodera, \href{http://link.aps.org/doi/10.1103/PhysRevLett.110.186807}{Phys. Rev. Lett. {\bf 110}, 186807 (2013)}. 

\bibitem{BlackSchaffer}
A.M. Black-Schaffer and J. Linder, \href{http://link.aps.org/doi/10.1103/PhysRevB.84.180509}{Phys. Rev. B {\bf 84}, 180509(R) (2011)}.

\bibitem{RLChu}
R.~L.~Chu, J.~Shi, and S.-Q.~Shen, \href{http://link.aps.org/doi/10.1103/PhysRevB.84.085312}{Phys. Rev. B {\bf 84}, 085312 (2011)}.
 
\bibitem{Honolka}
J. Honolka et al., \href{http://link.aps.org/doi/10.1103/PhysRevLett.108.256811}{Phys. Rev. Lett. {\bf 108}, 256811 (2012)}.

\bibitem{Scholz}
M.~R.~Scholz, J.~S\'{a}nchez-Barriga, D.~Marchenko, A.~Varykhalov, 
A.~Volykhov, L.V.~Yashina, and O.~Rader, 
\href{http://link.aps.org/doi/10.1103/PhysRevLett.108.256810}{Phys. Rev. Lett. {\bf 108}, 256810 (2012)}.

\bibitem{Lunde}
A.~M.~Lunde and G.~Platero, \href{http://link.aps.org/doi/10.1103/PhysRevB.88.115411}{Phys. Rev. B {\bf 88}, 115411 (2013)}.

\bibitem{PD}
T.~Paananen and T.~Dahm, \href{http://link.aps.org/doi/10.1103/PhysRevB.87.195447}{Phys. Rev. B {\bf 87}, 195447 (2013)}.
 
\bibitem{Nakada}
K.~Nakada, M.~Fujita, G.~Dresselhaus and M.~S.~Dresselhaus, \href{http://link.aps.org/doi/10.1103/PhysRevB.54.17954}{Phys. Rev. B {\bf 54}, 17954 (1996)}. 

\bibitem{Machida}
T.~Mizushima, M.~Sato, and K.~Machida, \href{http://link.aps.org/doi/10.1103/PhysRevLett.109.165301}{Phys. Rev. Lett. {\bf 109}, 165301 (2012)}. 

\bibitem{Silaev} M.~A.~Silaev and G.~E.~Volovik,
\href{http://link.aps.org/doi/10.1103/PhysRevB.86.214511}{Phys. Rev. B {\bf 86}, 214511 (2012)}.

\bibitem{SchnyderTimmPRL}
A.~P.~Schnyder, C.~Timm, and P.~M.~R.~Brydon, \href{http://link.aps.org/doi/10.1103/PhysRevLett.111.077001}{Phys. Rev. Lett. {\bf 111}, 077001 (2013)}. 

\bibitem{BrydonNJP}
P.~M.~R.~Brydon, C.~Timm, and A.~P.~Schnyder, \href{http://dx.doi.org/10.1088/1367-2630/15/4/045019}{New J. Phys. {\bf 15}, 045019 (2013)}. 

\bibitem{PALee} C.~L.~M.~Wong, J.~Liu, K.~T.~Law, and P.~A.~Lee,
\href{http://link.aps.org/doi/10.1103/PhysRevB.88.060504}{Phys. Rev. B {\bf 88}, 060504 (2013)}.

\bibitem{Tewari} J.~D.~Sau and S.~Tewari,
\href{http://link.aps.org/doi/10.1103/PhysRevB.86.104509}{Phys. Rev. B {\bf 86}, 104509 (2012)}.

\bibitem{Lau} A.~Lau and C.~Timm, \href{http://de.arxiv.org/abs/1305.1770}{arXiv:1305.1770 (2013)}.

\bibitem{Hu}  
C.-R.~Hu, \href{http://link.aps.org/doi/10.1103/PhysRevLett.72.1526}{Phys. Rev. Lett. {\bf 72}, 1526 (1994)}.

\bibitem{Tanaka} Y. Tanaka and S. Kashiwaya, \href{http://link.aps.org/doi/10.1103/PhysRevLett.74.3451}{Phys. Rev. Lett. {\bf 74}, 3451 (1995)}.

\bibitem{Kashiwaya} S. Kashiwaya and Y. Tanaka, \href{http://link.aps.org/doi/10.1088/0034-4885/63/10/202}{Rep. Prog. Phys. {\bf 63}, 1641 (2000)}.

\bibitem{Golovik}
G.~E.~Volovik, \href{http://dx.doi.org/10.1134/S0021364011020147}{JETP Lett. {\bf 93}, 66 (2011)}.  

\bibitem{Volovik} T.~T.~Heikkil\"a, N.~B.~Kopnin, and G.~E.~Volovik, 
\href{http://dx.doi.org/10.1134/S0021364011150045}{JETP Lett. {\bf 94},
  233 (2011)};
G.~E.~Volovik,
\href{http://dx.doi.org/10.1007/s10948-013-2221-5}{J.~Supercond.~Nov.~Magn. 
{\bf 26}, 2887 (2013)}.

\bibitem{RyuHatsugai} S.~Ryu and Y.~Hatsugai,
\href{http://link.aps.org/doi/10.1103/PhysRevLett.89.077002}{{Phys. Rev. Lett.} {\bf 89}, 077002 (2002)}.

\bibitem{Sato} M.~Sato, \href{http://link.aps.org/doi/10.1103/PhysRevB.73.214502}{Phys. Rev. B {\bf 73}, 214502 (2006)}.

\bibitem{Fogelstroem} M. Fogelstr\"om, D. Rainer, and J.A. Sauls, \href{http://link.aps.org/doi/10.1103/PhysRevLett.79.281}{Phys. Rev. Lett. {\bf 79}, 281 (1997)}.
\bibitem{Walter} H.~Walter, W.~Prusseit, R.~Semerad, H.~Kinder, W.~Assmann,
  H.~Huber, H.~Burkhardt, D.~Rainer, and J.~A.~Sauls, \href{http://link.aps.org/doi/10.1103/PhysRevLett.80.3598}{Phys. Rev. Lett. {\bf 80}, 3598 (1998)}.

\bibitem{Aprili} M.~Aprili, E.~Badica, and L.~H.~Greene,
\href{http://link.aps.org/doi/10.1103/PhysRevLett.83.4630}{{Phys.~Rev.~Lett.} {\bf 83}, 4630 (1999)}.

\bibitem{Krupke} R.~Krupke and G.~Deutscher,
\href{http://link.aps.org/doi/10.1103/PhysRevLett.83.4634}{{Phys.~Rev.~Lett.} {\bf 83}, 4634 (1999)}.

\bibitem{Chesca1} B.~Chesca, M.~Seifried, T.~Dahm, N.~Schopohl, D.~Koelle, 
          R.~Kleiner, A.~Tsukada, \href{http://link.aps.org/doi/10.1103/PhysRevB.71.104504}{Phys. Rev. B {\bf 71}, 104504 (2005)}.

\bibitem{Chesca2} B.~Chesca, D.~D\"onitz, T.~Dahm, R.~P.~Huebener, D.~Koelle,
          R.~Kleiner, Ariando, H.-J.~H. Smilde, H.~Hilgenkamp,
\href{http://link.aps.org/doi/10.1103/PhysRevB.73.014529}{Phys. Rev. B {\bf 73}, 014529 (2006)}.

\bibitem{Iniotakis} C.~Iniotakis, T.~Dahm, and N.~Schopohl, \href{http://link.aps.org/doi/10.1103/PhysRevLett.100.037002}{Phys. Rev. Lett. {\bf 100},
  037002 (2008)}.

\bibitem{Graser} S.~Graser, C.~Iniotakis, T.~Dahm, and N.~Schopohl, 
\href{http://link.aps.org/doi/10.1103/PhysRevLett.93.247001}{Phys. Rev. Lett. {\bf 93}, 247001 (2004)}.

\bibitem{Barash} Yu.~S.~Barash, M.~S.~Kalenkov, and J.~Kurkij\"arvi, \href{http://link.aps.org/doi/10.1103/PhysRevB.62.6665}{Phys. Rev.
B {\bf 62}, 6665 (2000)}.

\bibitem{Zare} A.~Zare, T.~Dahm, and N.~Schopohl, \href{http://link.aps.org/doi/10.1103/PhysRevLett.104.237001}{Phys. Rev. Lett. {\bf 104},
  237001 (2010)}.

\bibitem{Zhuvarel} A.~P.~Zhuravel, B.~G.~Ghamsari, C.~Kurter, P.~Jung,
  S.~Remillard, J.~Abrahams, A.~V.~Lukashenko, A.~V.~Ustinov, and
  S.~M.~Anlage, \href{http://link.aps.org/doi/10.1103/PhysRevLett.110.087002}{Phys. Rev. Lett. {\bf 110}, 087002 (2013)}.

\bibitem{Matsuura}
S.~Matsuura, P.-Y.~Chang, A.~P.~Schnyder, and S.~Ryu, \href{http://dx.doi.org/10.1088/1367-2630/15/6/065001}{New J. Phys. {\bf 15}, 065001 (2013)}.

\bibitem{Wan}
X.~Wan, A.~M.~Turner, A.~Vishwanath, and S.~Y.~Savrasov, \href{http://link.aps.org/doi/10.1103/PhysRevB.83.205101}{Phys. Rev. B {\bf 83},
205101 (2011)}.

\bibitem{Balents}
L.~Balents, \href{http://link.aps.org/doi/10.1103/Physics.4.36}{Physics {\bf 4}, 36 (2011)}.

\bibitem{Burkov}
A.~A.~Burkov and L.~Balents, \href{http://link.aps.org/doi/10.1103/PhysRevLett.107.127205}{Phys.~Rev.~Lett. {\bf 107}, 127205 (2011)}.

\bibitem{Li:NPhys10}
R.~Li, J.~Wang, X.-L.~Qi, and S.-C.~Zhang, \href{http://dx.doi.org/10.1038/nphys1534}{Nature Phys. {\bf 6},  284  (2010)}.

\bibitem{Zhang:NPhys09}
H.~Zhang, C.-X.~Liu, X.-L.~Qi, X.~Dai, Z.~Fang, and S.-C.~Zhang, 
\href{http://dx.doi.org/10.1038/nphys1270}{Nature Phys. {\bf 5},  438  (2009)}.

\bibitem{Liu:PRB10}
C.-X.~Liu, X.-L.~Qi, H.~Zhang, X.~Dai, Z.~Fang, and S.-C.~Zhang,
\href{http://link.aps.org/doi/10.1103/PhysRevB.82.045122}{Phys. Rev. B {\bf 82}, 045122 (2010)}.

\bibitem{Shan}
W.-Y.~Shan, H.-Z.~Lu, and S.-Q.~Shen, \href{http://dx.doi.org/10.1088/1367-2630/12/4/043048}{New J. Phys. {\bf 12}, 043048 (2010)}.

\bibitem{Hao:PRB11}
L.~Hao and T.~K.~Lee,
\href{http://link.aps.org/doi/10.1103/PhysRevB.83.134516}{Phys. Rev. B {\bf 83}, 134516 (2011)}.

\bibitem{Brouwer}
P.~G.~Silvestrov, P.~W.~Brouwer, and E.~G.~Mishchenko, \href{http://link.aps.org/doi/10.1103/PhysRevB.86.075302}{Phys. Rev. B {\bf 86},
075302 (2012)}.

\bibitem{Schnyderpt}
A.~P.~Schnyder, S.~Ryu, A.~Furusaki and A.W.W.~Ludwig, \href{http://link.aps.org/doi/10.1103/PhysRevB.78.195125}{Phys. Rev. B {\bf 78}, 195125 (2008)}.

\bibitem{Ryu}
S.~Ryu, A.~P.~Schnyder, A.~Furusaki and A.~Ludwig,
\href{http://dx.doi.org/10.1088/1367-2630/12/6/065010}{New J. Phys. {\bf 12}, 065010 (2010)}.

\bibitem{Altland} A.~Altland, B.~D.~Simons, and Z.~R.~Zirnbauer, 
\href{http://dx.doi.org/10.1016/S0370-1573(01)00065-5}{Phys. Rep. {\bf 359}, 283 (2002)}.

\bibitem{Schnyder} A.~P.~Schnyder and S.~Ryu,
\href{http://link.aps.org/doi/10.1103/PhysRevB.84.060504}{{Phys. Rev. B} {\bf 84}, 060504(R) (2011)}; P.~M.~R.~Brydon,
A.P.~Schnyder, and C.~Timm, \href{http://link.aps.org/doi/10.1103/PhysRevB.84.020501}{{Phys. Rev. B} {\bf 84}, 020501(R) (2011)};
A.P.~Schnyder, P.~M.~R.~Brydon, and C.~Timm, \href{http://link.aps.org/doi/10.1103/PhysRevB.85.024522}{{Phys. Rev. B} {\bf 85}, 024522 (2012)}.

\bibitem{Bruene}
C.~Br\"une, A.~Roth, H.~Buhmann, E.~M.~Hankiewicz, L.~W.~Molenkamp,
J.~Maciejko, X.-L.~Qi, and S.-C.~Zhang,
\href{http://dx.doi.org/10.1038/nphys2322}{Nature Phys. {\bf 8},  485  (2012)}.

\bibitem{Murakami}
S.~Murakami, \href{http://dx.doi.org/10.1088/1367-2630/9/9/356}{New
  J. Phys. {\bf 9}, 356 (2007)}.

\bibitem{Gerber} H.~Gerber, Bachelor thesis, Bielefeld University (2013).

\end{thebibliography}
\end{document}